# Machine learned Force-Fields for an ab-initio Quality Description of Metal-Organic Frameworks

Sandro Wieser and Egbert Zojer

*Institute of Solid State Physics, Graz University of Technology, NAWI Graz, Petersgasse 16, 8010 Graz, Austria*

## Abstract

Metal-organic frameworks (MOFs) are an incredibly diverse group of highly porous hybrid materials, which are interesting for a wide range of possible applications. For a reliable description of many of their properties accurate computationally highly efficient methods, like force-field potentials (FFPs), are required. With the advent of machine learning approaches, it is now possible to generate such potentials with relatively little human effort. Here, we present a recipe to parametrize two fundamentally different types of exceptionally accurate and computationally highly efficient machine learned potentials, which belong to the moment-tensor and kernel-based potential families. They are parametrized relying on reference configurations generated in the course of molecular dynamics based, active learning runs and their performance is benchmarked for a representative selection of commonly studied MOFs. For both potentials, comparison to a random set of validation structures reveals close to DFT precision in predicted forces and structural parameters of all MOFs. Essentially the same applies to elastic constants and phonon band structures. Additionally, for MOF-5 the thermal conductivity is obtained with full quantitative agreement to single-crystal experiments. All this is possible while maintaining a high degree of computational efficiency, with the obtained machine learned potentials being only moderately slower than the extremely simple UFF4MOF or Dreiding force fields. The exceptional accuracy of the presented FFPs combined with their computational efficiency has the potential of lifting the computational modelling of MOFs to the next level.

# Introduction

Metal-organic frameworks (MOFs) are highly porous hybrid materials consisting of inorganic metal oxide nodes connected by organic linkers. Since their first realization, a wide range of potential applications has been discussed including catalysis[1–3], gas storage and separation[4–6], electronic devices[7–9], as well as drug encapsulation and delivery[10,11]. In view of the seemingly limitless number of possible MOFs, finding the material with properties best suited for a specific application is a sizable challenge. The situation is further complicated by the frequently encountered difficulties when trying to measure the relevant materials parameters, e.g., due to often small crystallite sizes or the inclusion of foreign molecules in the MOF pores. Therefore, computational methods provide an excellent tool to not only predict properties but also to support the development of dependable structure-to-property-relationships. Due to the large number of atoms in MOF unit cells compared to conventional crystalline materials, ab-initio methods like density-functional theory (DFT) are frequently too computationally expensive for that task. This is particularly true when modelling properties at elevated temperatures or for properties that depend on the dynamics of the MOF constituents, like thermal expansion or thermal conductivity. Then it is necessary to resort to lower levels of theory, like classical force field potentials (FFPs), which can speed up calculations by many orders of magnitude[12–16].

The most commonly applied and most easy to use force field potentials are transferable force-fields, which can be straightforwardly applied to modelling the majority of the elements in the periodic table. These include the Dreiding[17] force field or the universal force field (UFF)[18], where for the latter a variant exists that is specifically adapted for the description of MOFs (UFF4MOF)[19–21]. Especially UFF4MOF has been frequently used in literature for modelling MOFs[22–25], as it is readily available and particularly convenient for rapid structure prediction[20]. The highly transferable potentials are, however, not designed for an accurate description of dynamical properties[20] and usually result in sizable errors. In the case of MOFs, poor agreement between UFF4MOF-based simulations and DFT-based methods were explicitly demonstrated, for example, for the elastic properties of MIL-53[26] and for the thermal conductivity of MOF-5. In the latter case, UFF4MOF overestimated the single-crystal derived experimental data by a factor of 2.6[27]. More advanced potentials focus on a more accurate description of a particular materials class, like GAFF[28] or COMPASS[29] for organic molecules or BTW-FF[13] for MOFs. However, even these potentials with reduced transferability tend to have difficulties in accurately predicting vibrational properties[30], which crucially influence many materials parameters and which are notoriously challenging to predict[31].

A higher degree of accuracy (at the prize of a further reduced transferability) is observed for potentials individually parameterized for specific molecular fragments. This, for example, applies to part of the MOF-FF family of potentials, which are parameterized for specific organic and inorganic MOF building blocks[12,32]. Potentials like MOF-FF[12,27,33] or QuickFF[15] can also be parameterized fully system-specifically giving up on transferability to achieve an even higher degree of precision. The functional form of these force fields is designed relying on chemical intuition, for example, distinguishing between bonding and non-bonding interactions, which is not always ideal for MOFs, in which the dynamic adsorption and desorption of guest molecules can be important. This triggered the development of more flexible potentials like ReaxFF[34], but often at the prize of an overall reduced accuracy combined with problems regarding energy conservation and thermal stability[35]. Another (major) disadvantage that we experienced when system-specifically parametrizing classical force fields like MOF-FF was that for achieving the required level of accuracy (compared to DFT and experimental data) a careful tailoring of the chosen set of potential terms, reference data and fitting algorithms is often crucial[27,33], which makes the parametrization process extremely cumbersome.

With increasing computational power and advances in machine learning approaches, more convenient to use and often more accurate machine learned potentials (MLPs) emerged. In recent years, they started to see widespread use for successfully describing many materials[36–39]. So far, for MOFs MLP approaches have been largely limited to neural network potentials, which already show a promising performance in terms of precision[26,40–45]. They were employed to reliably describe computationally expensive quantities like the thermal conductivity of several MOFs[44]. However, many of the said investigations required a relatively large number of DFT reference data for training[40,45] and the neural network potentials are still computationally less efficient than traditional FFPs when tested on the same computer architecture[40,46]. In this context it should be noted that some of the implementations of these potentials like GPUMD[47] and DeePMD-kit[48] feature parallelization on graphical processing units, making them also an attractive option for large simulations[44,47]. Other types of machine learned potentials like Gaussian approximation potentials[49] or moment tensor potentials (MTPs)[50] have seen widespread use for conventional materials[51–53], but their performance has, to the best of our knowledge, not yet been evaluated systematically for MOFs. Importantly, the sizable amount of DFT reference data required for machine learning based approaches can be cut down significantly by either only training molecular fragments of a system[40,41,54] or by employing strategies to more efficiently sample phase space[26,55].

## The Computational Approach

A particularly promising strategy for such an efficient sampling are active learning approaches, which, for example, have recently been implemented in conjunction with a kernel-based MLP in the Vienna ab-initio Simulation Package (VASP) and which manage to severely cut down the computational effort of generating reference data. So far the application of active learning approaches to MOFs has been limited[26] or the training still demanded a rather high number of DFT calculations[45]. The active learning approach in VASP trains a kernel-based machine learned potential (similar to a Gaussian approximation potential) during a molecular dynamics (MD) run. The potential is built on a basis set depending on local reference configurations (atoms with their surrounding neighborhoods), which is dynamically expanded during the simulation[55–57]. In short, this is done via a Bayesian error estimation, which decides whether or not the concurrent force field is accurate enough for performing the next molecular dynamics time step or whether that step requires DFT-calculated forces. The DFT calculated structures and their properties are continuously added to the reference data set and the MLP is expanded. This approach has already been successfully employed to a range of materials[55–57]. However, Gaussian approximation potentials are known to be relatively slow compared to other options[46]. Smart settings for the inclusion of only relevant basis function can substantially increase the speed of the potentials, but even then, other types of potentials usually tend to be computationally significantly more efficient. In recent implementations of VASP (version 6.4.0 and upward) the computational efficiency of the kernel-based potentials has, however, been improved further, making them a highly competitive option for describing even materials as complex as MOFs.

As a complementary approach, the aforementioned moment tensor potentials have been shown to lead to a substantially improved accuracy-cost ratio in force-field based simulations[46]. For them, the number of terms in the potential does not depend on the size of the reference data set, but rather is predetermined before the training, allowing to adjust the accuracy and speed of the MTP[50,58]. They are, however, not as compatible with the above-described active-learning strategy, as their continuous reparameterization would be very time consuming.

In the presented approach, we will begin by generating reference data sets using the active learning approach available in VASP. This will directly lead to the creation of kernel-based machine learned potentials (VASP MLPs). To test the impact of the improvements of more recent versions of the VASP code and to investigate how one can scale the performance of the VASP MLPs, we will subsequently retrain the potentials using various settings. These potentials will then be analyzed for their accuracy and computational efficiency. On the same reference data generated via the efficient VASP active learning, also MTPs will be trained as a computationally particularly efficient and also highly accurate alternative to the VASP MLPs. Our main aim is to provide a recipe to train highly accurate and computational efficient machine learned potentials by using a reasonable amount of reference data. To make this approach as widely applicable as possible, we will exclusively use tools that are publicly available and that can be used out-of-the-box with little human development effort required. We will test both, the VASP MLP and the MTP approaches on several MOFs extensively described in literature and perform rigorous benchmarking based on a number of properties like unit-cell parameters, energies, forces, stresses on strained cells, elastic constants, phonons, thermal expansion coefficients, and thermal conductivities. A major focus will be on the description of phonons, where we will not limit ourselves just to the Γ-point but will rather study phonons in the entire first Brillouin zone. This is important, as many phonon based properties, like thermodynamic stability[59], as well as heat and charge-transport properties crucially depend on off-Γ phonons[60]. The benchmarking will be done primarily via a comparison to reference data generated by DFT and only secondarily via a comparison to experimental data (e.g., when a quantity of interest is not accessible to DFT due to its computational complexity). The reason for that is that the goal of the present study is to assess the quality of the parametrization of the potentials and the suitability of their functional form. A comparison to experiments would additionally - and potentially even primarily - depend on the performance of the specific DFT methodology used in the simulations. A general DFT benchmarking for MOFs is, however, not in the focus of the present study. In fact, if alternative DFT-based approaches (relying, e.g., on different functionals or van der Waals correction) were found to be more suitable for describing certain quantities, one could easily adopt them in the suggested parametrization process, as long as they are efficient enough for the active learning strategy. Other complications with experimental data are that (i) for many of the benchmarked quantities they do not exist and (ii) that at realistic conditions MOF-based materials contain an often-unknown concentration of guest molecules and/or defects. Moreover, they typically consist of crystallites of limited size. This frequently makes the experimental determination of truly intrinsic materials properties of pristine MOF materials highly challenging.

To put the obtained agreement between the machine-learned potentials and the DFT data into perspective, for selected systems we will also include a comparison to results obtained with UFF4MOF[18–20], as an example for an off-the-shelf transferrable force fields widely used for modelling MOFs.

## Materials of Interest

The systems of interest for the present investigation share two features: first, they are widely studied due to their high thermal stability and/or promising properties and, secondly, their unit cells are small enough that they are accessible to DFT simulations with properly converged numerical settings on contemporary supercomputers. Moreover, the chosen MOFs fundamentally differ in their topologies, shapes, dimensionalities, and flexibilities of the pores, and in the nature of the metal ions. Regarding the latter, we focus on closed-shell metal ions to avoid spin-order phenomena as an additional complication. The structures of the systems are shown in Fig. 1.

The first system is MOF-5[61] consisting of $Zn_4O$ nodes and 1,4-benzene-dicarboxylate (bdc) linkers. It forms a stable face-centered cubic structure with space group Fm-3m (225) containing 106 atoms in its primitive unit cell. Being one of the first published MOFs[61], it has been intensively studied in the past, and is also included here as it is one of the very few MOFs for which reliable, single crystal thermal conductivity measurement are available[62]. The second material is UiO-66[63]. It consists of 12-coordinated ZrO based nodes connected by bdc linkers. Despite its different topology, the system still shows the same general symmetry as MOF-5 with space group Fm-3m (225). UiO-66 is one of the most commonly investigated MOFs[64] primarily due to its exceptionally high thermal stability[65]. Another system of interest is MOF-74[66], consisting of 1D extended metal-oxide pillar nodes connected by dobdc (2,5-dioxido-1,4-benzenedicarboxylate) linkers. They are aligned such that a honeycomb structure of hexagonally shaped 1D-extended pores is formed. This leads to a rhombohedral crystal structure with space group R-3 (148) with 54 atoms in the primitive unit cell. The system is known for its exceptional $CO_2$ uptake[67]. MOF-74 exists for various metals including Ca, Mg, Zn, Mn, Fe, Co, Cu and combinations thereof[68]. Here, we picked the prominently investigated Zn variant. MOF-74 is particularly interesting due to its pronounced anisotropy. The fourth and final system is MIL-53[69], a MOF consisting of metal-oxide node pillars in an octahedral arrangement connected by bdc linkers in the perpendicular direction forming a "wine rack" shaped structure. This MOF is known to occur in two primary phases: a low temperature "narrow" pore phase and a high temperature "large" pore phase[70]. The large pore phase is orthorhombic with space group Imma (74), while the narrow pore phase is a monoclinic system with space group Cc (9) [69]. For the purpose of this work, we will consider both phases as separate systems, referring to them as MIL-53 (lp) and MIL-53 (np). MIL-53 has been synthesized for a wide range of different metals. Here, we study the Al version of MIL-53, as one of the most commonly investigated variants[71].

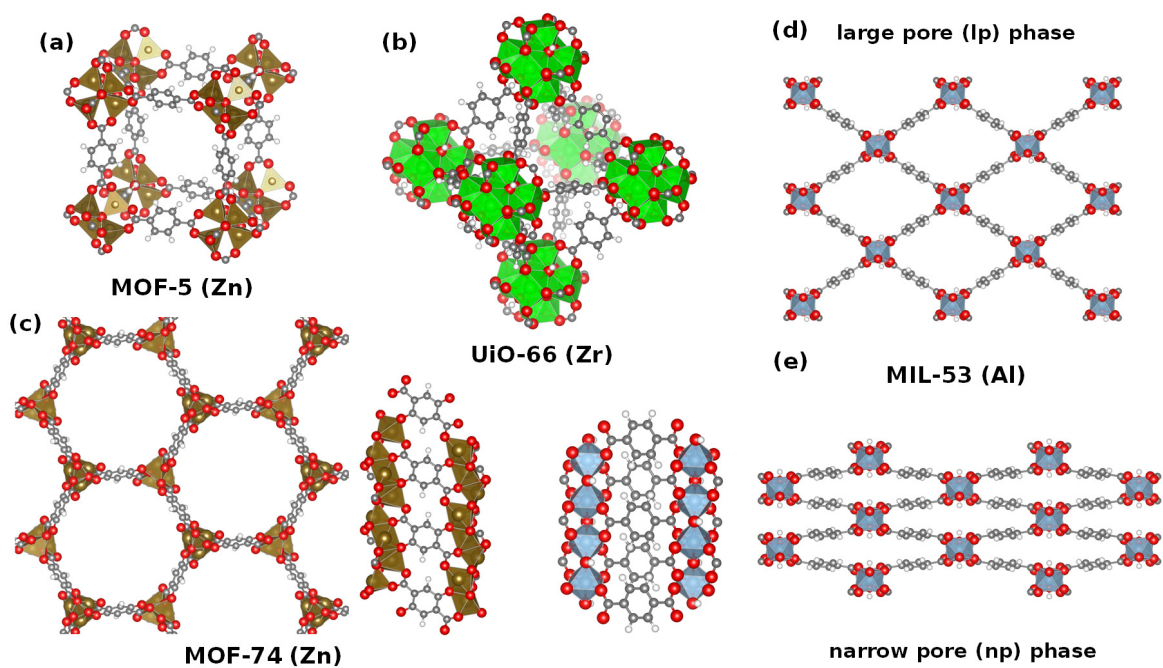

**Fig. 1: Atomic structures of the investigated metal-organic frameworks.** The following systems are considered: **a** MOF-5 (Zn), **b** UiO-66 (Zr), **c** MOF-74 (Zn), **d** large and **e** narrow pore phase of MIL-53 (Al). For the anisotropic systems MOF-74 and MIL-53 (np) side- and top-views are provided. The structures were visualized using the VESTA package[72]. Color coding: Zn: brown, Zr: green, Al: blue, C: grey, H: white, O: red.

# Results

## Learning the Force Field Potentials

The reference data required for the training of the machine learned potentials were obtained using forces, stresses and energies calculated by density functional theory (DFT), as described in the method section. The necessary reference structures were generated employing the on-the-fly machine-learning force field methodology as implemented in VASP. This approach is described in depth in ref. [55,73] and the details of the procedure applied here (e.g., regarding the chosen temperature intervals and simulation length) were optimized in prior tests based on MOF-5[61]. In short, the 0 K optimized structures of the respective systems containing between 54 and 114 atoms in the unit cell were used as starting geometries. A molecular dynamics (MD) simulation was initiated using a Langevin barostat[74] at zero pressure and a Langevin thermostat[75] starting at 50 K and heating the system up to 900 K over the course of 50,000 time steps of 0.5 fs. Despite the temperature of interest being around room temperatures, we chose the rather high maximum temperature of 900 K, because considering also the larger atomic displacement amplitudes at higher temperatures is beneficial for machine learned potentials. This a consequence of such potentials performing well in an interpolation regime, while showing a poor ability to extrapolate to unknown situations[58]. VASP decides whether any given time step is calculated using DFT, based on whether or not the estimated Bayesian error is above a certain threshold. The threshold is adjusted dynamically over the course of the simulation, such that it steadily increases with temperature as the atomic forces become larger. This is required when training with a temperature gradient, as a fixed threshold would add too few configurations at low temperatures and too many configurations at high temperatures. Using this approach, between 739 and 998 reference configurations were computed with DFT for each system in the initial training, as summarized in Table 1. This table also lists the Bayesian error threshold at the end of the training. It is lower in MOF-5 compared to the other systems, which indicates a somewhat higher degree of accuracy when describing the forces for this system. The number of reference configurations is in a similar range as for other accurate neural network potentials[26,44] for MOFs that employ an incremental learning approach[26] or after the selection of training configurations from an ab-initio MD pre-screening with less accurate DFT settings[44].

**Table 1: Total number of generated reference structures for the on-the-fly training of the VASP machine learned potentials.** The numbers are given for each system after the initial training run (heating of the system from 50 to 900 K over 25 ps), $\#_{references}^{initial}$, and after the extension of the reference data set (training at constant temperature of 300 K over 50 ps with a fixed Bayesian error threshold of 0.02 eV/Å), $\#_{references}^{final}$; additionally, the numbers of atoms in the unit cell used for the training, $\#_{atoms}^{UC}$, and the last Bayesian error threshold in the initial training, $error_{Bayesian}^{last}$ (given in eV/Å), are shown.

| system | $\#_{atoms}^{UC}$ | $\#_{references}^{initial}$ | $error_{Bayesian}^{last}$ | $\#_{references}^{final}$ |
|---|---|---|---|---|
| MOF-5 | 106 | 974 | 0.055 | 1110 |
| UiO-66 | 114 | 739 | 0.077 | 1373 |
| MOF-74 | 54 | 998 | 0.089 | 2549 |
| MIL-53 (lp) | 76 | 783 | 0.080 | 2832 |
| MIL-53 (np) | 76 | 827 | 0.088 | 3073 |

To assess the potentially beneficial impact of increasing the size of the reference data sets, in certain cases additional learning runs were appended starting from the already obtained potential in an NPT ensemble at a constant temperature of 300 K for 100,000 time steps. Here, the Bayesian error threshold was fixed to 0.02 eV/Å for all systems. This value is slightly lower than the threshold value at 300 K during the training runs with continuous heating. In the following, we will refer to the combined reference data from both training runs as the "extended reference data set". As can be seen in Table 1, for MOF-5, where the automatically set Bayesian threshold was already relatively low after the initial training, only 136 additional reference structures were added in the extended training run. In contrast, the number of structures in the datasets increased by a factor of two to three for the other systems, where the error threshold was higher initially. For more information regarding the training including the time evolution of the error threshold and the errors of the VASP MLPs compared to the training set, see the Supplementary Information section S2.

The active learning runs were carried out with VASP version 6.3.0. Considering several improvements in the recently released versions of VASP (including improved default settings of the machine learned potentials), the VASP MLPs for all systems were re-fitted based on the previously obtained DFT reference data using VASP version 6.4.1. To accomplish this, "ML_MODE=select" was set to reselect the local reference configurations forming the basis set of the potential. In a second step, another re-fitting using "ML_MODE=refit" was performed to yield the final, computationally particularly efficient and more accurate VASP MLPs used in the comparisons in the following sections unless explicitly noted otherwise. The impact of these refitting steps will be discussed in detail in "Evaluating the efficiency of the potentials".

As mentioned before, as a complementary approach, the DFT reference data from the VASP on-the-fly learning approach were used to obtain moment tensor potentials[50]. For that step, the MLIP (machine learning interatomic potential) package[58] was employed. In this context it should be mentioned that MLIP would support its own on-the-fly active learning approach based on molecular dynamics simulations. However, at least in our simulations for MOFs performing the active learning directly using MLIP turned out to be inconveniently slow. We primarily attribute that to the different concepts of the two employed machine learned potentials: in the case of the VASP MLPs, dynamically adding additional reference data to an existing potential is straightforward as the inherent basis set is built from local configurations of atoms and their neighborhood. However, for the MTPs, which have a fixed basis set with a fixed number of parameters, all parameters must be retrained even when only one additional reference structure is appended. This would lead to a substantially increased training overhead, which is especially inconvenient for MOFs with their relatively large numbers of atoms (and atom types) in the unit cells. Consequently, for complex materials like MOFs, the combined approach relying on an easy to modify machine learned potential for the active learning and a fit of the MTPs to the thus-generated reference data appears more efficient. In passing we note that a similar approach has already been applied successfully to interfaces to benefit from the high computational efficiency of MTPs[76]. However, in this context it should be mentioned that beginning with VASP version 6.4.1 the performance gap between VASP and MTP has been strongly narrowed, as will be discussed explicitly below. A further benefit of using the same reference data sets for both the MTPs and VASP MLPs is that then the performance of the different machine learned potentials can be assessed more rigorously without any bias from differently generated sets of reference data. Due to the stochastic nature of the initialization of the MTP fits, we performed several fits for each data set to analyze the deviations and to choose the optimal potential for the respective system based on its description of a validation set of reference structures. These aspects are described in more detail in the Methods section.

In the context of force-field parametrization it is also important to mention that (both for the VASP MLPs and the MTPs) during training we did not only determine separate parameters for each element contained in the material, but rather employed the concept of atom types prevalent in conventional force field potentials: i.e., atoms with a similar immediate neighborhood, determining, e.g., their hybridization, were considered as an independent atom type with its own force field parameters. This means that for example a C atom in the aromatic part of the linker is considered to be a different species than the C atom in the carboxylate connecting the linker to the node. System-specific details regarding the actually used atom types for each MOF are provided in section S2 of the Supplementary Information.

For the bulk of the simulations discussed below, we employed that separation of atom types approach, even though for the VASP MLPs increasing the number of atom types clearly increases the computational cost during production runs. Here, a benefit of using MTPs is that their computational efficiency depends only weakly on the number of atom types. Therefore, to better assess the tradeoff between performance and efficiency induced by the separation of atom types, we also performed parametrization runs choosing only a single atom type per element. Whenever the thus-obtained force fields are used, this will be explicitly mentioned. In fact, the impact of (skipping) the separation of atom types will be most prevalent when discussing the performance of the force fields in the concluding section of the paper.

All quantities discussed in the following were calculated for all considered MOFs with a VASP MLP, with an MTP of level 22 (and a radial basis-set size of 10), and with DFT whenever computationally possible. Additionally, for MOF-5 and MOF-74 they were also determined with the UFF4MOF force field to put the performance of the machine learned potentials into perspective. Explicitly showing all these results runs the risk of exploding the scale of the paper. Therefore, for most of the following quantities, we show in the main manuscript only the results for MOF-5 and MOF-74 as prototypical examples for an isotropic and an anisotropic MOF. The results for the other MOFs and the ensuing trends will typically be briefly mentioned in the discussion, but the corresponding data will be contained in the Supplementary Information (sections S3.1 and S3.5). Finally, it should be mentioned, that the bulk of the data shown in the plots of the main manuscript are based on force fields trained on the initial set of reference configurations. Only when including the additional reference data (300 K run, see above) significantly impacts the quality of the description of a physical observable, this will be discussed explicitly. The entirety of the data obtained when building on the extended training set are contained also in the Supplementary Information (sections S3.1 and S3.3).

## Crystallographic unit cells

As a first step, the performance of the machine learned potentials for predicting unit-cell parameters is assessed. We start with an analysis of the unit-cell volumes. Overall, the agreement of cell volumes between the VASP MLPs, the MTPs, and DFT is excellent, as evidenced by the relative deviations between force-field and DFT data plotted in Fig. 2 (a and b) on a linear and on a logarithmic scale. For the isotropic systems MOF-5 and UiO-66 the relative deviation between both machine learned potentials and the DFT calculated volumes is below 0.2 ‰. For MOF-5, the error of the VASP MLPs is reduced even further by approximately two orders of magnitude. This is in stark contrast to the situation for UFF4MOF, where deviations amount to up to 14 %. The errors of the machine learned force fields are somewhat larger for the anisotropic MOFs, but they are still comparably low (around 0.3 %). Here the errors of the UFF4MOF volumes are at least one order of magnitude larger. A similar trend emerges, when considering the unit-cell vectors with negligible deviations for the isotropic systems and somewhat larger errors for the c-parameters in MOF-74 and MIL-53 (see section S3.1.1 and Table S12 in the Supplementary Information).

When comparing the simulations to experimental unit-cell volumes, one has to keep in mind that the majority of the experiments for MOF-5[77–80], UiO-66[63,80–82], MOF-74[83–86], MIL-53 (lp)[70,87,88] and MIL-53 (np)[69,70,88] have been performed at or around room temperature (around 300K). Thus, for this comparison we did not merely optimize the geometries using the respective force fields but performed NPT simulations at 300 K for the machine learned potentials and for the UFF4MOF force field (for details see method section). In that case, the force fields not only need to be able to correctly predict "0 K" structures, but also need to correctly capture the thermal expansion of the MOF crystals. This is a particular challenge, as it is not really clear, how well the underlying DFT approach would perform for that task[83] (see also section on Future Challenges). Moreover, there usually is a rather significant spread in the experimental data. A comparison with the average experimental cell volumes is contained in Fig. 2c. The deviations of the experimental cell volumes of individual measurements from the average values are indicated by the horizontal lines. A summary of the experimental data sets and the simulated 300K lattice parameters can be found in Table S18. Again, for most systems the values obtained with the MTPs and VASP MLPs are close to the average values of the experiments (with deviations clearly below 2%). Only for MIL-53 (np) the situation appears to be somewhat worse with an underestimation of the average experimental volume by nearly 5%. In this context, one, however, has to note that also experimental values vary by ±4% around the average volume. Consistent with the comparison with the simulated unit cells, when using UFF4MOF significantly larger deviations are found for all systems.

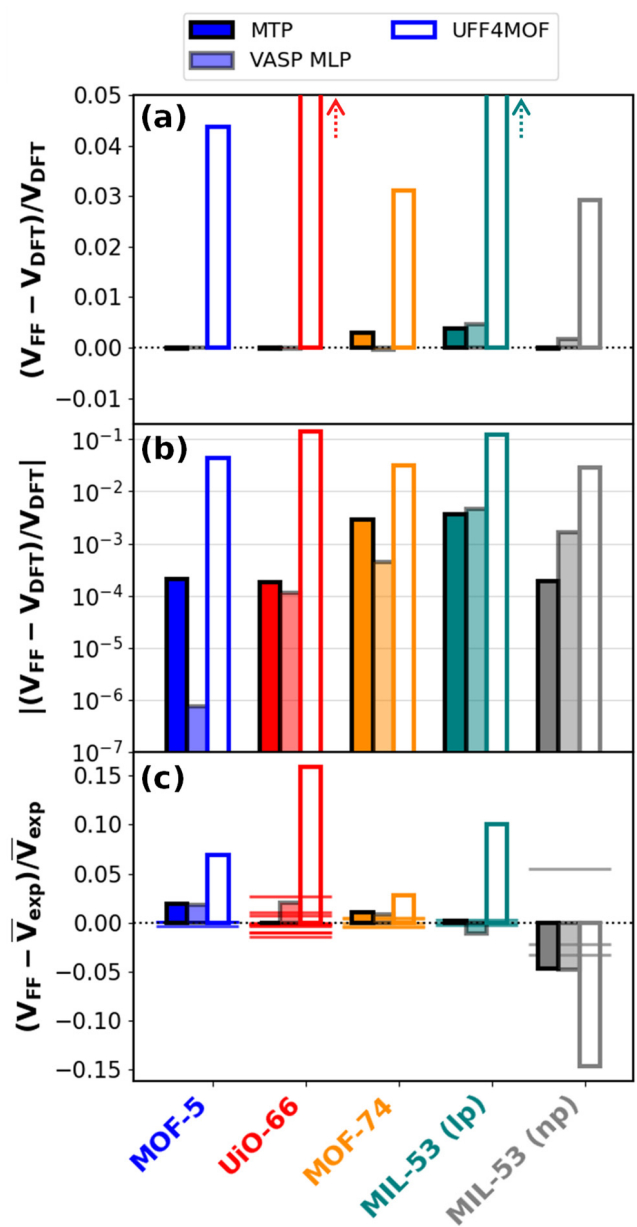

**Fig. 2: Relative deviations of the volumes obtained with the force field potentials compared to DFT and experimental reference data.** For the DFT comparison in **a** and **b**, 0 K optimized unit cells are considered, while for the comparison with experimental data in **c** the values at room temperature are

shown (see main text for details). For all systems (MOF-5: blue, UiO-66: red, MOF-74: yellow, MIL-53 (lp): teal, MIL-53 (np): grey) the comparison of the volumes is shown for machine learned potentials trained on the initial reference data set. MTP results are displayed by strongly colored bars, VASP MLPs results by lightly colored bars, and UFF4MOF results by empty bars. The simulated volumes in **c** were obtained via NPT simulations at 300 K. They are compared to the average of the experimentally obtained volumes. The individual experimental results for each system are indicated as vertical lines in the respective color.

## Prediction of energies, forces, and stresses for a validation set

The probably most important benchmark for the performance of the machine learned potentials is how well they describe total energies, forces, and stresses. To ensure that the potentials properly predict situations beyond the reference structures they were trained on, it is important to analyze their accuracy based on an independent set of validation data. Here, the validation set consists of 100 DFT computed displaced structures obtained via an active learning molecular dynamics simulation in VASP at a temperature of 300 K. The compared quantities comprise the total energies of the system, the atomic forces and the stresses acting on the cell. The deviations of the force-field predicted total energies, forces, and stresses from the DFT simulations are shown in Fig. 3 for MOF-5 and MOF-74. In neighboring panels, we compare MTP with VASP MLP results and MTP with UFF4MOF data, which requires very different scales. Corresponding plots for UiO-66, MIL-53 (lp), and MIL-53 (np) portraying a similar situation are contained in Fig. S15 In the Supplementary Information. In passing we note that (as described in the Methods section) the energies and forces were criteria for the choice of the MTP among multiple fitted ones with different random initialization seeds. Therefore, in Fig. 3 we not only report the RMSD values for the "ideal" MTPs, but in parentheses also provide the average RMSD values for all fitted MTPs for each system. The individual RMSD values for each potential can be found in Table S7 in the Supplementary Information.

For the energies of the displaced structures relative to the equilibrium structures (first line of plots in Fig. 3), one observes an excellent performance of both machine learned potentials (panels a and c) where for the MTPs and VASP MLPs almost all deviations amount to less than 1 % of the absolute values. Notably, the largest differences in total energies for MOF-5 are around 0.3 meV/atom for both machine learned potentials, which is well below the chosen convergence criterion for the underlying DFT calculations, which amounts to 1 meV/atom. In sharp contrast, for UFF4MOF the agreement is more than two orders of magnitude worse with deviations of up to 50 meV/atom (see Fig. 3b). For MOF-74 the MTP and VASP MLP errors are somewhat larger, but still two orders of magnitude below the UFF4MOF values. The trends for the maximum errors in energies are fully consistent with those of the corresponding RMSD values.

For the forces in the second row of Fig. 3, the distributions of the errors are symmetric around zero. Consistent with the situation for the energies, the spread for the VASP MLPs is again very similar to that of the MTPs. Also for the forces, the mediocre performance of traditional, transferrable force fields becomes evident. For example, for MOF-5 the root mean square deviation (RMSD) of the forces calculated with UFF4MOF amounts to 1.03 eV/Å, while it is only 0.02 eV/Å for the MTP and the VASP MLP. In view of the particular relevance of force predictions, it is worthwhile to consider also other force field types in the present comparison: the Dreiding force field as an example for a traditional potential ignorant of MOFs in its parametrization performs even worse than UFF4MOF with an RMSD of 2.13 eV/Å for MOF-5. These errors are severe considering that the mean absolute value of all the forces in the validation set for MOF-5 amounts to only 0.59 eV/Å. However, this does not mean that all traditional force field potentials perform poorly. A traditional, but system-specifically parameterized MOF-FF potential[27] for MOF-5 displays a substantially improved RMSD of 0.09 eV/Å, which is an order of magnitude lower than the UFF4MOF value (albeit still larger than the values for the machine learned potentials). This suggests that the most important step in improving the force-field performance is their system-specific parametrization.

For the errors of the stresses in the last row of Fig. 3, we observe slightly larger values for the VASP MLPs than for the MTPs, while the error increases by two orders of magnitude for UFF4MOF. A peculiarity are the significantly larger stress errors for MOF-74 in Fig. 3k compared to MOF-5 in Fig. 3i for all force fields. This can at least in part be attributed to the absolute stress values being distinctly higher in MOF-74.

The worse performance for the anisotropic systems like MOF-74 coincides with the larger Bayesian error threshold during the training. Notably, using the extended reference data set leads to a significant reduction of the force and stress errors of VASP MLPs such that the initial RMSDs for MOF-74 of 0.27 eV/Å and 0.17 kbar decreased to 0.22 eV/Å and 0.12 kbar, respectively. Conversely, for the MTPs the use of the extended reference data sets has only a minor impact on the errors for forces, energies and stresses (see Table S11).

Concerning the other investigated systems (see Fig. S15 and discussion in section S3.1), the situation of the force, stress and energy errors for the cubic UiO-66 MOF is comparable to the MOF-5 case, while the anisotropic MIL-53 behaves more like MOF-74 with somewhat higher deviations for forces and stresses.

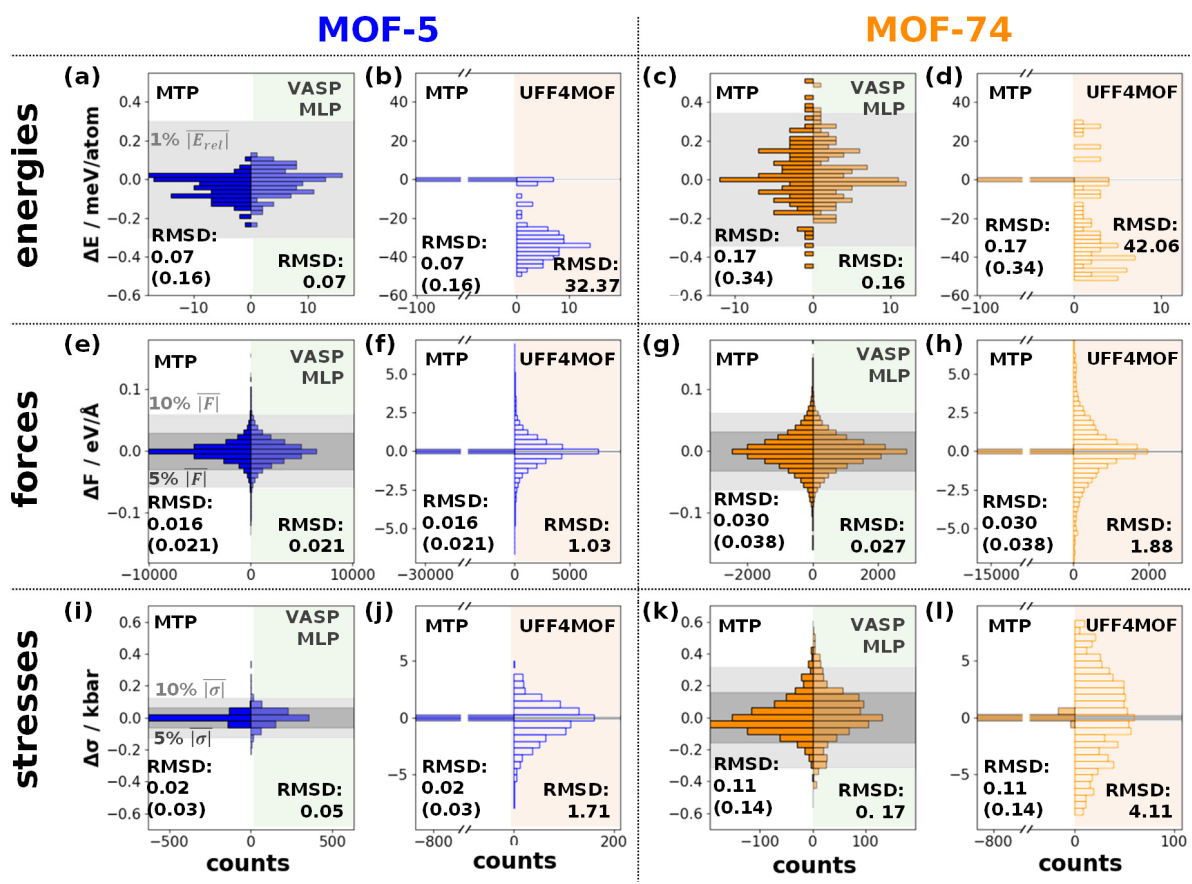

**Fig. 3: Histograms detailing the deviations of the energies, forces, and stresses.** Quantities obtained with the MTP (dark shading, all panels), VASP MLP (light shading; panels **a**, **c**, **e**, **g**, **i**, **k**) and UFF4MOF (no shading; panels **b**, **d**, **f**, **h**, **j**, **l**) have been subtracted from DFT reference data for the validation set. Deviations in energies, ΔE, relative to the situation at equilibrium geometry are shown in panels **a**, **b**, **c**, and **d**, differences in the components of the forces on individual atoms, ΔF, are shown in panels **e**, **f**, **g**, **h**, and differences in the components of the stresses Δσ are contained in panels **i**, **j**, **k**, **l**. The areas shaded in gray indicate the ranges with errors within 5% and 10% of the absolute DFT values for forces and stresses and within 1 % for the energy differences. The RMSD values of the various FFs compared to DFT are noted in the panels, where for the MTPs the RMSD values for the ideal force field and the average RMSD values of all parametrized force fields (in parentheses) are provided. The figure contains only the data for MOF-5 and MOF-74, with the data for all other studied MOFs contained in Fig. S15.

## Vibrational properties of the studied MOFs

After investigating the quality of the properties directly provided by the machine learned potentials, it is useful to assess derived quantities. Here, we will focus on vibrational properties, which for crystalline materials are determined by phonons. They crucially impact phononic heat transport, charge transport (via the strong electron-phonon coupling), and elastic properties. As a starting point, Fig. 4 a-d compares the low frequency phonon band structures obtained with DFT and the MTPs for MOF-5, UiO-66, MOF-74 and MIL-53 (lp) and in Fig. 4 e, the same comparison between VASP MLP and DFT for MOF-5 is shown. No comparison is shown for MIL-53 (np), as for this system it has been challenging to find a phonon band structure without imaginary modes when using DFT (see section S1.4 for details).

The low frequency region is insofar particularly interesting, as these phonon modes determine the above-mentioned transport properties as well as the relative thermal stability of different phases. It is immediately evident from the plots that the MTPs and the VASP MLPs (see also Fig. S24) reproduce the DFT-calculated phonon band structures almost perfectly. Especially for MOF-5 the largest deviations for both force fields amount to less than 5 cm$^{-1}$, with this excellent agreement persisting for UiO-66 for the MTPs. For MOF-74 and MIL-53 (lp), we see some larger deviations for individual phonon bands, however, the agreement is still extremely good and outperforms the results for any traditional force field potentials that we are aware of by a large margin. When comparing with the VASP MLPs in Fig. S24, the agreement for MIL-53 (lp) and UiO-66 is slightly worse than for the MTPs, but still excellent. However, as also shown in the Supplementary Material section S3.5, the description of phonons in the VASP MLPs can be substantially improved using the extended reference data set.

The good agreement between force field and DFT-calculated phonon band structures is lost when employing UFF4MOF, as shown for MOF-5 in Figure 4f. The situation is actually even worse for MOF-74, as illustrated in Figure S26. UFF4MOF substantially overestimates the dispersion of the acoustic modes, which results in too high group velocities. This is one of the origins of the severely overestimated thermal conductivity calculated for MOF-5 when employing UFF4MOF[27]. Additionally, virtually all optical modes show large shifts to higher frequencies.

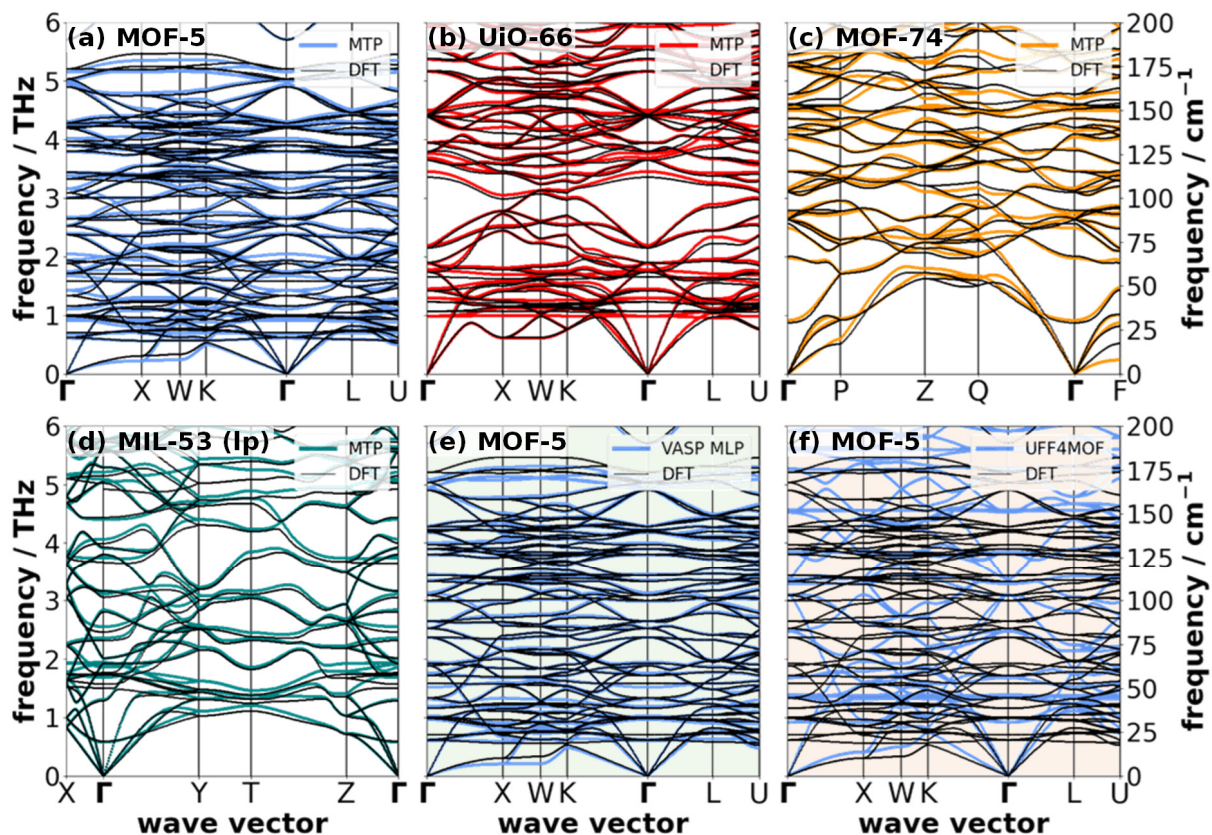

**Fig. 4: Comparison of low frequency phonon band structures.** They were computed using the "ideal" MTPs trained on the initial reference data set (colored lines) and employing DFT (black lines) for the following systems: MOF-5 in **a**, UiO-66 in **b**, MOF-74 in **c** and MIL-53 (lp) in **d**. The comparison between the VASP MLP and MOF-5 is given in panel **e**. Panel **f** contains the MOF-5 comparison between UFF4MOF and DFT. High symmetry point names correspond to the space group dependent convention detailed in ref. [89].

While the qualitative agreement of the phonon band structures is very reassuring, Fig. 4 displays only the situation for the low frequency phonons. As plotting the band structures over an extended frequency range would not be useful due to the sheer number of bands, in Fig. 5 the phonon densities of states (DOSs) calculated with the force fields are compared to those obtained with DFT for MOF-5 and MOF-74 (data for the other MOFs can be found in Fig. S16). In the left panels, DOSs in the low frequency region are displayed, while the right panels show them for the entire frequency range in which phonons exist in the studied materials. As expected from the preceding discussion, for both the MTPs and the VASP MLPs in the low frequency region one sees mostly minor deviations from the DFT data, which are somewhat more significant for MOF-74 than for MOF-5. Interestingly, this trend also prevails, when plotting DOSs for the full frequency range. The situation deteriorates for UFF4MOF, where it becomes virtually impossible to correlate specific features in the DOSs calculated with the force field with those calculated with DFT. Additionally, UFF4MOF massively underestimates the magnitude of the frequency gap found in the DFT simulations between 50 THz and 90 THz.

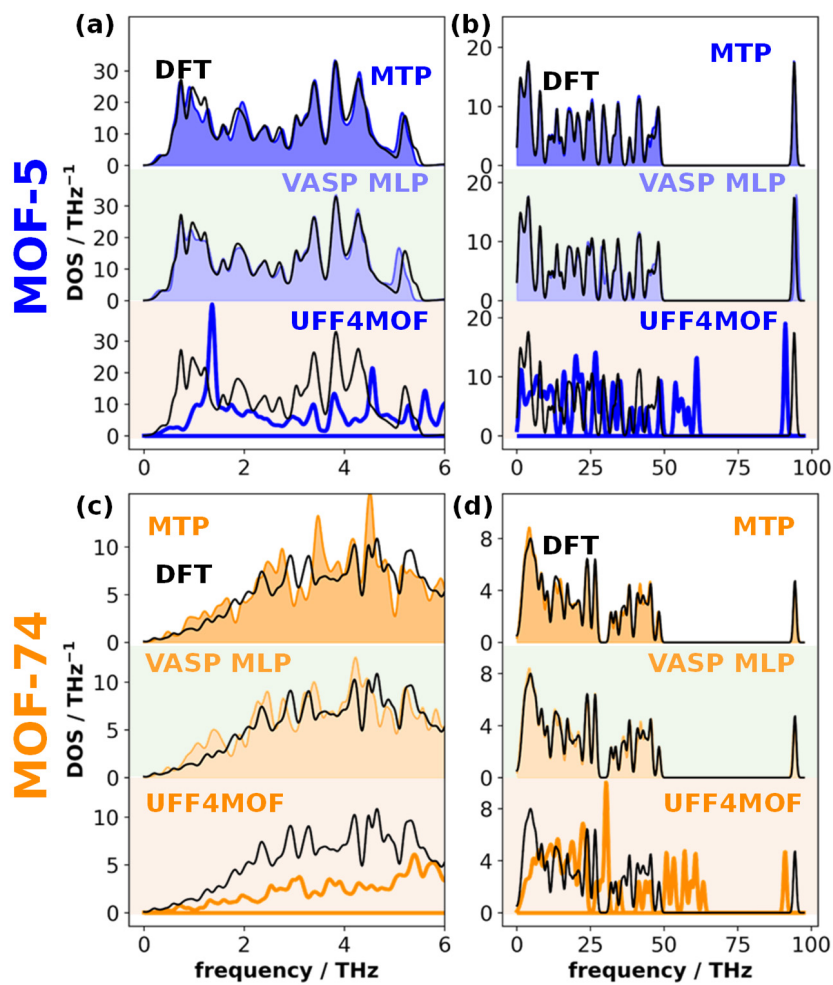

**Fig. 5: Phonon densities of states (DOSs) calculated with DFT and various force fields.** Comparisons of phonon DOSs are shown for DFT calculations (black lines) and the force field potentials (shaded areas/colored lines) for the low frequency phonon modes in **a**, **c** and for the entire phonon spectrum in **b**, **d**. The plots contain the comparisons for MOF-5 (a, b, blue) and for MOF-74 (c, d, orange) for the MTP, the VASP MLP, and the UFF4MOF. The DOSs were obtained for 11×11×11 q-point meshes used to sample the first Brillouin zone and a Gaussian smearing is applied to broaden the modes. The width of that smearing is set to 0.05 THz for the low frequency region in **a**, **c** and to 0.5 THz for the entire frequency range in **b**, **d**.

To quantify the quality of the description of phonons by the three types of force fields, Table 2 lists the RMSDs of the $\Gamma$-point frequencies for all materials for the low-frequency and for the full frequency range. Additionally, it compares the results of the machine-learned potentials obtained for the initial and for the extended reference data sets. To show the benefit of separating the chemical elements in additional subtypes for the VASP MLPs, the errors for a potential trained with only 4 atom types (i.e., one atom type per element) is also included. Finally, for MOF-5 Table 2 also contains the results for the Dreiding and the MOF-FF force fields already mentioned when discussing the forces. Especially in the low frequency region, an excellent agreement with the DFT data is achieved for all machine learned potentials with only single digit wavenumber errors. For MTPs and VASP MLPs all RMSDs are between 1.3 and 3.9 cm$^{-1}$ when considering different atom types for elements in different chemical environments. Even for the full frequency range and when using only the initial training set, MTP RMSD values for most systems are around an amazing 3 cm$^{-1}$ with a maximum of 7.6 cm$^{-1}$ for MIL-53 (np). For the VASP MLPs in this case deviations in the entire frequency range are somewhat larger ranging from 4.2 to 10.4 cm$^{-1}$. However, the VASP MLPs can benefit substantially from an extension of the reference data set, as can be seen in particular for MOF-74, where the full frequency RMSD drops by 30% to 2.9 cm$^{-1}$ becoming similar to the MTP situation. When using the extended reference data set, in the case of the VASP MLPs also for the other systems the situation consistently improves, albeit with varying significance. Conversely, for the MTPs extending the training set yields only minor changes with no clear trend. This being said, for both training sets the RMSDs for the machine learned potentials are always more than an order of magnitude lower than for the traditional UFF4MOF and Dreiding potentials, where average deviations can reach hundreds of wavenumbers. Notably, the frequency RMSD values for the conventional, system-specifically parametrized MOF-FF potential are also an order of magnitude lower than for UFF4MOF, but still substantially higher than for the machine learned potentials.

**Table 2: Root mean square deviations (RMSDs) for the phonon frequencies of the investigated systems.** The values are given for $\Gamma$-point phonons obtained by the MTPs and VASP MLPs trained on the initial and the extended reference data set for the full frequency range, $RMSD_{full}$, and for the low frequency modes up to 200 cm$^{-1}$ (= 6 THz), $RMSD_{200}$. The values for VASP MLPs trained on the initial reference data set are shown for different numbers of atom types (with 4 atom types referring to one atom type per element). Additionally, values for UFF4MOF[19,20], Dreiding[17] and a system-specifically parameterized MOF-FF[32,33] are provided for selected systems. Indicated in bold are the atom typed potentials trained on the initial reference data set, which are primarily used throughout this work.

| System | Method | types | RMSD$_{full}$ / cm$^{-1}$ | RMSD$_{200}$ / cm$^{-1}$ |
|---|---|---|---|---|
| MOF-5 | VASP MLP | 4 | 8.0 | 1.3 |
| | **VASP MLP** | **7** | **9.0** | **1.5** |
| | VASP MLP extended | 4 | 7.4 | 1.5 |
| | **MTP** | **7** | **3.3** | **1.6** |
| | MTP extended | 7 | 3.7 | 1.4 |
| | UFF4MOF | 5 | 203.2 | 51.0 |
| | Dreiding | 5 | 183.2 | 46.6 |
| | MOF-FF[33] | 7 | 14.1 | 7.8 |
| UiO-66 | VASP MLP | 4 | 11.9 | 9.0 |
| | **VASP MLP** | **9** | **10.4** | **2.7** |
| | VASP MLP extended | 4 | 7.8 | 3.0 |
| | **MTP** | **9** | **3.7** | **1.6** |
| | MTP extended | 9 | 2.8 | 1.5 |
| MOF-74 | VASP MLP | 4 | 5.7 | 1.4 |
| | **VASP MLP** | **8** | **4.2** | **1.1** |
| | VASP MLP extended | 4 | 4.8 | 1.2 |
| | VASP MLP extended | 8 | 2.9 | 0.7 |
| | **MTP** | **8** | **3.1** | **1.4** |
| | MTP extended | 8 | 2.9 | 1.2 |
| | UFF4MOF | 4 | 242.8 | 116.9 |
| MIL-53 (lp) | VASP MLP | 4 | 12.5 | 3.3 |
| | **VASP MLP** | **8** | **8.8** | **3.2** |
| | VASP MLP extended | 4 | 7.4 | 2.7 |
| | **MTP** | **8** | **5.2** | **2.8** |
| | MTP extended | 8 | 5.0 | 2.5 |
| MIL-53 (np) | VASP MLP | 4 | 11.4 | 4.5 |
| | **VASP MLP** | **8** | **6.9** | **2.5** |
| | VASP MLP extended | 4 | 10.8 | 7.7 |
| | **MTP** | **8** | **7.6** | **3.9** |
| | MTP extended | 8 | 6.4 | 3.3 |

## Elastic properties of the MOFs

Another relevant quantity related to the properties of acoustic phonons (via the Christoffel equation[90]) is the elastic stiffness tensor, C. The symmetry-inequivalent elements of C obtained from DFT and from the force field potentials are shown in Fig. 6 for MOF-5 and MOF-74. For the MTPs the agreement is essentially ideal for both materials with maximum deviations of 1 GPa. For the VASP MLPs the situation in MOF-5 is similar but in MOF-74 the agreement is slightly worse. This is presumably caused by the more demanding description of the elastic constants in the highly anisotropic material[91]. Nevertheless, the agreement is still very good also for the VASP MLPs, especially when using the extended reference data set. For the MTPs, similar to the situation discussed above for the phonon frequencies, extending the training set has virtually no effect on the performance for any of the systems.

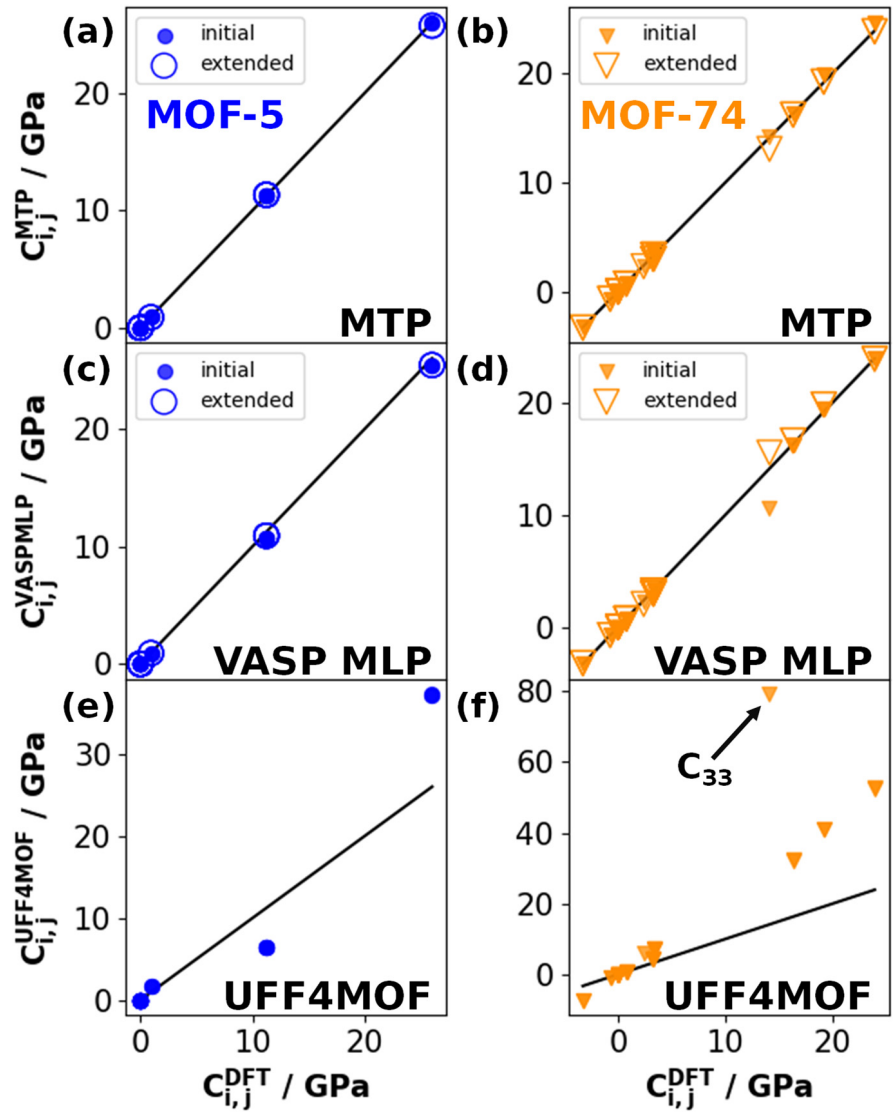

**Fig. 6: Comparison of the elements of the elastic stiffness tensor.** Shown are the tensor elements $C_{i,j}$ calculated with DFT and with the MTPs in **a**, **b**, with DFT and with the VASP MLPs in **c**, **d**, as well as with DFT and UFF4MOF in **e**, **f**. The plots shown here contain data for MOF-5 in **a**, **c**, **d** (blue) and MOF-74 in **b**, **d**, **f** (orange) with equivalent plots for the other studied MOFs provided in Fig. S17. In the upper and central panels, small filled symbols indicate values obtained with potentials trained on the initial reference data set and large empty symbols refer to the values for potentials trained on the extended reference data set. The solid black line pass through the origin and have a slope of 1 such that they indicate an ideal agreement between force field and DFT simulations.

Not unexpectedly, the performance of UFF4MOF is much worse with the extreme example of the $C_{33}$ constant in MOF-74 being more than 5 times as high as in the DFT reference (see Fig. 7f). Also, for MOF-5 one observes substantial deviations. Interestingly, the bulk modulus for MOF-5 calculated via the Voigt average form the components of the stiffness tensor is rather similar when calculated with UFF4MOF (16.8 GPa) and with DFT (16.1 GPa). These values are consistent with the literature[14], but considering that the individual components of the stiffness tensor differ substantially between the two approaches, the seeming agreement for UFF4MOF has to be attributed to a fortuitous cancellation of errors.

The absolute values of the tensor components as well as plots for UiO-66 and MIL-53 are contained in section S3.1.4 of the Supplementary Information. The situation for the machine learned potentials is again comparable to that in MOF-5 and MOF-74 with deviations being largest for MIL-53 (np) with an RMSD for all tensor elements of 1.2 GPa. However, this is still quite very good, given the higher absolute values of the tensor elements in that material.

## Thermal conductivity

As a final benchmark quantity, we consider the thermal conductivity. It is again intimately related to phonon properties, where now also anharmonic effects play a decisive role. As calculating the thermal conductivity of MOFs employing DFT is far beyond present computational possibilities, the comparison is restricted to experimental data and here to MOF-5, as to the best of our knowledge this is the only materials amongst the ones studied here for which high-quality single crystal data exist. According to ref. [62], the room temperature value of the thermal conductivity of MOF-5 amounts to 0.32 W/mK and it has already been shown that the UFF4MOF and the Dreiding force fields fail in reproducing that value (yielding thermal conductivities of 0.847 W/mK and 1.102 W/mK, respectively), while the system-specifically parametrized MOF-FF variant provides a sensible value of 0.29 W/mK[27]. Here, non-equilibrium molecular dynamics simulations for MOF-5 were performed using a level 18 MTP. Notably, for these simulations we used a lower level of the MTP to reduce the computational cost of correcting for finite size effects, which requires simulations for various supercell lengths[92,93]. This is, however, not expected to have a major impact, as the difference in the uncorrected thermal conductivities for individual cell lengths between level 22 and level 18 MTPs is minimal as shown in the supplementary information section S4.2. The resulting thermal conductivity obtained from NEMD simulations for cell lengths ranging from 208 to 416 Å perfectly reproduces the experiment, yielding a value of 0.32 W/mK. As a technical detail, it should be mentioned that this result has been obtained applying the traditional approach of determining the temperature gradient in the NEMD simulation in the region of a linear temperature decrease. When instead employing the method of Li et al., determining the temperature gradient from the temperature difference between the hot and cold thermostats[92] (which we used

previously[27,33]), a somewhat lower value of 0.26 W/mK is obtained. Therefore, we also employed the "approach to equilibrium molecular dynamics", AEMD, method as a complementary molecular dynamics-based strategy for obtaining thermal conductivities. This again yielded a value of 0.32 W/mK, confirming the excellent agreement of the MTP simulations to the experimental data. A full discussion und further details regarding the heat transport simulations are provided in section S.4.2 of the supplementary information. At the time of the calculations for this paper the NEMD simulations in VASP were not accessible, hence no comparison between the VASP MLPs and MTPs was done for the thermal conductivity.

## Future challenges

While the above discussion highlights the superior performance of the considered machine learned potentials for predicting various observables, we will next discuss thermal expansion as a process for which the performance of the machine learned potentials appears to be less satisfactory (as summarized in Table S19 with a much more extended discussion of the simulations in supplementary information section S4.1). In the case of thermal expansion, one again needs to compare the calculated thermal expansion coefficients to experimental data, as simulating the thermal expansion of MOFs with DFT is extremely challenging and prone to errors (see discussion in the Supporting Information of ref. [83]). As far as the available experimental data are concerned, some of the issues were already mentioned in the context of the crystallographic unit cells at 300 K for the narrow pore phase of MIL-53, where the experimental results also show a substantial spread. The situation is also problematic for MOF-74, where only very minor changes of the unit-cell volume with temperature have been found experimentally[83,94].

In fact, for MOF-74, the MTPs/VASP MLPs predict a very small positive ($1.6 \cdot 10^{-6}$ K$^{-1}$ and $1.2 \cdot 10^{-6}$ K$^{-1}$/$2.0 \cdot 10^{-6}$ K$^{-1}$ and $6.9 \cdot 10^{-6}$ K$^{-1}$) thermal expansion coefficient in linker direction instead of a very small negative value ($-2 \cdot 10^{-6}$ K$^{-1}$) suggested by the experiments. Due to generally limited accuracy of the experimental and computational methods this difference might still be acceptable. What is, however, more concerning is that along the pore direction, the MTPs/VASP MLPs predict a thermal expansion value of $31.0 \cdot 10^{-6}$ K$^{-1}$/$25.3 \cdot 10^{-6}$ K$^{-1}$, which is nearly an order of magnitude larger than the experimental result of approximately $4 \cdot 10^{-6}$ K$^{-1}$ [83]. For UiO-66 in the linker direction the situation is analogous to that in MOF-74 with an MTP/VASP MLP-calculated small positive thermal expansion coefficient ($2.2 \cdot 10^{-6}$ K$^{-1}$/ $1.8 \cdot 10^{-6}$ K$^{-1}$) compared to a slightly negative one in the experiments ($-5.6 \cdot 10^{-6}$ K$^{-1}$)[80].

There are, however, also cases in which the agreement between simulations and experiments is improved: for example, for MOF-5 the simulated thermal expansion coefficients of $-11.3 \cdot 10^{-6}$ K$^{-1}$ for the MTP and $-11.5 \cdot 10^{-6}$ K$^{-1}$ for the VASP MLP are rather close to the experimental values, which range between -13.1·$10^{-6}$ $K^{-1}$ and -15.3·$10^{-6}$ $K^{-1}$ [78,79,95]. Also for MIL-53 (lp) at least the qualitative trends are consistent for the MTP, with simulated anisotropic expansion values of -15.5, 4.4 and 40.2·$10^{-6}$ $K^{-1}$ compared to -14.5, -1.0 and 24.2·$10^{-6}$ $K^{-1}$ from experiment[87]. Conversely, the VASP MLPs predict qualitatively different linear thermal expansion coefficients of 9.3, 0.5 and -24.9·$10^{-6}$ $K^{-1}$ for MIL-53 (lp). Given the inconsistent quality of the agreement for thermal expansion coefficients, it is at this point unclear whether these discrepancies are a consequence of the parametrization of the machine-learned potentials, or whether the problems are at least in part inherent to the DFT approach used for the reference data generation. To answer that question, a more much more thorough investigation involving a larger number of systems would have to be performed. This would need to be combined with concerted efforts to reliably simulate thermal expansion coefficients using DFT. Especially the latter will be a sizable challenge, as ab-initio MD based approaches would be numerically extremely expensive and approaches based on the quasi-harmonic approximation would suffer from serious numerical issues, as already exemplarily shown for MOF-74[83]. An indication that the chosen DFT methodology does have a significant impact on calculated thermal expansion coefficients is that for MOF-74 it has actually been observed that when applying the numerically more stable mode Grüneisen theory of thermal expansion, the sign of the thermal expansion coefficient in linker direction would still depend on the chosen functional (PBE vs. PBEsol)[83].

The other major challenge in particular when applying the MTP approach to MOFs is to train the potential such that a force field is obtained that is fully stable for the targeted simulation conditions. Stable here means that the system in question remains structurally intact also over hundreds of thousands (or even millions) of molecular dynamics time steps that one would, for example, apply when performing NEMD simulations. A structural disintegration of the material would not be possible for simple force field potentials characterized, e.g., by harmonic bonding potentials, but for mathematically more complex models like machine learned potentials this can become a problem. This is especially true when in the stochastic description of the atomic motion at finite temperatures large displacements occur for which the potential has not been sufficiently well trained.

Improving the long-term stability of the force fields was one of the main motivations for increasing the temperature to up to 900 K in the on-the-fly training runs in the first place. Nevertheless, not all force fields we trained turned out to be stable over hundreds of thousands of time steps even at room temperature. This is discussed in detail for the MTPs in the Supplementary Information section S2.2. The MTPs for MIL-53 were particularly problematic, as the material features a comparatively flat energy landscape. One possibility to mend this problem was to train several MTPs using the same reference data. In that case, due to variations originating from the stochastic nature of the initialization of the training procedure, some potentials turned out to be substantially more stable than others. However, this is not an ideal solution, as additional extensive MD test runs would be required for each MTP to judge the thermal stability and training a large number of accurate MTPs is computational costly. Unfortunately, the extended reference data set obtained based on the molecular dynamics simulations at 300 K did not substantially improve the situation. However, adding 1009 DFT reference structures based on active learning runs at 400 K at a reduced error threshold did result in MTPs for MIL-53 (lp) that were all stable at least up to room temperature.

These results suggest that at least under certain circumstances adding more strongly displaced reference structures obtained by higher-temperature learning steps helps overcoming the problem of long-term stability of the force field. Unfortunately, this is associated with significantly increasing the learning effort, which for additional reference data beyond the first 500-1000 structures does not substantially improve the accuracy of the MTPs. Therefore, for MTPs, it will be useful in the future to explore more efficient options for improving the force-field stability, like further improving the data sampling approach or augmenting the force fields with potential terms that prevent a structural disintegration of the studied system.

The situation here is much less problematic for the VASP MLPs, where based on our stability tests in an NPT ensemble, all potentials trained on the initial reference data set including atom type separation were stable up to 700 K, which is the maximum temperature included in the tests as shown in the Supplementary Information section S4.1.2. Admittedly, at very high temperatures some most likely unphysical cell deformations did take place for some of the systems, but at least the structure never disintegrated. This implies that the VASP MLPs represents a more robust framework for modeling especially flexible MOFs with a flat energy landscape, like MIL-53.

## Evaluating the efficiency of the potentials

So far, the discussion focused on the precision of the machine learned potentials, demonstrating that they show an excellent performance when compared to DFT data. What has not been touched upon so far is an evaluation of the computational efficiency of the potentials. Here, the MTPs allow a straightforward adjustment of the number of parameters of the potentials and, thus, of their numerical efficiency. Based on our experience, especially for complex systems like MOFs the "level" of the MTP has the most profound impact on both precision and performance (compared, e.g., to the chosen number of radial basis functions, which was, thus, always set to 10). The other setting of the MTPs that substantially affects the computational efficiency is the cutoff for the interactions. However, if one deviates too much from the default value of 5 Å chosen also here, the precision of the potentials deteriorates severely. Therefore, using the same reference data as above we systematically varied the level of the MTPs from 10 to 24 and compare the accuracy (in terms of force and frequency RMSDs) and the speed of the force fields in Fig. 7. In panel a, the computational speed of the machine learned potentials is correlated with their accuracy and compared to more traditionally force fields for MOF-5. For this comparison we use three different types of conventional potentials: a system-specifically parameterized MOF-FF potential for MOF-5 used in our previous work[32,33], the transferable universal force field[18] extended for MOFs (UFF4MOF)[19,20], and the transferable Dreiding[17] potential. Here, for the VASP MLPs we show a variant refitted for VASP 6.4.1 for the initial reference data set with and without a separation of atom types based on the atom's chemical environment. The computational speed was evaluated for NPT simulations on 4×4×4 conventional supercells of MOF-5 carried out on 64 cores of a dual socket AMD EPYC 7713 (Milan) node of the supercomputer VSC-5. It should be mentioned that differences in the speeds of the force fields are caused by differences in scaling with the number of nodes (see section S4.1.4) and can be affected by using different libraries when compiling the VASP and the LAMMPS codes. Additionally, we stress that the tests performed here are far from comprehensive and focus on the force fields considered in this study and that there are other options, like, for example, the GPUMD[47] force fields for which a very high speed at good accuracy has been reported for MOFs[44].

The traditional UFF4MOF and Dreiding potentials are the most computationally efficient options, but they also suffer from a poor description of the forces with RMSD values above 1 eV/Å. As discussed already above, the system-specifically parameterized MOF-FF potential hugely improves the situation with an RMSD value of 0.09 eV/Å. However, the more complex functional form including cross terms and an optional, more accurate treatment of long-range electrostatic interactions via a particle-particle particle-mesh solver[96] slows down MOF-FF by a factor of ca. 5.

Significantly more accurate are MTPs, where the highest levels show the best degree of accuracy among the tested potentials for MOF-5. The force RMSDs range between 0.033 eV/Å and 0.016 eV/Å with a change in computational speed by more than an order of magnitude. The fastest, level-10 MTPs even clearly outperforms the MOF-FF in terms of accuracy, while being computationally only twice as expensive as UFF4MOF. For level 22, we also tested the performance of an MTP without separating the atom types. As can be clearly seen in Fig. 7 a, this hardly has any impact on the computational efficiency, but slightly decreases the degree of accuracy (increasing the force RMSD from 0.016 eV/Å for the force field with 7 atom types to 0.021 eV/Å with 4 atom types).

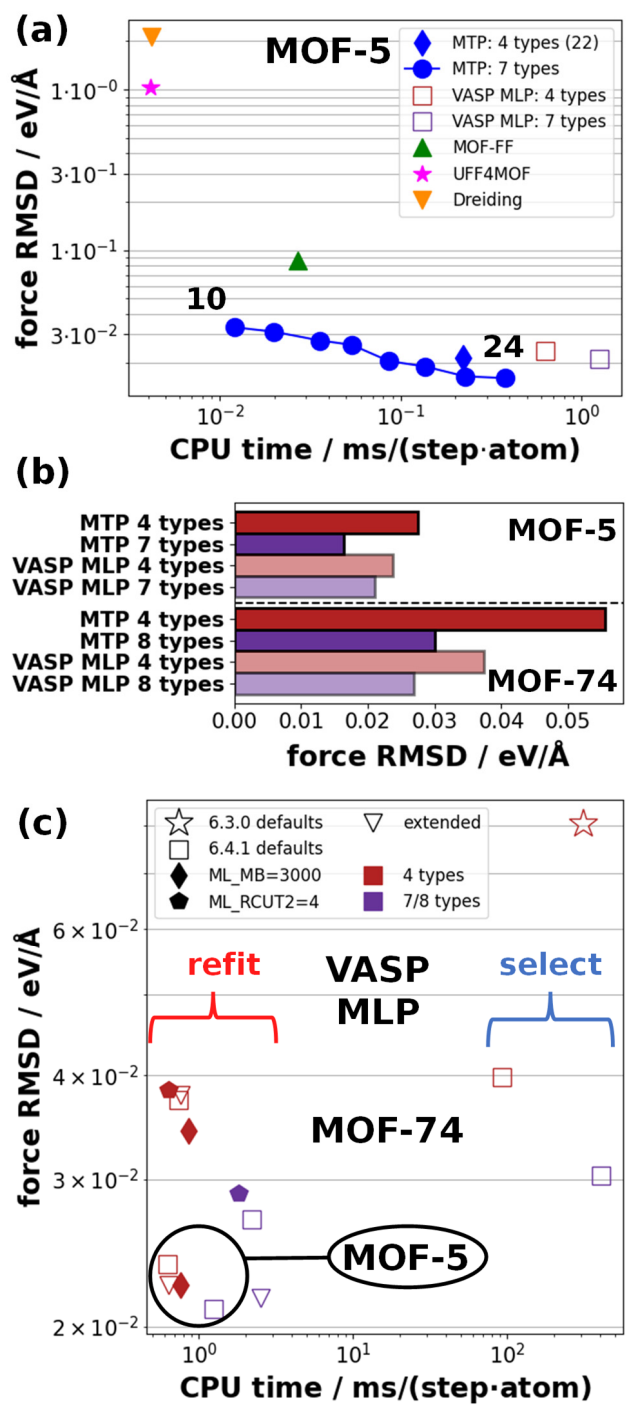

**Fig. 7: Evaluation of the accuracy and speed of different variants of the tested force field potentials obtained for different numbers of atom types.** In **a** the RMSDs of the forces in the validation set between DFT and various force-field approaches are shown as a function of the CPU time used per time step and atom for MOF-5. Data points are included for: MTPs at various levels trained with 7 atom types (blue squares), a level 22 MTP trained with only 4 atom types (blue diamond), two variants of the VASP MLP trained with 4 atom types (red square) and with 7 atom types (purple square) (both refitted with VASP 6.4.1, see main text), a system-specifically trained MOF-FF variant[33] (green triangle), and the transferable UFF4MOF[19] (pink star) and Dreiding[17] force fields (orange down

triangle). The speed was obtained by performing an MD simulation in the NPT ensemble for a 4×4×4 conventional supercell of MOF-5 (27136 atoms) for at least 2,000 time steps. The calculations were performed on 64 cores on a dual socket AMD EPYC 7713 (Milan) (64 cores each) node on the VSC-5 supercomputer (see https://vsc.ac.at/systems/vsc-5/). In **b**, the impact of including the separation of atom types on the validation set force RMSD is shown for the MTP and the VASP MLP for MOF-5 and MOF-74. Panel **c** contains the differences in force RMSD and CPU time between different variants of the VASP MLPs for MOF-5 and MOF-74. The color coding refers to the number of used atom types when training the potential (purple symbols: 7 atom types for MOF-5 and 8 for MOF-74; red symbols: 4 atom types). Inverted triangles indicate potentials trained on the extended reference data set, all other potentials were trained on the initial reference data set. Empty symbols indicate MLPs obtained with the default settings for the respective VASP version. The result of the VASP MLP obtained during active learning in version 6.3.0 (star) is compared to the potentials retrained using VASP 6.4.1 using the "ML_MODE=select" option (squares on the right side) with a subsequent refit using "ML_MODE=refit" (symbols on the left side). Additional variation in the settings comprise: increasing the maximum number of local reference configurations from 1500 to 3000 for the 4 typed potential (ML_MB, diamond) and reducing the cutoff of the angular descriptors from 5 to 4 (ML_RCUT2, pentagons). In the lower left circle, results for MOF-5 are shown, while the rest of the panel only contains data for MOF-74. The CPU time used was tested for a 31104 atom sized supercell of MOF-74 with otherwise the same settings as for MOF-5.

Of a similar accuracy as the high level MTPs are the VASP MLPs, which show a force RMSD of only 0.021-0.024 eV/Å. They are, however, a factor of 2-3 slower than the highest level MTPs. However, it should be noted that, as discussed in the Supplementary Information section S4.1.4, this is strongly correlated with the rather unfavorable scaling of the parallelization of the molecular dynamics part of VASP (which we expect to be improved in future implementations). For example, when using only up to 16 cores, the VASP MLP with 4 atom types is only about 50% slower than the level 22 MTP. It is also evident from Figure 7a that the VASP MLP with 7 atom types is about a factor of 2 slower than the one with 4 types while only offering a moderately increased degree of accuracy, so in the case of MOF-5 the separation of atom types is of questionable usefulness.

However, already the frequency errors in Table 2 suggest that this is highly system dependent. This is also illustrated in Fig. 7 b, which shows the force RMSDs in the validation set for MTPs and VASP MLPs for MOF-5 and MOF-74 trained for different numbers of atom types. For MOF-5 and the VASP MLP the error only drops by 11% between 7 and 4 atom types, while for MOF-74 the improvement between 4 and 8 types amounts to 28 %. In passing we note that a possible reason for larger impact of the separation of atom types in MOF-74 is that there the linker contains two chemically inequivalent types of O atoms, which is not the case for MOF-5. For the MTPs, increasing the number of atom types significantly reduces the force RMSDs for both MOFs (by 40 % in MOF-5 and even by 46 % in MOF-74). This in combination with the negligible cost associated with increasing the number of atom types for MTPs (in the case of MOF-5 the potential with 7 types is 2 % slower than the one with 4 types) suggests that for MTPs it is always useful to explicitly distinguish between atoms in chemically different environments.

As far as the VASP MLPs are concerned, there is a large number of different settings that can be chosen when learning the MLPs. In the following, we will discuss a few of them that were observed to have beneficial effects (with additional information in the Supplementary Information section S2.1 and S3.7). The corresponding data on force RMSD and required CPU times are shown in Fig. 7c. Here, we primarily focus on MOF-74, as for that system changing the number of atom types had the most profound impact. For the sake of comparison, we also included the most relevant data for MOF-5. What is most striking is the comparatively poor performance both in terms of cost and accuracy of the MLP that we generated using VASP 6.3.0 defaults in the course of the active learning run (red star in Figure 7c). The reason for the somewhat lower degree of accuracy is primarily the rather low number of local reference configurations extracted from the reference data set. To improve the situation the same reference data set was used to generate a VASP MLP using the improved default settings in VASP 6.4.1 including the "ML_MODE=select" feature (details on the changed settings can be found in section S3.7). For MOF-74 this reduces the force error by a factor of two while increasing the number of local reference configurations from 1877 to 4830. Despite this increase, for the same number of atom types (c.f., red star and red square in Figure 7c), this even increases the speed of the VASP MLP due to optimizations in the more recent VASP versions. As a next step for improving the performance, one can then use the option "ML_MODE=refit" when reparametrizing the VASP MLP. This leads to a two orders of magnitude increase in speed by disabling the error estimation of the force field, which is needed only in the active learning. Additionally, the accuracy of the force-fields is improved by refitting using a singular value decomposition[97]. When comparing the VASP MLPs with 4 and 7/8 atom types (red vs purple symbols), one sees that the latter are still a factor of ~3 slower. One of the big differences between including and not including the separation of atom types is the different number of local reference configurations. The maximum number of local reference configuration is by default limited to 1500 for each atom type. Therefore, it is of interest whether a similar improvement of accuracy can be achieved when increasing this maximum for the potential with only 4 types to 3000 (filled red diamond; ML_MB=3000). This yielded about a drop in the force RMSD by 10 % combined with a similar relative increase in computational cost, but clearly fell short of the accuracy of the VASP

MLP comprising 8 atom types suggesting that it is actually the distinction of different chemical environments that is most relevant here. A possibility for accelerating the potential would be to reduce the cutoff radius for the angular descriptors (ML_RCUT2=4). This, indeed leads to a minor increase of the efficiency, but at the cost of an (also minor) increase of the force RMSD.

The above considerations suggest that the most accurate VASP MLP for MOF-74 should be obtained by refitting the MLP using the ML_MODE=select feature in VASP 6.4.1 in combination with 8 atom types and (following the insights from the frequency RMSD comparison in Table 2) employing the extended reference data set. This is indeed the case, as shown by the purple down triangle in Figure 7 c. Notably, this resulted in a negligibly small force RMSD similar to the situation in MOF-5 at only marginally increased computational costs compared to the potential trained on the initial reference.

## Discussion

In this work we describe a strategy for obtaining machine learned potentials that allow the description of the properties of MOFs at essentially DFT quality with hugely reduced computational costs. This strategy uses the active learning approach of VASP[55,98] for sampling the configuration space and for generating reference structures. On these reference data, machine learned potentials can already be trained in VASP, which are highly accurate and efficient (especially when using the latest releases of the code). The resulting force fields are nearly two orders of magnitude more accurate, but also significantly slower than traditional potentials like UFF4MOF. If speed is the priority, the effective reference data generation procedure of VASP can be combined with the numerical efficiency and also high accuracy of moment tensor potentials (MTPs)[50,58]. They allow a flexible adjustments to their speed and accuracy by manually adjusting the size of the basis set leading to potentials, where the faster variants are only less than an order of magnitude slower than traditional transferable potentials while still maintaining an exceptionally high degree of accuracy. As all necessary codes (the VASP and MLIP software packages) are readily available, a general use of the proposed strategy should be comparably straightforward.

A systematic comparison of predicted forces, energies, stresses, phonon bands-structures, and thermal transport coefficients shows that the such-trained MTPs and VASP MLPs yield a truly amazing agreement with DFT reference data (as the primary benchmark quantities) and with experimental data in cases where DFT data are not available. The said tests were performed for a representative selection of isotropic as well as anisotropic MOFs including MOF-5, MOF-74, UiO-66, and the narrow and large pore phases of MIL-53. Compared to transferrable force fields like UFF4MOF the improvements in prediction quality is truly amazing (for some quantities prediction errors drop by several orders of magnitude) and this at only very moderately increased computational costs. Only when attempting to describe thermal expansion processes, distinct deviations between the machine learned potentials and experiments are observed, where there are some indications that in that case the reference DFT methodology might be at fault. Notably, in the case of MTPs for some systems also the thermal stability of the force fields needs to be further improved. This can become an issue when performing extended molecular dynamics runs comprising hundreds of thousands of time steps, but the problem can be mediated by including reference data from dedicated high temperature runs. The VASP potentials suffer substantially less from issues of thermal stability. This makes them a highly promising option especially for flexible materials with flat potential energy surfaces, particularly when computational speed is not the main priority.

Overall, the data presented here and other recent advancements in MOF force field development[26,44] clearly point towards machine-learned potentials as becoming a game-changer for the modeling of this complex class of materials. This particularly applies to the simulation of dynamical MOF properties, which are often computationally not accessible to ab-inito methods like DFT. Here, the use of VASP MLPs or MTPs can speed up the simulations by many orders of magnitude keeping essentially the accuracy level of the reference method used in their parametrization.

## Methods

In this section, we will discuss the most important settings and methods used in this work. More in-depth details and convergence tests are contained in the supplementary information.

### DFT approach

For the required reference ab-initio calculations, density functional theory (DFT) simulations were performed with the Vienna ab-initio Software Package (VASP)[99–104] employing the Perdew-Burke-Ernzerhof (PBE) functional[105,106] and applying Grimme's D3 correction with Becke-Johnson damping[107,108] to account for dispersive forces. For all training runs a plane-wave energy cutoff of 900 eV was chosen after careful convergence tests based on total energies, vibrational frequencies and elastic moduli. The energy convergence for the self-consistent field approach was set to $10^{-8}$ eV. The structures for each system were optimized until the maximum absolute force component in the system reached at least less than $10^{-3}$ eV/Å using a quasi-Newton optimization algorithm. Further details on the VASP settings including the system-specifically chosen k-space sampling mesh are provided in the supplementary information in Section S1.

### On-the-fly training of the VASP MLPs

All molecular dynamics simulations were carried out using a time step of 0.5 fs. The friction coefficient in VASP for the Langevin thermostats and barostats was chosen to be 10 ps$^{-1}$ and a fictious mass of 1000 amu was assigned for all systems as required for the Langevin barostat method of Parrinello and Rahman[109]. In the initial training runs, the temperature was set to gradually increase from 50 K to 900 K over 50,000 time steps. For the training of the VASP MLPs, the maximum number of local reference configurations (basis functions) in the machine learned force field approach was set to 3000 for each atomic species. The remaining machine learned settings were left at the default values as set in VASP version 6.3.0. This means that the Bayesian error threshold, which decides whether or not additional time steps are performed using DFT and added to the reference data, starts at 0.002 eV/Å and is allowed to adjust dynamically such that it can be different for the different systems and at different temperatures. To generate the independent test sets to validate energies, atomic forces and stress tensors of the individual force field potentials, another set of completely independent on-the-fly machine learning force field runs were performed from scratch in an NPT ensemble at 300 K with a time step of 0.5 fs. They were run until 100 DFT calculations had been performed, which then served as system-specific test sets. The machine learning approach was used in this case to improve the sampling of phase space compared to pure ab-initio MD.

Since the start of the project, new VASP versions, 6.4.0 and 6.4.1, have been released, which substantially improve the computational speed and also the accuracy of the VASP MLPs by optimizing the code and by improving default settings of the training. Therefore, for a true analysis of the VASP MLPs, we retrained all potentials in VASP version 6.4.1 for all system using "ML_MODE=select", which once again chooses the local reference configurations to be included in the potential with improved settings. The potentials were subsequently refitted without error estimation and employing a singular value decomposition with Tikhonov regularization[97] using "ML_MODE=refit". This substantially enhances their computational efficiency. This was done for both the initial and extended reference data sets using 4 atom types. For the initial reference data set potentials were also trained using a separation into several atom types based on their atomic neighborhood. The latter potentials are used for generating the displayed results unless explicitly stated otherwise, as they show improved accuracy with no extra training overhead. Moreover, they are more comparable to the MTPs. For MOF-74, we additionally investigated the impact of the extended reference data set when including the separation of atom types.

## Functional Form of the MTPs

The MTPs describe the atomic energies based on a fixed basis set with a static number of parameters that has to be defined by the user. The basis sets of the MTPs are built from moment tensor descriptors consisting of a radial and an angular part. The radial part is represented by a series of polynomials up to a certain order while the angular part is represented by a tensor up to a certain rank containing information about the geometric arrangement. Combinations of those moment tensors then define the basis functions for which the parameters are trained. How many such combinations are included in the potential, and therefore to what extent many-body interactions are considered, is defined by an arbitrarily constructed maximum "level" parameter, which represents the primary way to adjust the speed and accuracy of an MTP. Additionally, the radial basis set size defining the maximum order of the used polynomials can also be adjusted and affects the number of parameters in the MTP. Increasing these parameters will lead to an increased accuracy but also a reduced computational efficiency. Since we wish to show how well the MTPs reproduce the reference data, we use a relatively large basis set as a baseline with a level of 22 and a radial basis set size of 10. However, tests were also performed varying the level of the MTP between 10 and 24 for each system (see section S3.6 for detailed results) to show the impact of modifying the number of parameters contained in the MTPs. The MTPs were generally fitted including the separation of atom types since there is barely any extra computational cost involved. However, to test the importance of introducing this additional chemical knowledge and of increasing the number of parameters, MTPs at level 22 were also trained when only using 4 atom types for MOF-5 and MOF-74 (i.e., when each chemical element present in the MOF is represented by only a single atom type).

## MTP training

To train the MTPs, the reference data generated during the VASP on-the-fly learning runs were used. For determining the MTP parameters, it is necessary to minimize a cost function built from the energies, forces and stresses in the training set[58]. The minimization was performed using a BFGS (Broyden-Fletcher-Goldfarb-Shanno)[110] algorithm. The initial parameters were chosen randomly. Then an initial training was performed for 75 iterations. Subsequently the main minimization procedure was started and performed until the differences of the cost function in the previous 50 iterations reached less than $2 \cdot 10^{-3}$. Said cost function is built from a weighted sum over the square deviations of forces, energies and stresses. The weights in the cost function for stresses and energies were chosen to be 1.0 and for forces to be 0.01. These are the default values for forces and energies. For stresses, the weight was increased compared to the default value of 0.001, as we found that it improves the accuracy of the stresses without substantially impacting the precision of the forces or energies.

For most of the training runs, a few thousand iterations were sufficient to reach the convergence criterion. However, it should be noted that complications occurred for some of the resulting MTPs, where the minimization algorithm managed to escape the lowest minimum found during the process to ultimately reach a local minimum with a larger cost function. This can be remedied by either choosing looser convergence criteria or by choosing the intermediate solution with the lowest cost function. The latter approach was employed in this work for the cases where this issue was observed and no further minimum was found within 10000 optimization steps. For additional details and convergence curves for the MTP training, see the supplementary information section S2.2.

Due to the stochastic nature of the initialization of the training procedure, multiple potentials were trained for each system and reference data set. A possible strategy for choosing the final MTP would have been to pick the one with the smallest cost function. However, it is common practice to judge the quality of a fitted potential based on quantities independent of the parametrization procedure, like the properties of a validation set[111]. In this way, one can actually judge the performance of the interpolation and not just the description of the training set, which could also be described well by the MTP due to overfitting. Therefore, for the further analysis and for the calculation of physical properties, the potentials were chosen that best managed to reproduce the forces, energies and stresses in the validation set and that were stable up to at least 300K. In this context it should be mentioned that for high level MTPs (like the level 22 MTPs used for most parts of this manuscript), the variations in the properties by the multiply fitted potentials are typically rather small and picking the MTPs with the smallest cost functions would have made an only marginal difference for the force, energy and stress errors and would not have impacted any of the trends discussed here. The only exception is the thermal stability of the potentials, where some substantial deviations were found between the differently initialized MTPs. This is why training several MTPs can represent a workaround for issues related to the thermal stability. Corresponding details are shown in Table S8 in the supporting information. The data in that table also illustrate that it can happen that certain force fields provide a clearly worse description of the validation data without having a significantly increased cost function in the original parametrization. This supports the strategy of the picking the "ideal" potential based on its performance for a set of validation data.

## Evaluation of benchmark properties

When creating a new force field potential, proper validation is crucial and thus we aim to provide a large array of properties to benchmark based on DFT and experimental reference data. In general, the single point and MD simulations for the DFT references and the VASP MLP were performed directly using VASP and for the MTPs using the LAMMPS (Large-scale Atomic/Molecular Massively Parallel Simulator) package[112] interfacing MLIP[58].

Structure optimizations at 0 K for the machine learned potentials were carried out starting from the DFT relaxed structures and for the VASP MLPs the same optimizations settings were used as for the actual DFT calculations. For the MTPs an iterative process was used, where first the atomic positions and lattice parameters were optimized using a conjugate gradient algorithm. This was followed by an optimization of only the atomic positions and both steps were repeated 10 times. The end of either of the steps was defined as the situation, when either 1000 force evaluations had been performed (or technically also when the excessive convergence criterion of force changes falling to less than $10^{-8}$ eV/Å had been reached).

The elastic stiffness tensors were computed using a finite differences approach based on the stress-strain relationship, where the lattice was strained anisotropically by a fraction of 0.01. For each strain direction, an optimization of atomic positions was carried out with the MTP and the resulting stresses were used for the evaluation. For DFT and the machine learned potential in VASP, the strained cells were not individually optimized, but rather the contributions from relaxing the ions were computed from the second-order force constants in the harmonic approximations to save computation time (as implemented in VASP using the IBRION=6 setting).

Phonon band structures were computed using a supercell-based finite-differences approach as implemented in phonopy[113]. For the required atomic displacements, a displacement distance of 0.01 Å was used. The used supercells were carefully converged. Whenever possible, this was done based on DFT results. However, due to the large unit cells in MOFs, for some of the systems, supercell convergence tests were possible only based on the force-field results. For specific details regarding these tests and the supercells used for each system, view sections S1.4 and S3.5 in the supplementary information.

To evaluate the thermal expansion coefficients, molecular dynamics simulations in the NPT ensemble were performed for the MTPs at temperatures ranging between 100 K to 700 K. This was done over 100,000 time steps for supercells of the respective system (details can be found in the supplementary information, section S4.1). In addition to providing temperature dependent lattice parameters, this also allows a rough assessment of the thermal stability of the obtained force field potentials. To obtain the actual thermal expansion coefficients linear fits (sometimes in restricted temperature ranges) were performed over the averaged lattice parameters from the thermally stable MD simulations.

To obtain the thermal conductivity of MOF-5 for comparing the experimental data to the MTP simulations, non-equilibrium molecular dynamics (NEMD)[93] and "approach to equilibrium molecular dynamics" (AEMD)[114,115] simulations were performed. For NEMD, a supercell based on the lattice parameters at the desired temperature of 300 K obtained from the preceding NPT simulations was constructed, as detailed below. First, an equilibration at 300 K was performed in an NVT ensemble for 25 ps. Afterwards, the system was treated in an NVE ensemble while employing the Müller-Plathe algorithm[116], which periodically swaps the kinetic energies of the 16 highest energy atoms in the first slab of the supercell with the kinetic energies of the 16 lowest energy atoms in the center slab of the supercell every 4800 time steps. Here, each slab was chosen to be 12.98 Å long, encompassing one node and one linker of MOF-5. This leads to the creation of a hot and a cold region while maintaining the total energy in the system. As a consequence, a heat flux from the hot to the cold region emerges allowing the straightforward computation of the thermal conductivity using Fourier's law in combination with the temperature gradient. Finite size effects due to scattering at the thermostat boundaries were accounted for by computing the thermal conductivity for several different cell lengths and performing an extrapolation to the infinite size limit[92,93]. Perpendicular to the heat-flow direction, the supercell comprised 2 conventional unit cells For the AEMD simulations, we used a similarly shaped cell, but this time one half of the cell was set to a higher temperature of 350 K and the other half of the cell was set to a lower temperature of 250 K. This is realized by using a thermostat on one half of the supercell, while not solving the equations of motion for atoms in the other half. This step was repeated, but now equilibrating the other half of the supercell to its starting temperature, while the equations of motions were not solved for the already thermalized half of the supercell. Then the thermostats were turned off and the entire system was treated in an NVE ensemble. The temperature difference between both halves of the system was recorded as a function of time during the molecular dynamics simulation. The temperature decay can be fitted to a function containing the thermal diffusivity, which in turn leads to the thermal conductivity[115]. Similar to NEMD, AEMD is also prone to finite size effects, which can be corrected using a slightly different extrapolation approach than for NEMD derived from the Boltzmann transport equation[117]. For more information regarding thermal conductivity calculations (including a justifications for the used supercell sizes) see Supplementary Information section S4.2.

To evaluate the computational efficiency separate simulations in an NPT ensemble were computed at 300 K using large supercells comprising 27136 atoms for MOF-5 and 31104 atoms run for at least 2000 time steps. Here, care was taken not to include the initialization overhead and not to lose a significant amount of time to outputting the trajectories.

For obtaining observables based on the traditional force field potentials for MOF-5 and MOF-74, LAMMPS was employed using the same methods as specified for the MTPs. To assign the parameters for the Dreiding and UFF4MOF potentials to the structure files, the lammps-interface code written by Boyd et al. was used[14]. This code has been developed to benchmark various force fields for several MOFs like MOF-5, UiO-66 and IRMOF-10. For a more specific discussion and analysis regarding UFF4MOF, see the Supplementary Information section S3.1 and S3.5. To compare the machine-learned potentials to a system-specifically fitted traditional force fields, a MOF-FF type FFP was used. The parametrization of that force field is described in detail in our previous work[33].

## Data availability

The datasets generated and/or analyzed during the current study are available in the TU Graz Repository; https://doi.org/10.3217/wyc7s-8en40

## Code availability

VASP can be acquired from the VASP Software GmbH (see www.vasp.at/); LAMMPS is available at www.lammps.org; MLIP is available at mlip.skoltech.ru/download; the lammps-mlip interface is available at gitlab.com/ashapeev/interface-lammps-mlip-2; Phonopy is available at phonopy.github.io/phonopy; further scripts used to compute some of the properties of the MTPs, like the elastic stiffness tensors, are publicly available at https://doi.org/10.3217/wyc7s-8en40.

# Acknowledgements

The authors thank Lukas Legenstein, Lukas Reicht, Nina Strasser and Florian Lindner for insightful discussions. The authors also thank the group of Georg Kresse, in particular the machine learned force field team Ferenc Karsai, Andreas Singraber and Jonathan Lahnsteiner for their help in the training of the VASP MLPs. S.W. is a recipient of a DOC Fellowship of the Austrian Academy of Sciences at the Institute of Solid State Physics (26163). This project also received funding from also from the Graz University of Technology through the Lead Project Porous Materials @ Work (for Sustainability) (LP-03) and from the Austrian Science Fund (FWF) [P33903-N]. The computational results presented have been generated using the Vienna Scientific Cluster (VSC).

# Author Contributions

SW and EZ conceptualized the presented research. SW performed all simulations and continuously discussed the outcomes with EZ and his group (see Acknowledgements). The first draft of the manuscript was written by SW and was revised by him and EZ in several iterations including input from the Kresse group. All authors read and approved the final manuscript.

# Competing Interests

All authors declare no financial or non-financial competing interests.

# Machine learned Force-Fields for an ab-initio Quality Description of Metal-Organic Frameworks

Sandro Wieser[1], Egbert Zojer[1]

[1]Institute of Solid State Physics, Graz University of Technology, NAWI Graz, Petersgasse 16, 8010 Graz, Austria

# Table of Contents

# S1. DFT settings, convergence and computation of the benchmarking data

## S1.1 Overview of the chosen settings

To add more details regarding the used settings for the DFT simulations carried out by VASP, Table S1 shows the chosen k-point sampling and energy cutoff for each system. Detailed convergence tests for all the settings can be found in section S1.2. Since the accuracy of the forces is vital for many of the investigated properties, the precision-mode in VASP[1–3] was set to PREC=Accurate. The electronic minimization algorithm was set to ALGO=FAST for all systems except for UiO-66, where ALGO=NORMAL was required to consistently achieve SCF convergence. A Gaussian smearing was used for the partial occupancies of the orbitals with a width of SIGMA = 0.05 eV. To account for dispersive forces, the D3 scheme by Grimme et al. with Becke-Johnson damping was used[4,5]. The used pseudopotentials required for the projector-augmented wave (PAW)[6] method are listed in Table *S2* for each element. The evaluation of the projection operators was set to be performed in reciprocal space (LREAL=.FALSE.) for improved accuracy.

*Table S1: Chosen $\Gamma$-centered k-point mesh and energy cutoff for the investigated materials. The values were carefully chosen as described in section S1.2. For MIL-53 (lp) the same k-point sampling mesh as for MIL-53 (np) was used for the molecular-dynamics steps since a phase transition between lp and np would be technically possible. For MIL-53 (np) a denser k-point sampling mesh of 2×4×4 was used for the reduced unit cell containing only 38 atoms.*

| System | k-point sampling | Energy cutoff / eV |
|---|---|---|
| MOF-5 | 1×1×1 | 900 |
| UiO-66 | 1×1×1 | 900 |
| MOF-74 | 2×2×2 | 900 |
| MIL-53 (lp) | 1×2×1* | 900 |
| MIL-53 (np) | 1×2×2 (2×4×4) | 900 |

*Table S2: Used PAW pseudopotentials for each element. The VASP default sets were used. For the title, the names given after the "TITEL" flag in the respective POTCAR files is listed.*

| Element name | PAW PBE title |
|---|---|
| Zn | PAW_PBE Zn 06Sep2000 |
| Al | PAW_PBE Al 04Jan2001 |
| Zr | PAW_PBE Zr_sv 04Jan2005 |
| O | PAW_PBE O 08Apr2002 |
| C | PAW_PBE C 08Apr2002 |
| H | PAW_PBE H 15Jun2001 |

## S1.2  K-point and energy cutoff convergence and minimization of the structures

The k-point mesh and energy cutoff for all the systems were converged until the deviation in the total energy was below 1 meV/atom. However, this was not the only requirement, because it is far more important that the actually investigated properties are converged. Especially elastic tensor elements have been proven to require highly accurate DFT settings[7,8]. This is why we also provide convergence tests regarding the phonon properties and elastic constants.

For the relaxation procedure, symmetries were turned off in VASP and all constraints were removed. The energy was then minimized until the maximum absolute force component in the system was below $10^{-3}$ eV/Å. Due to the very flat energy landscape of MIL-53, the initial relaxation for each phase at 900 eV energy cutoff and the k-point mesh specified in Table *S2* was performed as a 2-step procedure. In this case, a second relaxation run starting from the final structure from the first run was appended to minimize the effect of Pulay stresses. In the end, the result did not change much using this approach due to the relatively large plane-wave energy cutoff which already decreases the impact of Pulay stresses. This approach and the full relaxation from an experimentally obtained structure was only performed for one of the settings, for the other settings an already pre-relaxed cell was used as the input to save computation time.

Fig. S1 shows the convergence of the DFT obtained total energy for each system for two cases: the single point energy for the same atomic positions without further optimizations using the tested settings and also the total energy after a full relaxation. The k-point mesh used for these tests corresponds to Table S1. It can be seen in (a), that beginning with an energy cutoff of 800 eV, the differences in energy are below 1 meV/atom which is one of our chosen convergence criteria. In (b) the situation looks relatively similar overall, but for the large pore phase of MIL-53 the energy at a cutoff of 700 eV also seems very similar to the higher settings. However, this is just an artifact of the relaxation run finding a different minimum structure. In fact, one of the lattice parameters reaches a value of 18.93 Å at 700 eV instead of the 17.26 Å obtained for the higher settings. This emphasizes the extremely flat energy landscape of MIL-53 and illustrates, why a careful convergence of the chosen settings is crucial.

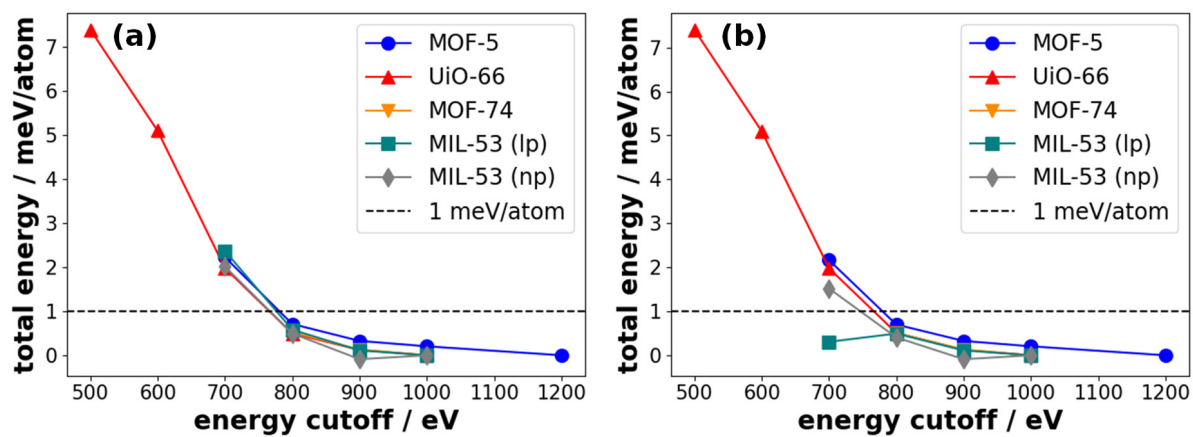

*Fig. S1: Convergence of the total energy obtained from DFT as a function of the energy cutoff for the same unit cell (a) and after a full relaxation (b) for each system. The origin of the energy axis was defined as the total energy calculated for the highest energy cutoff. The black dashed line indicates the chosen convergence criterion of the energy differences being smaller than 1 meV/atom.*

In Table S3 the total energies for various k-point sampling meshes are shown. It can be seen that the finally chosen k-point meshes are less than 1 meV/atom different compared to their denser counterparts for all of the systems. Another interesting aspect in this table is the small difference of the energies for the two phases of MIL-53 where the narrow pore phase shows a less than 3.8 meV/atom lower energy.

Regarding the convergence of MIL-53 (np), it should be noted that a smaller unit cell was found based on the system symmetries after relaxation. This smaller cell contains only 38 atoms. The considerations in this section apply to the larger cell with 76 atoms. However, the smaller cell was used in the phonon calculations and the evaluation of the elastic stiffness tensor. This will lead to a lower number of phonon modes and a differently oriented elastic tensor. This smaller cell shows a different orientation with a=10.42 Å, b=10.42 Å, c=6.66 Å, α=103.4°, β=103.4°, γ=36.6° and has space group C2/c (15). The difference between the two symmetry-equivalent cells is visualized in Fig. S2. Due to the smaller extent and the different orientation of the new cell, a different k-point sampling had to be used. Here, we chose a denser 2×2×4 grid to ensure convergence.

*Table S3: Total energies of the investigated systems for various k-point meshes at an energy cutoff of 900 eV. Total energies are provided for both a single point calculation of the same atomic positions and lattice parameters for each setting and also for the final setting-specific relaxed geometry. The k-point mesh reaching the desired maximum energy difference of 1 meV/atom compared to the highest tested setting is indicated in bold.*

| System name | k-point mesh | Energy cutoff / eV | Total energy / meV/atom | Total energy after relaxation / meV/atom |
|---|---|---|---|---|
| MOF-5 | **1×1×1** | 900 | -6850.6 | -6850.6 |
| | 2×2×2 | 900 | -6850.6 | -6850.6 |
| UiO-66 | **1×1×1** | 900 | -7525.0 | -7525.0 |
| | 2×2×2 | 900 | -7525.3 | -7525.3 |
| MOF-74 | 1×1×1 | 900 | -6980.7 | -6981.2 |
| | **2×2×2** | 900 | -6987.0 | -6987.0 |
| | 3×3×3 | 900 | -6987.1 | -6987.1 |
| MIL-53 (lp) | 1×1×1 | 900 | -7121.8 | -7130.9 |
| | **1×2×1** | 900 | -7133.1 | -7133.1 |
| | 2×3×2 | 900 | -7133.1 | -7133.1 |
| MIL-53 (np) | **1×2×2** | 900 | -7136.9 | -7136.9 |
| | 2×4×4 | 900 | -7136.7 | -7136.7 |
| | 2×6×6 | 900 | -7136.7 | -7136.7 |

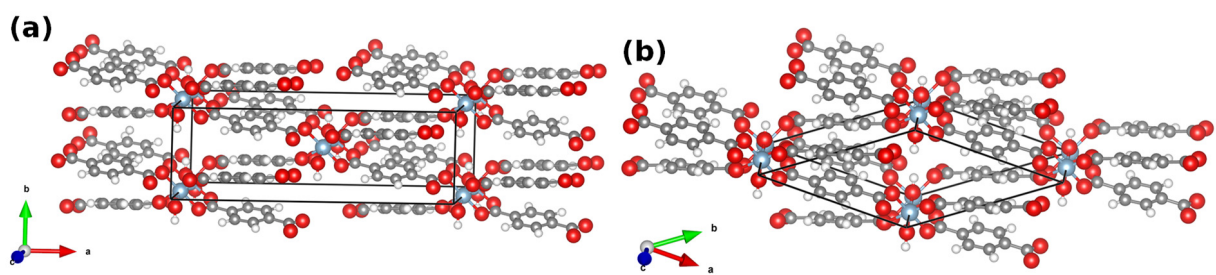

*Fig. S2: Unit cell of MIL-53 (np) with 76 atoms (a) and the smaller but equivalent 38 atom unit cell (b).*

Another oddity that occurred is the breaking of the orthorhombic symmetry during the relaxation of MIL-53 (lp) leading to a more energetically favorable structure with space group Cc (9). In this case, slight deviations from the 90° angles were observed for the lattice parameters amounting to ± 1° for two equivalent minima. To visualize this, a comparison of the orthorhombic structure and non-orthorhombic structure is shown in Fig. S3. Additionally depicted is the effect of the hydrogens in the nodes bending in a way that they do not align with the orientation of the node octahedra any more. The origin of that bending is that the straight geometry (central panel) represents an energetic saddle point. The structure becomes energetically favorable when the hydrogens bend either way. This is shown in Fig. S4 by calculating the total energy of the system for displacing the structure along the relevant phonon eigenvector obtained for the structure with the central hydrogens (see Fig. S3 (a)). This leads to imaginary phonon frequencies for that structure. However, it should be stressed that the large pore phase of MIL-53 usually occurs only at higher temperatures, while the data in Fig. S4 represent tiny energy differences at 0 K. This means that in reality the experimentally measureable phase would be the average between the two minimum configurations. We confirmed this using our trained force field potentials where the slight tilt angle disappeared on average for molecular dynamics simulations for temperatures above 150 K, as shown in detail in section S4.1. This implies that what is observed here is merely a low temperature effect, but it is still important, as the lower degree of symmetry, for example, increases the number of non-zero elastic tensor elements compared to what has been reported in other publications[9,10].

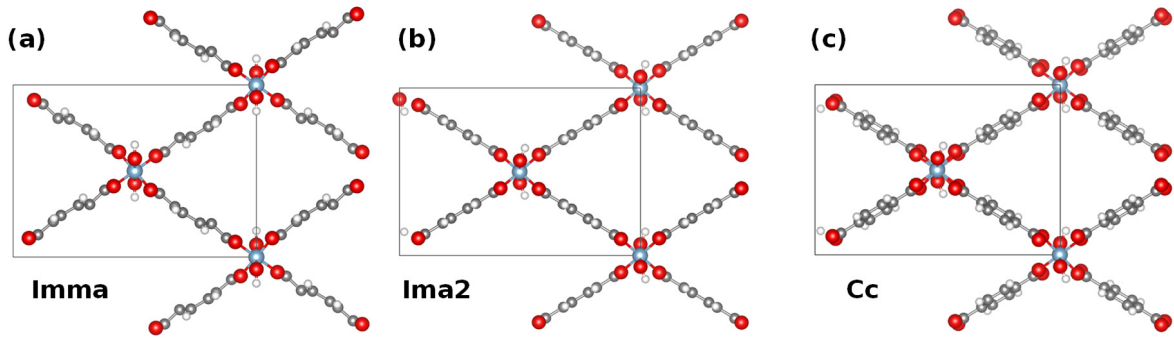

*Fig. S3: MIL-53 (lp) at various stages of the energy minimization procedure. In (a) one sees the structure with space group Imma (that has also been reported in experiments[11,12]) with hydrogens aligned centrally between the linkers. In (b) a lower energy structure after further minimization is shown, where the hydrogens are bent in a consistent direction reducing the space group to Ima2 and in (c) a structure with even lower energy is shown, with a monoclinic angle of 91° and rotated linkers reducing the space group to Cc.*

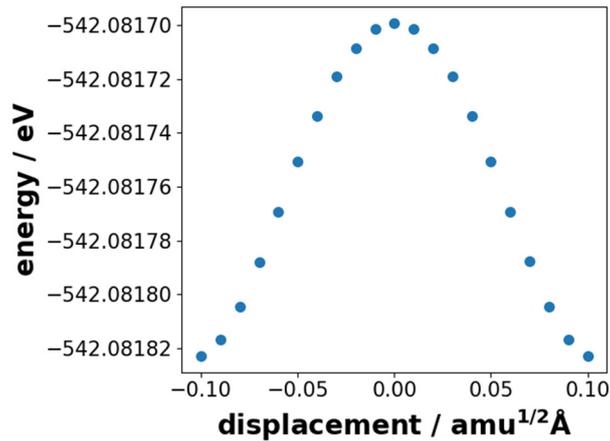

*Fig. S4: Total energy as a function of displacement distance for the Imma structure of MIL-53 displaced along the imaginary phonon mode describing the bending motion of the central hydrogens.*

### S1.3 Computation of the elastic tensor elements

In the following, convergence tests regarding the computation of the elastic tensor elements will be provided. The relevant quantity to converge is the applied fractional strain distance which was chosen as 0.01 in this work. The distance should, on one hand, be as small as possible so that the finite difference approach can still be applied but, on the other hand, should not be too small not to seriously suffer from numerical noise. Fig. S5 shows the elastic tensor elements in MOF-74 as a function of the strain distance. As can be seen, a reduction of this distance below 0.01 in the strain leads to no visible differences. The tests for slgihtly larger strain distances up to 0.02 were already performed in [7] and were also found to be inconsequential. Technical note: the test provided here only investigates the impact of the strain distance and not the displacement distance to obtain the interatomic force constants. When computing elastic constants as implemented in VASP (IBRION=6), the same parameter (POTIM) controls both the strain and interatomic displacement distances. If one changes this parameter, severe numerical issues occur at a displacement distance of 0.001 Å, while at 0.005 Å no significant differences were observed compared to 0.01 Å. Conclusively, for all displacement-based calculations of the force constants (which are also required to compute the phonons) in this work, we used a displacement distance of 0.01 Å. For all computations of elastic constants a relative strain of 0.01 was used.

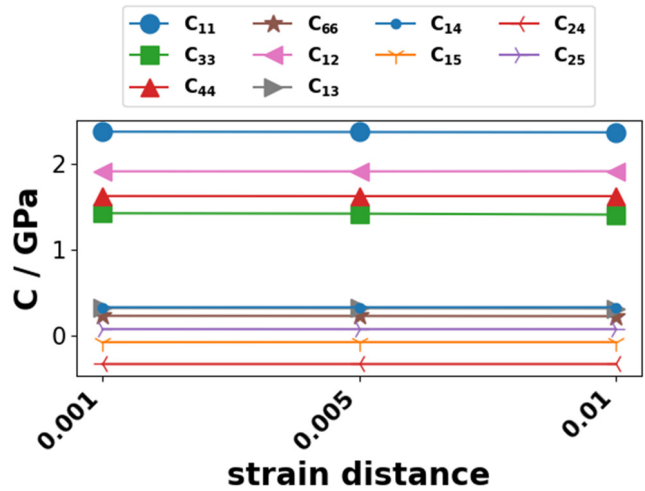

*Fig. S5: Comparison of all unique elastic tensor elements in MOF-74 for different strain distances.*

Additionally, we provide the energy cutoff convergence tests for the elastic tensor elements for each system in Fig. S6. This is done, despite the total energy changing only insignificantly after reaching 800 eV as shown in the previous section, because the objective is to converge the actual properties of interest. The guideline of converging energies down to less than 1 meV/atom is somewhat arbitrary and it depends on the objective of the calculation how severe the impact of the inaccuracy is. Here, the evaluation was performed using the optimized cell for the respective setting. In Fig. S6, we see that especially for MIL-53 (lp) and MOF-74 very high energy cutoffs of at least 900 eV are required to obtain properly converged elastic constants. For MOF-5, UiO-66 and MIL-53 (np) the difference for the different energy cutoffs is not that significant. However, small differences are still apparent even beyond a cutoff of 900 eV. So, if one would require extremely accurate elastic properties, going to even higher energy cutoffs might be necessary. For our purposes, however, we limit ourselves to 900 eV for each system as our main objective is not to provide the most accurate values possible, but rather to reproduce the DFT reference data with the machine learned potentials. In passing we note that in [7] it is shown for MOF-74 that 900 eV provide rather well converged values for the elastic tensor but that even higher cutoffs are required for converging the compliance tensor.

Fig. S7 shows the differences between the tested k-point sampling meshes. As can be seen, for all systems the k-meshes chosen based on the energy convergence listed in Table S3 are also sufficient for converging the elastic constants. Here, it should be noted, that for MIL-53 (np) the smaller 38 atom unit cell was used for the computation of the elastic tensor elements, which is the reason why the k-point convergence is very different than for MIL-53 (lp).

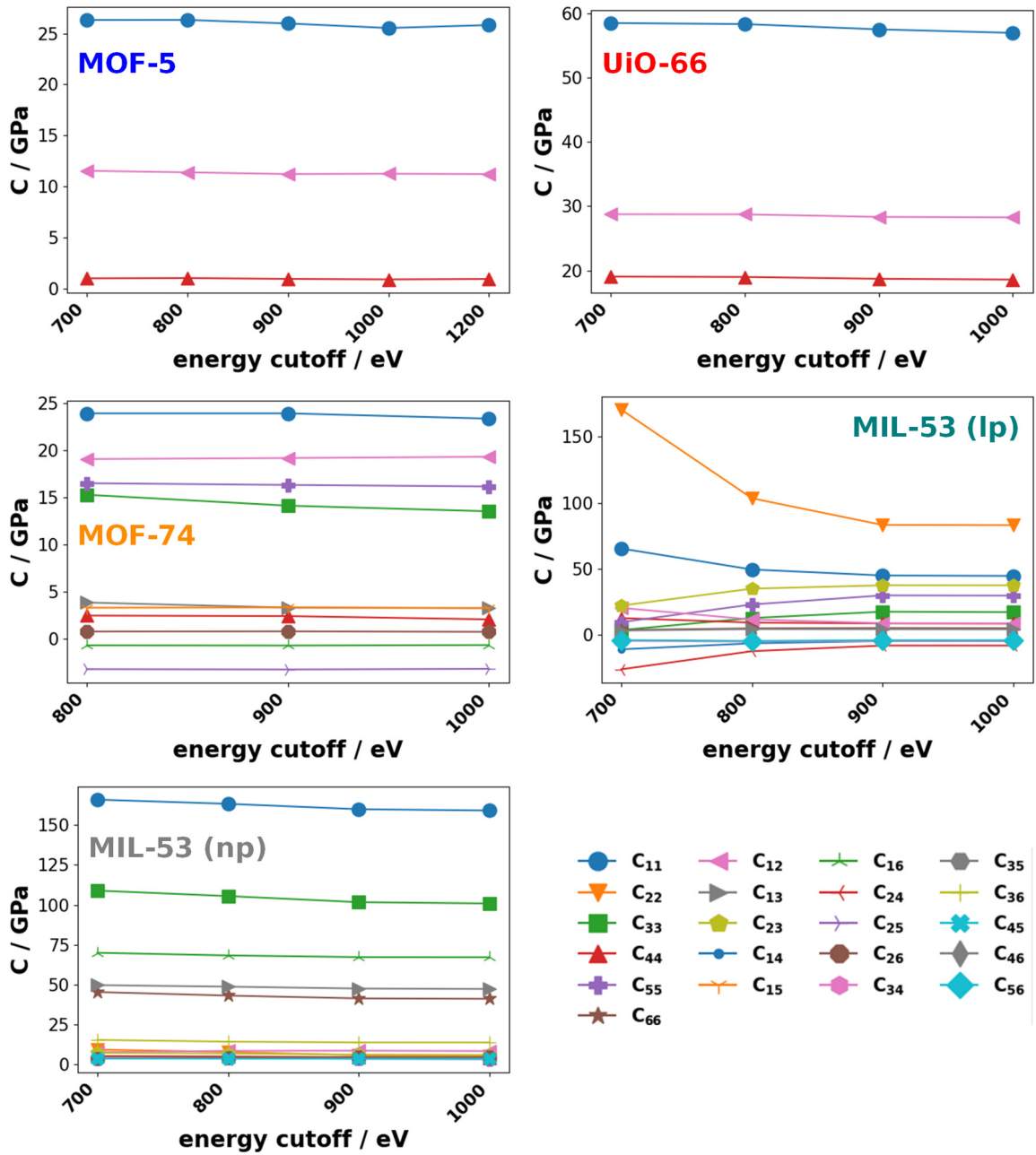

Fig. S6: Values of the unique elements of the elastic tensor for different energy cutoffs in the DFT calculations for each of the studied systems.

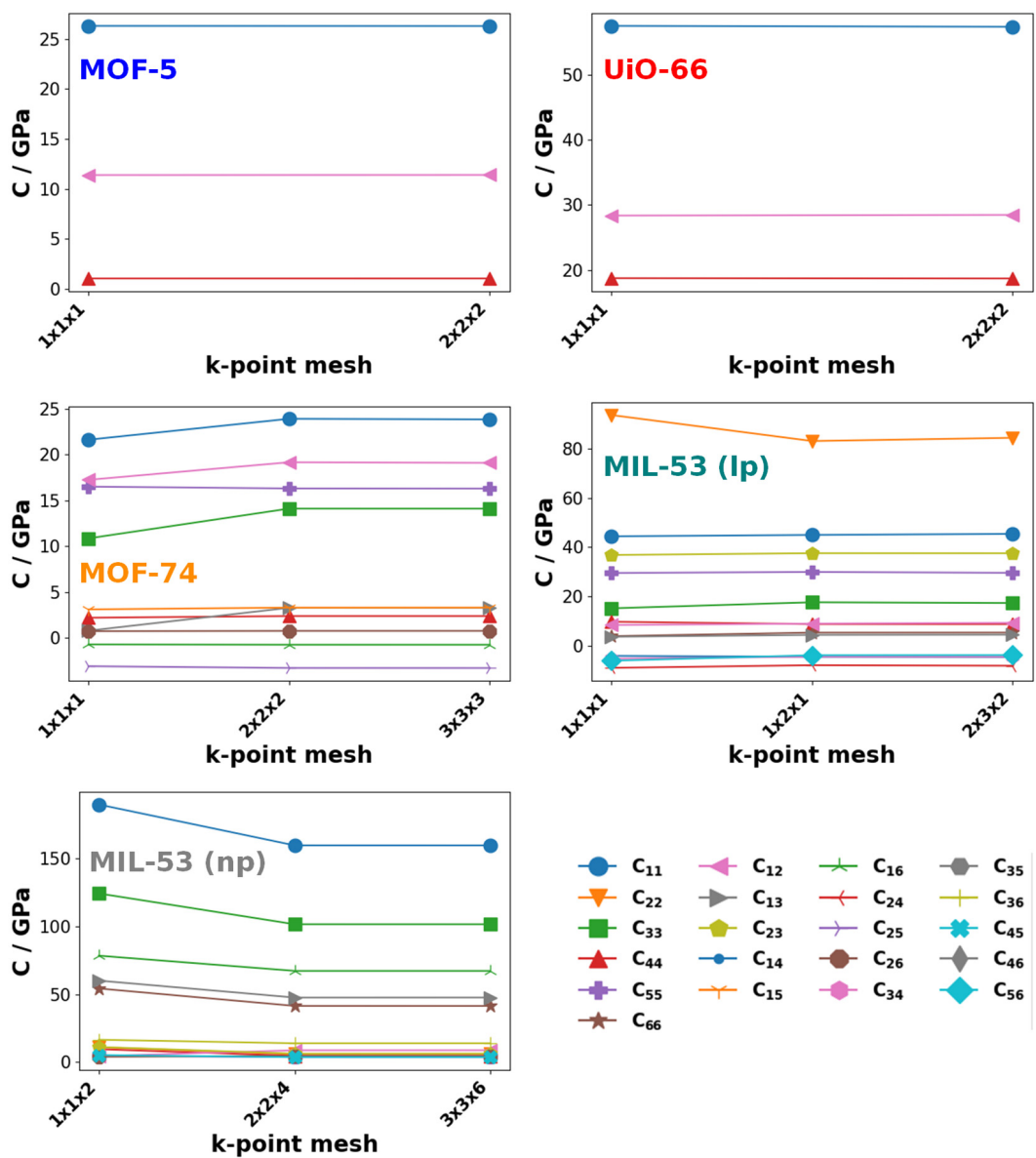

*Fig. S7:  Values of the unique elements of the elastic tensor for different k-point sampling meshes in the DFT calculations for each of the studied systems. For MOF-5, the elastic constants were computed for an energy cutoff of 800 eV and for all the other systems at 900 eV.*

For the precise numerical values of the elastic tensor elements refer to the comparisons with the machine learned potentials in section S3.1. When computing the elastic tensor elements with DFT, the harmonic force constants are also provided. These can also be used to compute the phonon frequencies at the Γ point. So, we can also immediately provide the convergence tests for the vibrational frequencies. Due to the large number of modes in the MOFs, we will only report the RMSD values compared to the highest used energy cutoff in Table S4. Additionally, tests for denser k-point sampling meshes are provided in Table S4. It can be seen that for all systems after reaching an energy cutoff of 900 eV and also for the denser k-point meshes than those that were ultimately used for the results, the RMSD values fall below a very small value of less than 1 cm$^{-1}$.

*Table S4: Deviations in the phonon frequencies at Γ for various k-point meshes and energy cutoffs. The RMSD values for results obtained with the same k-point sampling mesh (k-point mesh that was actually used for the presented results) are given compared to the entry in bold, which is for the highest energy cutoff used for the respective system. RMSD values for different k-point meshes are given compared to the results indicated with an asterisk, which is another calculation with the same energy cutoff for the k-point mesh that was actually used in the presented results.*

| System name | k-point mesh | Energy cutoff / eV | Γ-frequency RMSD / cm$^{-1}$ |
|---|---|---|---|
| MOF-5 | 1×1×1 | 700 | 1.91 |
| | 1×1×1 | *800 | 0.65 |
| | 1×1×1 | 900 | 0.36 |
| | 1×1×1 | 1000 | 0.38 |
| | **1×1×1** | **1200** | - |
| | 2×2×2 | 800 | 0.05 |
| UiO-66 | 1×1×1 | 700 | 1.40 |
| | 1×1×1 | 800 | 0.64 |
| | 1×1×1 | *900 | 0,16 |
| | **1×1×1** | **1000** | - |
| | 2×2×2 | 900 | 0.46 |
| MOF-74 | 1×1×1 | 900 | 2.82 |
| | 2×2×2 | 800 | 0.80 |
| | 2×2×2 | *900 | 0.16 |
| | **2×2×2** | **1000** | - |
| | 3×3×3 | 900 | |
| MIL-53 (lp) | 1×1×1 | 900 | 8.12 |
| | 1×2×1 | 700 | 4.02 |
| | 1×2×1 | 800 | 1.55 |
| | 1×2×1 | *900 | 0.12 |
| | **1×2×1** | **1000** | - |
| | 2×3×2 | 900 | 0.35 |
| MIL-53 (np) | 1×1×2 | 900 | 20.54 |
| | 2×2×4 | 700 | 3.86 |
| | 2×2×4 | 800 | 1.68 |
| | 2×2×4 | *900 | 0.09 |
| | **2×2×4** | **1000** | - |
| | 3×3×6 | 900 | 0.03 |

## S1.4 Computation of the phonon band structures

The phonon band structure calculations were performed employing phonopy[13] using a finite difference approach in a supercell of the respective material. The atomic displacement distance was chosen as 0.01 Å and the supercell sizes are shown in Table S5. Due to the large size of the supercells, the k-point mesh was set to Γ only for all systems. Due to the large computational expense of computing even larger supercells, convergence tests beyond these specified cells, were only carried out using the later trained MTPs which will be discussed in section S3.4. However, DFT includes longer range contributions to the forces, so the convergence is not entirely transferable. Therefore, some deviations might still occur. The remaining error is, however, expected to be small in view of the quite large supercells considered here. Additionally, we note that the supercell matrix chosen for MOF-5 and UiO-66 is used to create the cubic conventional unit cell from the primitive unit cell (for a visualization of the different unit cells for MOF-5, see the supporting information of [14]). The supercell for MOF-74 is used to create the hexagonal conventional unit cell from the primitive unit cell (for a more in-depth explanation and visualization of the different MOF-74 unit cells, see the supporting information of [7]). These differently shaped unit cells are then also used as a baseline to span larger supercells for the molecular dynamics simulations which were computed using the force field potentials.

*Table S5: Supercells used for the phonon band structure calculations for each of the investigated systems.*

| System name | Supercell |
|---|---|
| MOF-5 | $\begin{pmatrix} -1 & 1 & 1 \\ 1 & -1 & 1 \\ 1 & 1 & -1 \end{pmatrix}$ |
| UiO-66 | $\begin{pmatrix} -1 & 1 & 1 \\ 1 & -1 & 1 \\ 1 & 1 & -1 \end{pmatrix}$ |
| MOF-74 | $\begin{pmatrix} 1 & 0 & 3 \\ -1 & 1 & 3 \\ 0 & -1 & 3 \end{pmatrix}$ |
| MIL-53 (lp) | $\begin{pmatrix} 1 & 0 & 0 \\ 0 & 2 & 0 \\ 0 & 0 & 2 \end{pmatrix}$ |
| MIL-53 (np) | $\begin{pmatrix} 2 & 0 & 0 \\ 0 & 4 & 0 \\ 0 & 0 & 2 \end{pmatrix}$ |

For the large pore phase of MIL-53, some small imaginary frequencies close to the Γ-point could be observed. However, this was corrected by applying a non-analytical term correction (NAC) [15–17] using the born-effective charges computed from the relaxed unit cell. The difference of the phonon bands with and without NAC correction is shown in Fig. S8. Beyond these imaginary frequencies, the low frequency phonon band structures with and without NAC are very similar deviate only slightly.

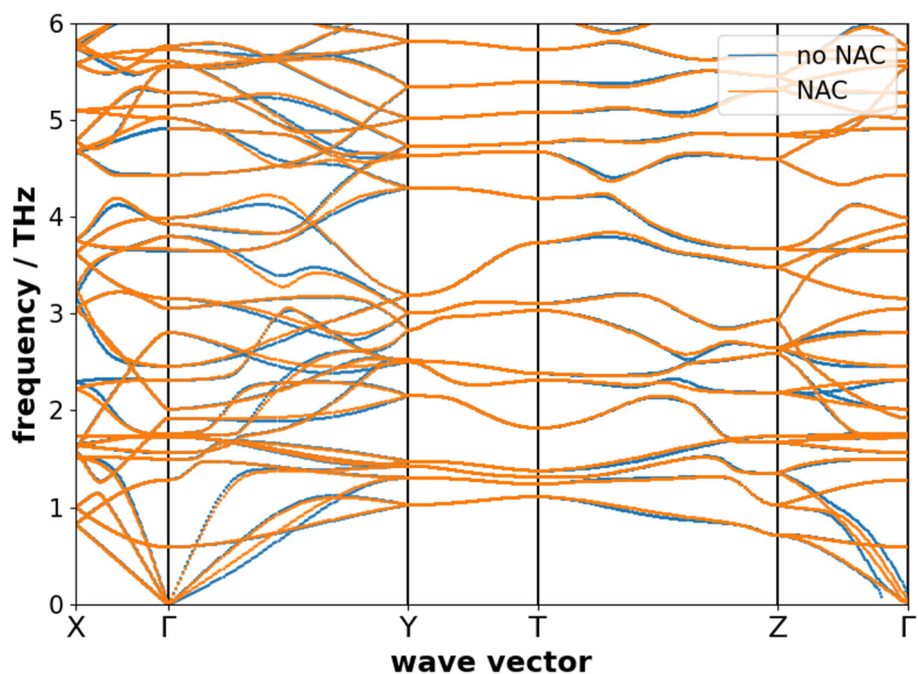

*Fig. S8: Comparison of MIL-53 (lp) low frequency phonon band structures without (blue) and with (orange) non-analytical term correction applied.*

For the narrow pore phase of MIL-53 (np), calculating the phonon band structures led to imaginary modes beyond the Γ point as can be seen in Fig. S9 for a 2×4×2 supercell. This means, imaginary modes occur at the X point which for the chosen supercell should be exact, as it is commensurate with the used supercell. This could mean that actual crystal has a larger unit cell than the originally chosen one. So, we also tested a larger unit cell for building the supercells. This was the usually reported cell containing 76 atoms. After another relaxation and a new phonon computation, a new set of phonon bands was obtained, which once again showed imaginary modes at the X point. The procedure was repeated for a supercell of this larger cell – which was then used as the base unit cell – which did not resolve the issue either. Unfortunately, due to mounting computational cost it was deemed too expensive to consider even larger unit cells. A possible origin of the complications could be that the narrow-pore phase of MIL-53 is not stable at zero K and becomes a minimum of the free-energy landscape only at elevated temperatures. However, studying such aspects would go far beyond the scope of the present paper.

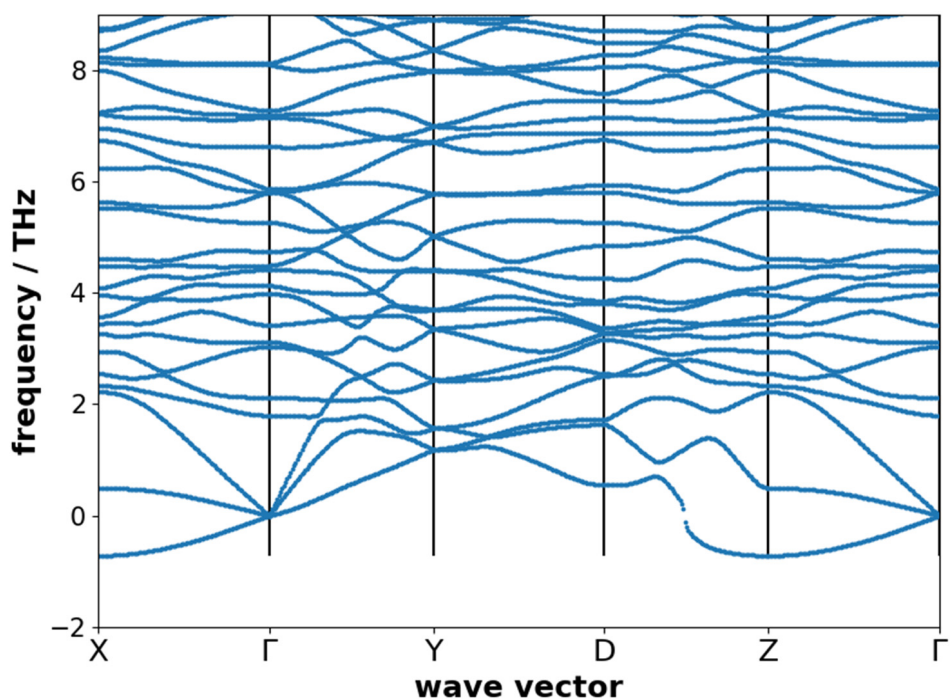

*Fig. S9: Phonon band structure of MIL-53 (np) obtained from PBE forces showing imaginary phonon frequencies (here presented as negative frequencies) in particular at the X and Z points.*

## S2. Training of the machine learned potentials

For the training of the machine learned potentials (except during the VASP active learning runs) the elements in the materials are further split into atom types depending on their neighborhood. For this, by utilizing a set of common covalent radii of the contained atoms, a list of immediate ("bonded") neighbors was created and atoms with an equivalent list were assigned an atom type each. For the specific types used, refer to Table S6.

*Table S6: Atom types used for the machine learned potentials for each system. The types are given in a form, where the first symbol indicates the atom type element and the subsequent number specifies the number of atoms in its immediate neighborhood. After the underscore a list of element symbols is given followed by how many atoms of this element are in the base atom's immediate neighborhood.*

| System name | Metal types | O types | C types | H types |
|---|---|---|---|---|
| MOF-5 | Zn4_O4 | O4_Zn4 | C3_C1O2 | H1_C1 |
| | | O2_C1Zn1 | C3_C2H1 | |
| | | | C3_C3 | |
| UiO-66 | Zr8_O8 | O4_H1Zr3 | C3_C1O2 | H1_C1 |
| | | O3_Zr3 | C3_C2H1 | H1_O1 |
| | | O2_C1Zr1 | C3_C3 | |
| MOF-74 | Zn4_O4 | O3_C1Zn2 | C3_C1O2 | H1_C1 |
| | | O2_C1Zn1 | C3_C2O1 | |
| | | | C3_C2H1 | |
| | | | C3_C3 | |
| MIL-53 | Al6_O6 | O2_Al1C1 | C3_C1O2 | H1_C1 |
| | | O3_Al2H1 | C3_C2H1 | H1_O1 |
| | | | C3_C3 | |

## S2.1 Performing the learning with VASP

As mentioned in the main manuscript, for the initial training runs the force threshold criterion during active learning was allowed to adjust dynamically over the course of the simulation based on the average Bayesian error over the previous 10 training steps. In Fig. S10 the value of this threshold criterion is shown as a function of time step. As can be seen, the value starts between 0.01-0.03 eV/Å and shows a steady increase over time (and, thus, temperature) for all systems. This is due to the larger displacements present at higher temperatures, which also show larger absolute force values making it more difficult to maintain the same absolute errors. It is also notable, that MOF-5 shows a substantially lower threshold criterion than all the other systems. A possible explanation for this is that MOF-5 is the most simple system considered here, without displaying a pronounced anisotropy like MOF-74 or MIL-53 and also with a smaller inorganic node than UiO-66. Here it should be noted that compared to the situation in MOF5, the larger node of UiO-66 contains more types of interactions that have to be modeled accurately. At the target temperature of 300 K, we see that the threshold criterion reaches values between 0.025 eV/Å and 0.035 eV/Å which is the main motivation why the fixed criterion was set to 0.02 eV/Å for the extended training runs. A lower criterion has to be chosen to add additional reference configuration to the machine learned potential, but it should also not be chosen too low, such that the MD run would be performed nearly exclusively with DFT.

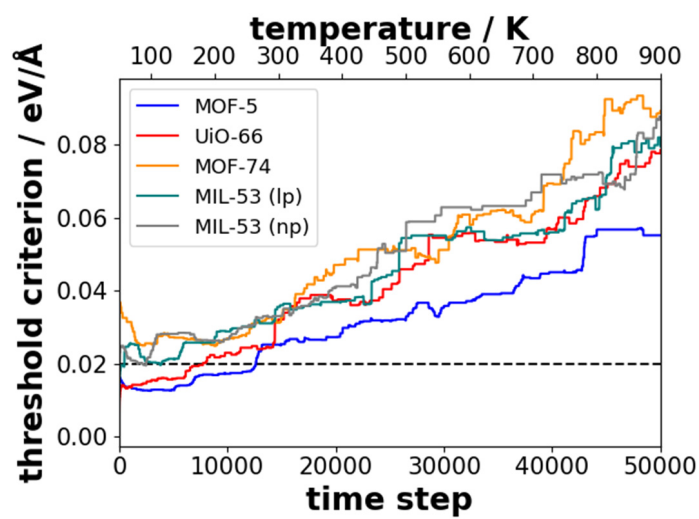

*Fig. S10: Evolution of the dynamically adjusted threshold criterion during the initial VASP active learning runs for each investigated systems as a function of time step. The current system temperature is indicated which results from the heating during training. The black dashed line indicates the fixed threshold criterion for the extended training runs. In the course of the run, the temperature of the system has been increased steadily from 50 to 900 K.*

Further details regarding the initial training procedure for the investigated systems are shown in Fig. S11. In panels a,b, and c the evolutions of the force/energy/stress RMSDs are shown as a function of time step. It can be seen that similar to the increase in threshold criterion, the RMSDs rise as more reference configurations are added at higher temperatures. While the trends are similar, for energy and forces, the anisotropic systems show a higher error than the isotropic MOF-5 and UiO-66. For the stresses more pronounced differences between the systems can be observed, which is largely a result of differences in the absolute values of the stresses experienced by the various systems. In Fig. S11 d the cumulative number of time steps computed with DFT is shown. All systems show a strong increase in added reference data in the early stage of the training compared to the later stages. Notably, for MOF-5 and MOF-74 the rate at which new reference structures are added at higher temperatures is larger than in the other systems. In Fig. S11 e the number of local reference configurations (basis function) used by the VASP MLP is shown. A large portion of the configurations is added immediately after the start of the training to achieve a reasonable initial force field potential. Afterwards, we can observe a steady increase of added configurations for all systems. The different systems show a different number of configurations that were added during the initial training phase, where MOF-5 and MOF-74 show the lowest initial number. However, at the end of the run all systems show a relatively similar number of total local reference configurations.

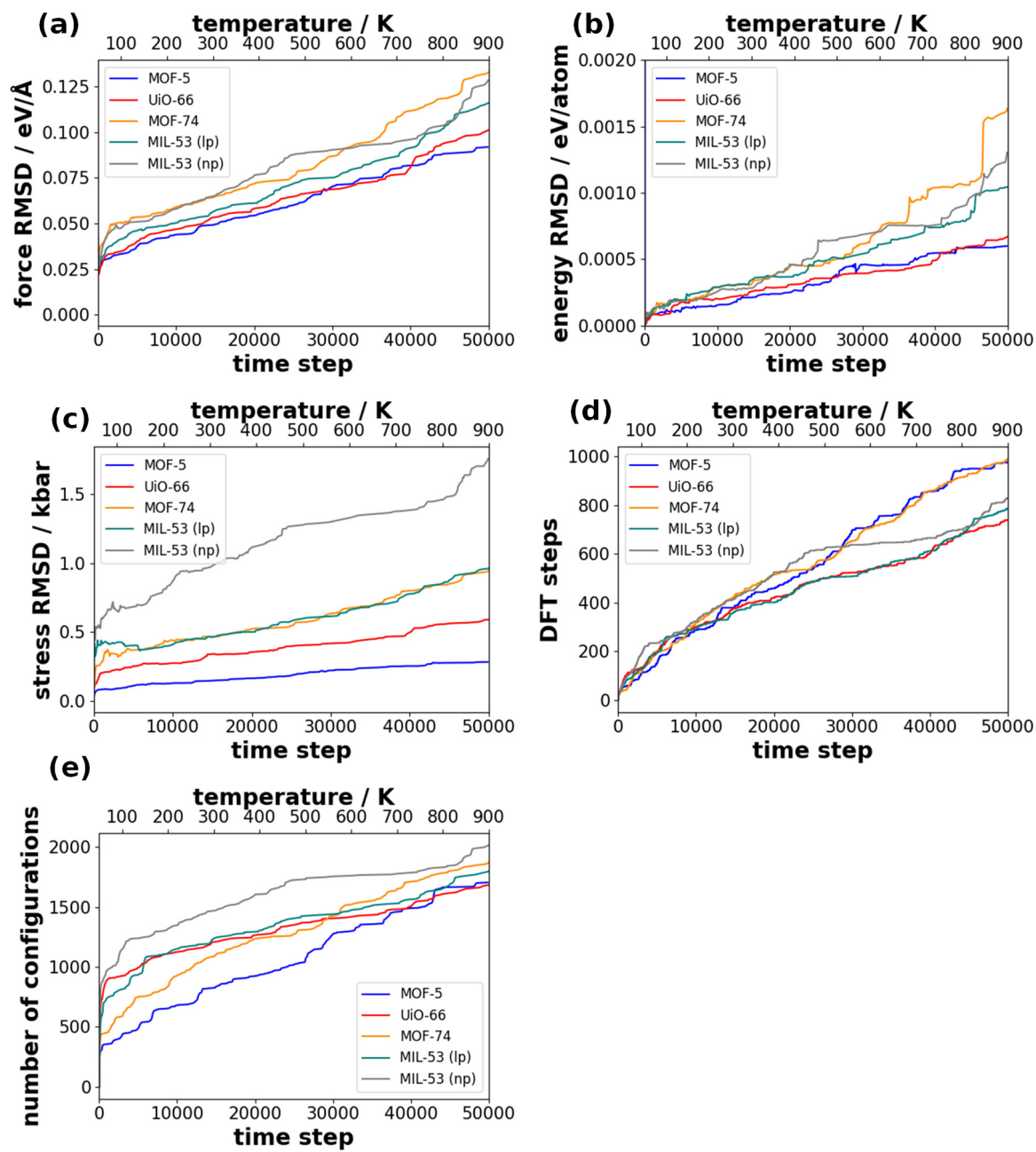

Fig. S11: Overview of the progress during the initial training runs for each of the investigated systems. The VASP MLP RMSDs for forces a, energies b and stresses c are shown as a function of simulation time step based on the reference data generated up to that point. In panel d the cumulative number of DFT steps computed over the course of the training is shown and panel e shows the total number of local reference configurations (basis function) for the concurrent VASP MLP in the training run.

Fig. S12 is the analogue to Fig. S11 for the extended training run, where the threshold criterion was fixed at 0.02 eV/Å and the temperature was kept constant at 300 K. For the force/energy/stress RMSDs in panels a, b, and c we see an immediate decrease of the errors followed by a slower drop for all systems except for MOF-5, where barely any changes can be seen. For UiO-66 the effect is also less pronounced. For MOF-74 and MIL-53 (np) a discontinuity can be observed in the RMSD values (especially in the forces), which happened when a restart of the training was initiated to increase the total number of reference configuration. For UiO-66, where a restart was also performed after 62,152 time steps, such a step cannot be discerned. It is also interesting that for MIL-53 and MOF-74 the force RMSD for the training set is actually lower than for UiO-66 and MOF-5 after the extended training set. For the number of DFT steps in Fig. S12 d we now see large differences between the systems, with almost no new DFT steps performed for MOF-5, while especially MIL-53 sees a substantial increase in added reference data. For the local reference configurations in Fig. S12 (e) we see a strong immediate increase for MIL-53 and MOF-74, while UiO-66 shows a much weaker effect. Especially the narrow pore phase of MIL-53 appends a large number of reference configurations immediately. This correlates with the most substantial reduction in the RMSD values across all systems. The reason why proportionally fewer additional reference configurations were added in the later stage of the simulation is because the maximum number of reference configurations per atomic species was set to 3000. This limit was reached for the hydrogen atoms meaning that existing configurations were replaced with newer ones instead of merely appending new ones to the potential. Such a limit was initially imposed because it is known that lighter atom types, like hydrogen, tend to accumulate more configurations than required.

It should be noted, that for materials that undergo large deformations of the lattice while heating, it could be required to perform the heating over several individual simulations with restarts in between. This serves to prevent basis set errors as VASP only builds the basis at the beginning of a molecular dynamics run. In the investigated materials, the deformations of the cells during the MD runs were minor and basis set errors are less significant. This is not only a consequence of the limited extend of cell deformation over time but is also due to choosing a relatively high plane-wave energy cutoff.

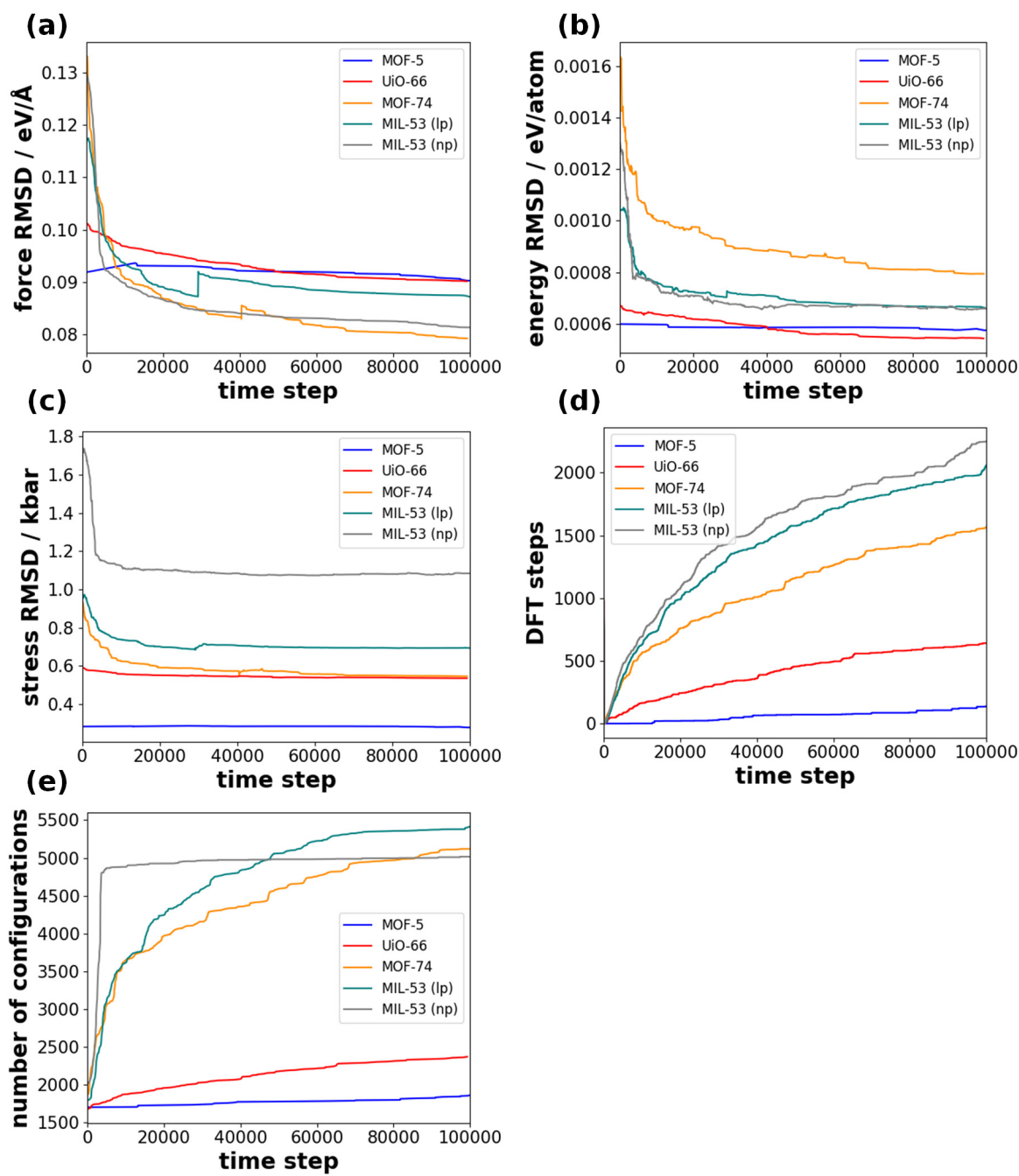

*Fig. S12: Overview of the progress during the extended training runs for each of the investigated systems. The RMSDs for forces a, energies b, and stresses c are shown as a function of the simulation time step for the reference data generated up to that point. In panel d the cumulative number of DFT steps computed over the course of the training is shown. Panel e shows the total number of local reference configurations (basis function) for the concurrent VASP MLP in the training run.*

After the reference data had been obtained during the active learning run, new potentials were trained using a newer VASP version 6.4.1 (see methods section in main manuscript for details). This was done for the initial reference data set with and without including separation of atom types as described in the beginning of this section. For the extended reference data set, the training was for the most part only performed without further separation of atom types. An exception to this is MOF-74, where we also investigated the atom separated extended VASP MLPs alongside with many other settings, which are detailed in section S3.7. To discuss the impact of the retraining on the basis sets, Table S7 lists the number of local reference configuration for each of the potentials presented in this work. There, it can clearly be seen that the number of local reference configurations increased substantially after retraining with VASP version 6.4.1. Especially the number of configurations for the metal atoms was very low for most of the systems after active learning. This means, that potentials retrained using a new VASP version, should improve in accuracy. Note, that the maximum number of local reference configurations for each atom type was set to 1500 during the retraining. In Fig. 7 in the main manuscript, we already discussed based on the example of MOF-74 that increasing this limit does only have a minor impact on the accuracy. And since the computational cost increases when including a larger number of reference configurations, the limit was left at 1500 for the VASP MLPs we typically show throughout this work.

Table S7: Number of local reference configuration for each atom type of VASP MLPs from the active learning runs (with 6.3.0) and those retrained with VASP version 6.4.1 trained on the initial (init.) and extended (ext.) reference data sets. For VASP MLPS including separation of atom types, multiple lines are given.

| System | Training type | Metal conf. | O conf. | C conf. | H conf. | Total conf. |
|---|---|---|---|---|---|---|
| MOF-5 | 6.3.0 | 146 | 1141 | 378 | 43 | 1708 |
| | 6.4.1 | 764 | 1500 | 1352 | 319 | 3935 |
| | 6.4.1 7 types | 1287 | 289 | 1317 | 289 | 5060 |
| | | | 1500 | 269 | | |
| | | | | 109 | | |
| | 6.3.0 ext. | 146 | 1287 | 387 | 43 | 1863 |
| | 6.4.1 ext. | 783 | 1500 | 1406 | 333 | 4022 |
| UiO-66 | 6.3.0 | 80 | 633 | 291 | 684 | 1688 |
| | 6.4.1 | 181 | 1500 | 850 | 1033 | 3564 |
| | 6.4.1 7 types | 840 | 770 | 176 | 544 | 6133 |
| | | | 544 | 236 | 562 | |
| | | | 1500 | 961 | | |
| | 6.3.0 ext. | 81 | 732 | 293 | 1271 | 2377 |
| | 6.4.1 ext. | 240 | 1500 | 1253 | 1500 | 4493 |
| MOF-74 | 6.3.0 | 54 | 543 | 342 | 938 | 1877 |
| | 6.4.1 | 525 | 1500 | 1500 | 1305 | 4830 |
| | 6.4.1 7 types | 830 | 1500 | 1156 | 1425 | 8942 |
| | | | 883 | 954 | | |
| | | | | 954 | | |
| | | | | 1240 | | |
| | 6.3.0 ext. | 54 | 1103 | 962 | 3000 | 5119 |
| | 6.4.1 ext. | 737 | 1500 | 1500 | 1500 | 5237 |
| | 6.4.1 ext. 7 types | 1047 | 1500 | 1500 | 1500 | 11350 |
| | | | 1303 | 1500 | | |
| | | | | 1500 | | |
| | | | | 1500 | | |
| | | | | 1500 | | |
| MIL-53 (lp) | 6.3.0 | 79 | 592 | 235 | 894 | 1800 |
| | 6.4.1 | 313 | 1500 | 837 | 1500 | 4150 |
| | 6.4.1 7 types | 346 | 597 | 659 | 1235 | 6038 |
| | | | 1500 | 264 | 788 | |
| | | | | 649 | | |
| | 6.3.0 ext. | 306 | 1855 | 253 | 3000 | 5414 |
| | 6.4.1 ext. | 446 | 1500 | 1478 | 1500 | 4924 |
| MIL-53 (np) | 6.3.0 | 44 | 517 | 360 | 1097 | 2018 |
| | 6.4.1 | 257 | 1500 | 1500 | 1500 | 4757 |
| | 6.4.1 7 types | 276 | 1500 | 1454 | 1500 | 8638 |
| | | | 400 | 1500 | 609 | |
| | | | | 1399 | | |
| | 6.3.0 ext. | 121 | 1263 | 631 | 3000 | 5015 |
| | 6.4.1 ext. | 545 | 1500 | 1500 | 1500 | 5045 |

## S2.2 Training the MTPs with MLIP

Since the starting parameters for the MTP training are initialized randomly, the trained potentials are also going to be slightly different. Fig. S13 shows the convergence of the training procedures for each system based on the initial training set. It is evident that all systems show a certain degree of variance in the value of the target function. However, this function is somewhat arbitrarily constructed out of differences in forces/stresses and energies, so that it should be more telling to investigate the differences based on the chosen benchmark criteria. A numerical comparison of the final cost function values, and comparisons based on the Γ phonon frequencies, the elastic tensor, the forces and energies of the validation set as well as a maximum temperature for which the respective potential was stable, can be seen in Table S8. The thermal stability was evaluated by performing molecular dynamics simulations up to a temperature of 700 K in 100 K steps for 100 ps. If the run was concluded, the MTP was considered to be "stable" at that temperature. Please note, that this is not a definitive criterion as unphysical behavior can occur at random when the simulation reaches regions not as well sampled by the reference data set. That means, that the testing duration might not have been sufficient to ensure thermal stability at that temperature with an absolute certainty. It is evident, that even though some reference data up to 900 K were included in the training runs, none of the trained potential is stable up to even 700 K. This is the reason why such a high initial target temperature was aimed for, even though the potentials are intended to be used at room temperature. However, despite this, some of the systems, especially MIL-53, features a fair number of trained potentials that are not even stable up to room temperature. This can be remedied by adding additional training data to the reference data set at slightly higher than the target temperature as will be shown later in this section. It is also apparent, that the thermal stability varies wildly with the differently trained MTPs and the maximum temperature and lowest errors for the other criteria does not always correlate with the lowest cost function during training which makes choosing the "best" potential somewhat challenging. Ideally more reference data can be added to make the potential more reliable, but this can be expensive depending on the system and the overall accuracy of these MTPs is already excellent for many applications. Alternatively, simulations in general can be performed at a reduced time step which also enhances the stability of the machine learned potentials. Such issues are the main drawback of machine learned approaches compared to conventional bonded force field potentials, which usually have well defined potential wells. This constitutes an area within the MTPs which that can still be improved upon.

Aside of the thermal stability, Table S8 shows that in some relatively rare cases poor potentials can emerge that show very high errors in the phonons or other properties. However, most of the other cases show relatively similar error values in all criteria. Indicated in bold in Table S8 are those potentials, which are used for the comparisons including the initial reference data set MTPs in this work. This was done based on a low cost function during training and a thermal stability up to at least 300 K.

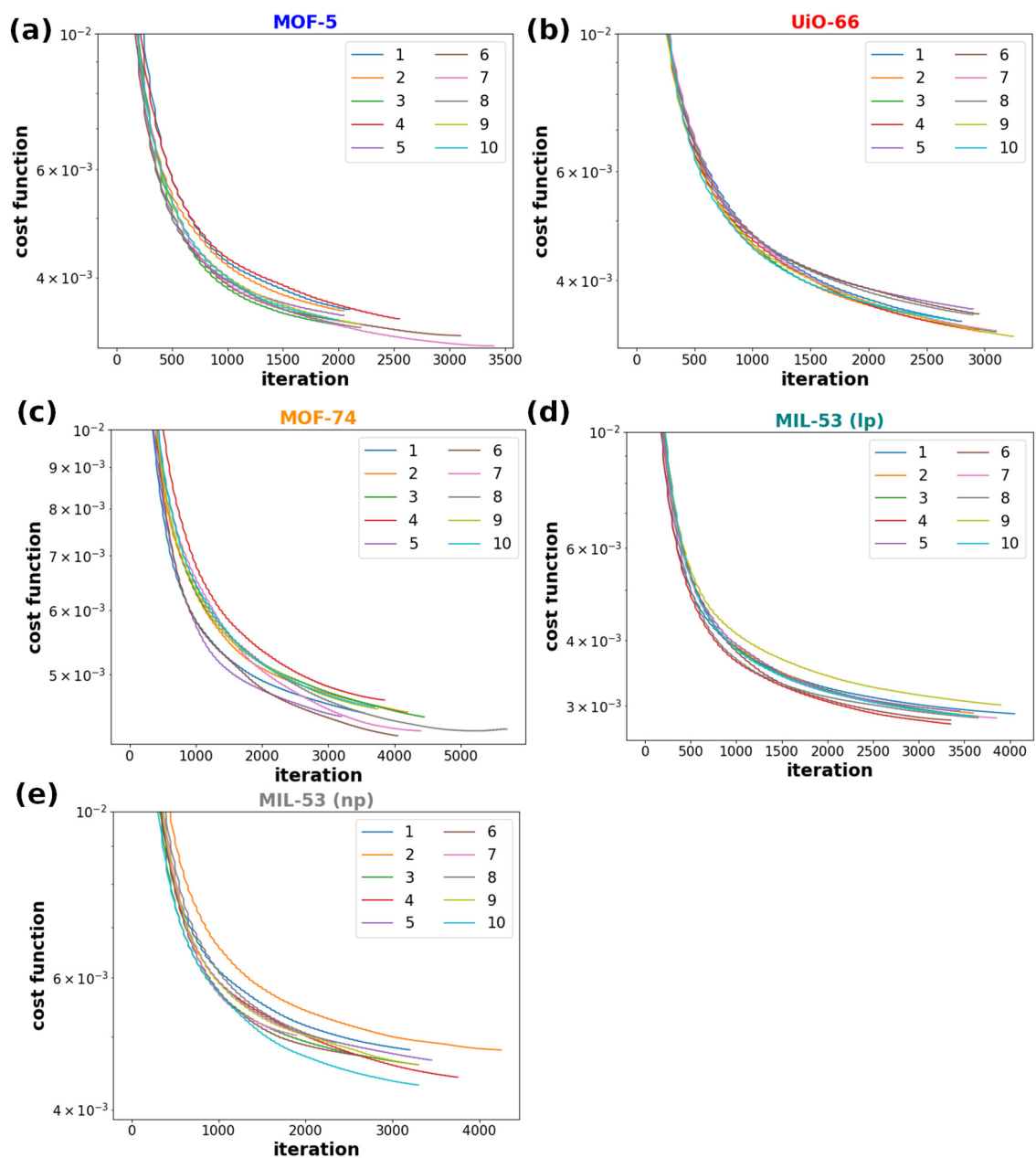

*Fig. S13: Cost functions below $10^{-2}$ for multiple MLIP training runs based on the initial reference data set as a function of the numbers of iterations until convergence Has been achieved: a MOF-5, b UiO-66, c MOF-74, d MIL-53 (lp), and e MIL-53 (np). The individual training runs are each assigned a number to ease identification.*

*Table S8: Variation of errors for multiple MLIP training runs performed for the initial training. The final values of the cost function, the RMSDs of the $\Gamma$-frequencies, the RMSDs of the elastic constants, and the RMSDs of the forces and energies based on the validation are listed. Also, the temperatures*

up to which the potentials were stable in the NPT simulations are provided. The potentials used for the remainder of the evaluation and for in-depth discussions in the main manuscript and in other chapters in the Supplementary Information are indicated in bold for each system.

| System name | cost fct. | Γ freq. RMSD / cm$^{-1}$ | $C_{i,j}$ RMSD / GPa | Force RMSD / eV/Å | Energy RMSD / meV/atom | Stable until / K |
|---|---|---|---|---|---|---|
| MOF-5 | 0.00354 | 3.5 | 0.03 | 0.017 | 0.08 | 500 |
| | 0.00353 | 3.8 | 0.10 | 0.017 | 0.10 | 600 |
| | 0.00333 | 17.1 | 2.24 | 0.045 | 0.62 | 400 |
| | 0.00342 | 6.9 | 0.04 | 0.024 | 0.21 | 400 |
| | 0.00347 | 3.7 | 0.07 | 0.017 | 0.08 | 400 |
| | 0.00321 | 5.7 | 0.05 | 0.021 | 0.15 | 400 |
| | **0.00309** | **3.3** | **0.02** | **0.016** | **0.07** | **500** |
| | 0.00331 | 4.1 | 0.12 | 0.018 | 0.11 | 500 |
| | 0.00335 | 3.6 | 0.08 | 0.017 | 0.08 | 400 |
| | 0.00341 | 3.5 | 0.12 | 0.017 | 0.07 | 400 |
| UiO-66 | 0.00344 | 4.4 | 0.20 | 0.017 | 0.07 | 300 |
| | 0.00330 | 3.9 | 0.22 | 0.016 | 0.05 | 200 |
| | 0.00331 | 3.7 | 0.13 | 0.016 | 0.07 | 300 |
| | **0.00333** | **3.7** | **0.18** | **0.016** | **0.07** | **400** |
| | 0.00360 | 4.7 | 0.30 | 0.017 | 0.06 | 200 |
| | 0.00353 | 4.3 | 0.29 | 0.017 | 0.08 | 300 |
| | 0.00332 | 3.6 | 0.13 | 0.016 | 0.08 | 300 |
| | 0.00352 | 3.4 | 0.29 | 0.017 | 0.09 | 400 |
| | 0.00325 | 3.8 | 0.11 | 0.016 | 0.06 | 300 |
| | 0.00346 | 3.6 | 0.25 | 0.017 | 0.07 | 300 |
| MOF-74 | 0.00449 | 3.0 | 0.61 | 0.031 | 0.20 | 400 |
| | 0.00450 | 3.3 | 0.74 | 0.030 | 0.24 | 200 |
| | 0.00444 | 3.7 | 1.33 | 0.031 | 0.27 | 500 |
| | 0.00465 | 5.0 | 0.68 | 0.037 | 0.33 | 300 |
| | 0.00444 | 3.1 | 0.51 | 0.030 | 0.19 | 200 |
| | 0.00421 | 3.4 | 0.53 | 0.029 | 0.18 | 300 |
| | 0.00427 | 4.2 | 0.55 | 0.032 | 0.25 | 200 |
| | **0.00428** | **3.1** | **0.30** | **0.030** | **0.17** | **600** |
| | 0.00454 | 3.2 | 0.77 | 0.031 | 0.28 | 300 |
| | 0.00457 | 16.7 | 3.50 | 0.096 | 1.30 | 500 |
| MIL-53 (lp) | 0.00290 | 24.1 | 3.14 | 0.052 | 0.40 | 0 |
| | 0.00291 | 5.8 | 1.11 | 0.022 | 0.14 | 400 |
| | 0.00287 | 4.9 | 0.30 | 0.021 | 0.16 | 300 |
| | 0.00277 | 5.9 | 0.52 | 0.024 | 0.18 | 200 |
| | 0.00293 | 5.8 | 0.96 | 0.022 | 0.17 | 200 |
| | 0.00282 | 6.4 | 0.67 | 0.023 | 0.16 | 400 |
| | 0.00285 | 5.5 | 0.48 | 0.021 | 0.21 | 200 |
| | 0.00285 | 5.3 | 0.50 | 0.022 | 0.16 | 300 |
| | 0.00301 | 5.5 | 1.04 | 0.026 | 0.30 | 200 |
| | **0.00287** | **5.2** | **0.52** | **0.022** | **0.17** | **300** |
| MIL-53 (np) | 0.00481 | 9.5 | 0.47 | 0.029 | 0.12 | 100 |
| | 0.00480 | 9.4 | 1.10 | 0.033 | 0.24 | 100 |
| | 0.00459 | 7.3 | 0.70 | 0.028 | 0.13 | 300 |
| | 0.00442 | 7.3 | 0.94 | 0.027 | 0.14 | 100 |
| | 0.00466 | 9.8 | 0.58 | 0.028 | 0.15 | 400 |
| | 0.00471 | 6.6 | 0.75 | 0.029 | 0.15 | 200 |
| | 0.00503 | 7.6 | 0.77 | 0.030 | 0.15 | 600 |
| | 0.00492 | 7.0 | 0.64 | 0.029 | 0.17 | 400 |
| | **0.00459** | **7.7** | **0.48** | **0.028** | **0.14** | **300** |
| | 0.00432 | 8.4 | 0.73 | 0.028 | 0.12 | 200 |

A similar selection was performed for the extended reference data set MTP, as can be seen in Table S9. Here, now no really "bad" potentials with exceptionally poor properties are present any more and overall the larger reference data set leads to more consistent and usually slightly lower error values among the different benchmark properties. Only the thermal stability is still quite inconsistent, especially for MIL-53 (lp), where the initial 5 performed training runs showed a quite poor stability. To investigate, whether this is simply a coincidence, 3 more MTP training runs with the same settings and different initializations were performed. All of them show an improved thermal stability. The randomness of the stability is somewhat concerning and even the extended reference data set did only slightly improve the situation for most systems. However, the extended reference data set was obtained for a constant temperature of 300 K and mostly served to analyze the improvements in accuracy at that particular temperature. To ensure the stability at a given temperature, one might expect that training the potential based on data generated at somewhat higher temperatures would be preferable to sample also more strongly distorted structures during the training. To test this, we, in particular, generated 1009 additional reference configurations at a constant temperature of 400 K for MIL-53 (lp). The training settings other than the temperature and the initial potential (which was now including the initial and extended reference data sets) were the same as the regular extended reference data set. The results can be seen in Table S10. All 5 of the trained MTPs on that reference data set were stable up to a temperature of 300 K. So, there definitely is an improvement. However, due to the variance involved in the stability issues and the relatively small sample size, we cannot be certain that such a training at 400K solves the issue entirely.

In general, the stability issues seem to be the primary weakness of the machine learned potentials considered here (they are also present for the VASP MLPs, as we discuss in section S4.1.2). In the interest of saving computation time, it would be beneficial to enhance the MTPs with some additional physical knowledge to make it less likely that the potential diverges in the unknown regime. Alternatively, different data sampling approaches could be used to reduce the total number of reference structures or one might consider adding auxiliary "solid-wall" type potentials to prevent a decomposition of the studied systems. However, this would compromise the possibility to model adsorption or desorption processes or chemical reactions when using the potentials.

*Table S9: Variation of errors for multiple MLIP training runs performed for the extended training. The final values of the cost function, the RMSDs of the $\Gamma$-frequencies, the RMSDs of the elastic constants, and the RMSDs of the forces and energies based on the validation are listed. Also, the temperatures up to which the potentials were stable in the NPT simulations are provided. The potentials used for the remainder of the evaluation and for in-depth discussions in the main manuscript and in other chapters in the Supplementary Information, whenever data for the extended reference data set are shown, are indicated in bold for each system.*

| System name | cost fct. | $\Gamma$ freq. RMSD / cm$^{-1}$ | $C_{i,j}$ RMSD / GPa | Force RMSD / eV/Å | Energy RMSD / meV/atom | Stable until / K |
|---|---|---|---|---|---|---|
| MOF-5 | 0.00359 | 4.8 | 0.05 | 0.020 | 0.16 | 500 |
|  | 0.00347 | 4.0 | 0.08 | 0.018 | 0.08 | 600 |
|  | 0.00497 | 4.3 | 0.16 | 0.020 | 0.11 | 400 |
|  | 0.00349 | 6.4 | 0.02 | 0.023 | 0.16 | 500 |
|  | **0.00318** | **3.7** | **0.09** | **0.017** | **0.08** | **500** |
| UiO-66 | **0.00289** | **2.8** | **0.32** | **0.016** | **0.05** | **400** |
|  | 0.00298 | 3.2 | 0.13 | 0.016 | 0.06 | 400 |
|  | 0.00293 | 3.8 | 0.21 | 0.016 | 0.08 | 400 |
|  | 0.00305 | 3.7 | 0.25 | 0.016 | 0.06 | 200 |
|  | 0.00281 | 3.6 | 0.19 | 0.016 | 0.08 | 300 |
| MOF-74 | 0.00356 | 2.8 | 0.54 | 0.028 | 0.21 | 500 |
|  | 0.00328 | 2.8 | 0.43 | 0.028 | 0.17 | 300 |
|  | 0.00337 | 2.9 | 0.62 | 0.028 | 0.20 | 400 |
|  | **0.00355** | **2.9** | **0.21** | **0.028** | **0.20** | **700** |
|  | 0.00356 | 2.9 | 0.61 | 0.028 | 0.15 | 700 |
| MIL-53   (lp) | 0.00242 | 5.1 | 1.36 | 0.024 | 0.17 | 200 |
|  | 0.00239 | 5.3 | 1.69 | 0.022 | 0.19 | 200 |
|  | 0.00226 | 5.0 | 0.62 | 0.020 | 0.12 | 100 |
|  | 0.00255 | 4.6 | 0.50 | 0.020 | 0.14 | 100 |
|  | **0.00254** | **4.1** | **0.56** | **0.021** | **0.16** | **500** |
|  | 0.00247 | 4.6 | 1.14 | 0.021 | 0.15 | 400 |
|  | 0.00216 | 5.2 | 0.46 | 0.019 | 0.13 | 300 |
|  | 0.00243 | 5.6 | 1.40 | 0.022 | 0.22 | 400 |
| MIL-53  (np) | **0.00422** | **6.4** | **0.41** | **0.027** | **0.14** | **500** |
|  | 0.00407 | 6.8 | 0.35 | 0.027 | 0.15 | 100 |
|  | 0.00419 | 7.8 | 0.40 | 0.028 | 0.16 | 400 |
|  | 0.00413 | 7.6 | 0.53 | 0.027 | 0.15 | 500 |
|  | 0.00396 | 7.2 | 0.59 | 0.027 | 0.15 | 400 |

*Table S10: Variation of errors for multiple MLIP training runs performed for the additionally extended reference data set based on 400 K MD reference data for MIL-53 (lp). The final values of the cost function, the RMSDs of the $\Gamma$-frequencies, the RMSDs of the elastic constants, and the RMSDs of the forces and energies based on the validation are listed. Also, the temperatures up to which the potentials were stable in the NPT simulations are provided.*

| System name | cost fct. | $\Gamma$ freq. RMSD / cm$^{-1}$ | $C_{i,j}$ RMSD / GPa | Force RMSD / eV/Å | Energy RMSD / meV/atom | Stable until / K |
|---|---|---|---|---|---|---|
| MIL-53 (lp) | 0.00271 | 5.11 | 1.62 | 0.021 | 0.19 | 400 |
| | 0.00257 | 4.57 | 1.42 | 0.020 | 0.17 | 300 |
| | 0.00281 | 5.51 | 1.05 | 0.021 | 0.17 | 400 |
| | 0.00275 | 5.13 | 1.52 | 0.021 | 0.17 | 400 |
| | 0.00271 | 4.98 | 1.40 | 0.021 | 0.17 | 600 |

For the training for some of the systems, especially at lower MTP levels, we also encountered the problem that the training algorithm escaped a previously found minimum without finding a better minimum in a reasonable time frame. This issue is visualized in Fig. S14 for the example of MOF-5 at level 10 and is the reason why the convergence of the cost function needs to be observed carefully during the training. To solve the issue, the MTP for the minimum cost function can be chosen from the intermediate results produced during training However, as it stands, MLIP does not support outputting all intermediate training results, so it is up to the user to track and store away the output files. This can easily be realized by running a script as a background process during training. As an example, we provide the used bash script in the following:

```bash
#!/bin/bash
touch $1
date=$(stat -c %y "$1")
while sleep 0.01;
do
  date2=$(stat -c %y "$1")
  if [[ $date2 != $date ]];
  then
      cat $1 >> "all_$1"
      echo -e "\n" >> "all_$1"
  fi
  date=$date2
done
```

It takes one argument, which is the file name of the concurrent potential output by MLIP when using the "--curr_pot_name=<file_name>" option. Note, that for very fast fits for small systems it is necessary to adjust the time for the sleep command. Then, all intermediate results are written in the file all_<file_name> from which the intermediate result with the lowest cost function can be extracted.

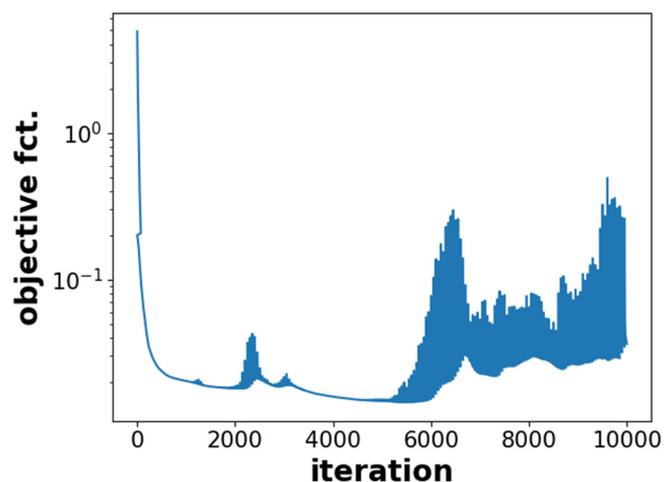

*Fig. S14: Example for the unsatisfactory convergence behavior of the cost function for training a level 10 MTP on the initial reference data set of MOF-5. The training was aborted after 10000 time steps.*

# S3. Analysis of the potentials

## S3.1 Full evaluation of the errors

To support the results given in Fig. 3, 5 and 7 in the main manuscript, we show here a quantitative analysis of the errors. Table S11 summarizes the root mean square errors in the energy-, force-, stress-, and $\Gamma$-point frequency calculations together with the root mean square errors of the elastic constants for all studied systems and for the initial and the extended training sets. Values are provided for the VASP MLPs and for the MTPs. An exception here is MOF-74, where the potential trained on the extended reference data set including atom typing shows similar errors compared to the MTP. However, in general it should be noted that the errors for the VASP MLPs are also low compared to most traditional potentials found in the literature.

*Table S11: RMSD values of the VASP MLPs and MTPs trained on the initial and extended reference data set with the DFT reference for various quantities: the phonon frequencies at the $\Gamma$-point; the components of the elastic stiffness tensor, the components of the forces in the validation set; the energy differences between individual total energies of the structures in the validation set minus the total energy of the minimum structure; and the components of the stress tensor for structures in the validation set. The VASP MLPs trained on the initial reference data set are given for the case of*

*including separation of atom types based on their neighborhood and for only 4 atom types as indicated by $n_{types}$ (which denotes the number of atom types).*

| System | Method | $n_{types}$ | $f_r$ RMSD / cm$^{-1}$ | $C_{i,j}$ RMSD / GPa | Force RMSD / eV/Å | Energy RMSD / meV/atom | Stress RMSD / kbar |
|---|---|---|---|---|---|---|---|
| **MOF-5** | VASP MLP | 4 | 8.00 | 0.20 | 0.024 | 0.15 | 0.05 |
| | VASP MLP | 7 | 8.96 | 0.27 | 0.021 | 0.07 | 0.05 |
| | MTP | 7 | 3.26 | 0.02 | 0.016 | 0.07 | 0.02 |
| | VASP MLP extended | 4 | 7.36 | 0.21 | 0.022 | 0.10 | 0.05 |
| | MTP extended | 7 | 3.70 | 0.09 | 0.017 | 0.01 | 0.02 |
| **UiO-66** | VASP MLP | 4 | 11.85 | 1.16 | 0.033 | 0.13 | 0.20 |
| | VASP MLP | 9 | 10.35 | 0.60 | 0.024 | 0.11 | 0.13 |
| | MTP | 9 | 3.71 | 0.18 | 0.016 | 0.07 | 0.06 |
| | VASP MLP extended | 4 | 7.82 | 0.29 | 0.025 | 0.11 | 0.15 |
| | MTP extended | 9 | 2.78 | 0.32 | 0.016 | 0.06 | 0.05 |
| **MOF-74** | VASP MLP | 4 | 5.67 | 1.26 | 0.037 | 0.25 | 0.24 |
| | VASP MLP | 8 | 4.72 | 0.62 | 0.027 | 0.20 | 0.17 |
| | MTP | 8 | 3.07 | 0.30 | 0.030 | 0.17 | 0.11 |
| | VASP MLP extended | 4 | 4.81 | 0.31 | 0.034 | 0.22 | 0.19 |
| | VASP MLP extended | 8 | 2.89 | 0.39 | 0.022 | 0.14 | 0.12 |
| | MTP extended | 8 | 2.91 | 0.21 | 0.028 | 0.20 | 0.09 |
| **MIL-53 (lp)** | VASP MLP | 4 | 12.51 | 0.91 | 0.033 | 0.28 | 0.27 |
| | VASP MLP | 8 | 8.83 | 0.78 | 0.027 | 0.19 | 0.19 |
| | MTP | 8 | 5.23 | 0.52 | 0.022 | 0.17 | 0.09 |
| | VASP MLP extended | 4 | 7.35 | 0.95 | 0.027 | 0.21 | 0.19 |
| | MTP extended | 8 | 4.98 | 1.40 | 0.021 | 0.17 | 0.06 |
| **MIL-53 (np)** | VASP MLP | 4 | 11.40 | 1.87 | 0.040 | 0.19 | 0.47 |
| | VASP MLP | 8 | 8.70 | 0.67 | 0.030 | 0.17 | 0.32 |
| | MTP | 8 | 7.64 | 0.48 | 0.028 | 0.14 | 0.24 |
| | VASP MLP extended | 4 | 10.78 | 1.53 | 0.038 | 0.21 | 0.44 |
| | MTP extended | 8 | 6.41 | 0.41 | 0.027 | 0.14 | 0.15 |

### S3.1.1 Evaluation of lattice parameters

As we only provide volume deviations for the crystallographic unit cells in the main paper, all parameters of the unit cells calculated with the different methos for all systems are shown in Table S12.

*Table S12: Optimized Lattice parameters for the investigated systems obtained for the various force field potentials and for the DFT reference. The angles not specified in the table are 90°. The number of atom types used to perform the calculations is indicated with $n_{types}$.*

| System | Method | $n_{types}$ | a / Å | b / Å | c / Å | γ / ° | V / Å³ |
|---|---|---|---|---|---|---|---|
| **MOF-5** | MTP | 7 | 26.066 | 26.066 | 26.066 | 90.00 | 17710.188 |
| | MTP extended | 7 | 26.065 | 26.065 | 26.065 | 90.00 | 17708.150 |
| | VASP MLP | 4 | 26.069 | 26.069 | 26.069 | 90.00 | 17715.051 |
| | VASP MLP | 7 | 26.068 | 26.068 | 26.068 | 90.00 | 17714.397 |
| | VASP MLP extended | 4 | 26.069 | 26.069 | 26.069 | 90.00 | 17716.105 |
| | DFT | 4 | 26.068 | 26.068 | 26.068 | 90.00 | 17714.265 |
| | UFF4MOF | 5 | 26.442 | 26.442 | 26.442 | 90.00 | 18487.701 |
| **UiO-66** | MTP | 9 | 20.898 | 20.898 | 20.898 | 90.00 | 9126.708 |
| | MTP extended | 9 | 20.898 | 20.898 | 20.898 | 90.00 | 9126.708 |
| | VASP MLP | 4 | 20.894 | 20.894 | 20.894 | 90.00 | 9121.881 |
| | VASP MLP | 9 | 20.898 | 20.898 | 20.898 | 90.00 | 9127.235 |
| | VASP MLP extended | 4 | 20.899 | 20.899 | 20.899 | 90.00 | 9128.174 |
| | DFT | 4 | 20.899 | 20.899 | 20.899 | 90.00 | 9128.019 |
| **MOF-74** | MTP | 8 | 26.138 | 26.138 | 6.704 | 120.00 | 3966.517 |
| | MTP extended | 8 | 26.139 | 26.139 | 6.704 | 120.00 | 3966.821 |
| | VASP MLP | 4 | 26.152 | 26.152 | 6.694 | 120.00 | 3965.438 |
| | VASP MLP | 8 | 26.136 | 26.136 | 6.690 | 120.00 | 3953.436 |
| | VASP MLP extended | 4 | 26.139 | 26.139 | 6.691 | 120.00 | 3959.878 |
| | VASP MLP extended | 8 | 26.133 | 26.133 | 6.689 | 120.00 | 3955.691 |
| | DFT | 4 | 26.078 | 26.078 | 6.718 | 120.00 | 3956.573 |
| | UFF4MOF | 4 | 26.293 | 26.351 | 6.798 | 120.01 | 4078.545 |
| **MIL-53 (lp)** | MTP | 8 | 17.206 | 6.666 | 12.360 | 91.04 | 1417.399 |
| | MTP extended | 8 | 17.308 | 6.670 | 12.217 | 90.92 | 1410.202 |
| | VASP MLP | 4 | 17.266 | 6.668 | 12.335 | 90.95 | 1417.025 |
| | VASP MLP | 8 | 17.310 | 6.689 | 12.250 | 90.16 | 1418.465 |
| | VASP MLP extended | 4 | 17.229 | 6.669 | 12.357 | 90.73 | 1419.833 |
| | DFT | 4 | 17.266 | 6.665 | 12.272 | 90.99 | 1412.025 |
| **MIL-53 (np)** | MTP | 8 | 19.818 | 6.664 | 6.547 | 104.37 | 837.592 |
| | MTP extended | 8 | 19.830 | 6.663 | 6.513 | 104.39 | 833.547 |
| | VASP MLP | 4 | 19.790 | 6.661 | 6.548 | 104.19 | 836.839 |
| | VASP MLP | 8 | 19.788 | 6.661 | 6.568 | 104.20 | 839.207 |
| | VASP MLP extended | 4 | 19.796 | 6.658 | 6.612 | 104.23 | 844.677 |
| | DFT | 4 | 19.783 | 6.659 | 6.561 | 104.16 | 838.052 |

## S3.1.2 Deviations of forces, energies and stresses based on the validation sets

In the main manuscript, we focused on the force, energy and stress error evaluations for MOF-5 and MOF-74. In Fig. S15 the error distributions for the remaining systems can be seen for the VASP MLPs and the MTPs. It is evident that the deviations for all systems are in a similar range. Only for the stresses, we see larger deviations for the anisotropic systems, which is related to higher absolute values of the stresses in these systems.

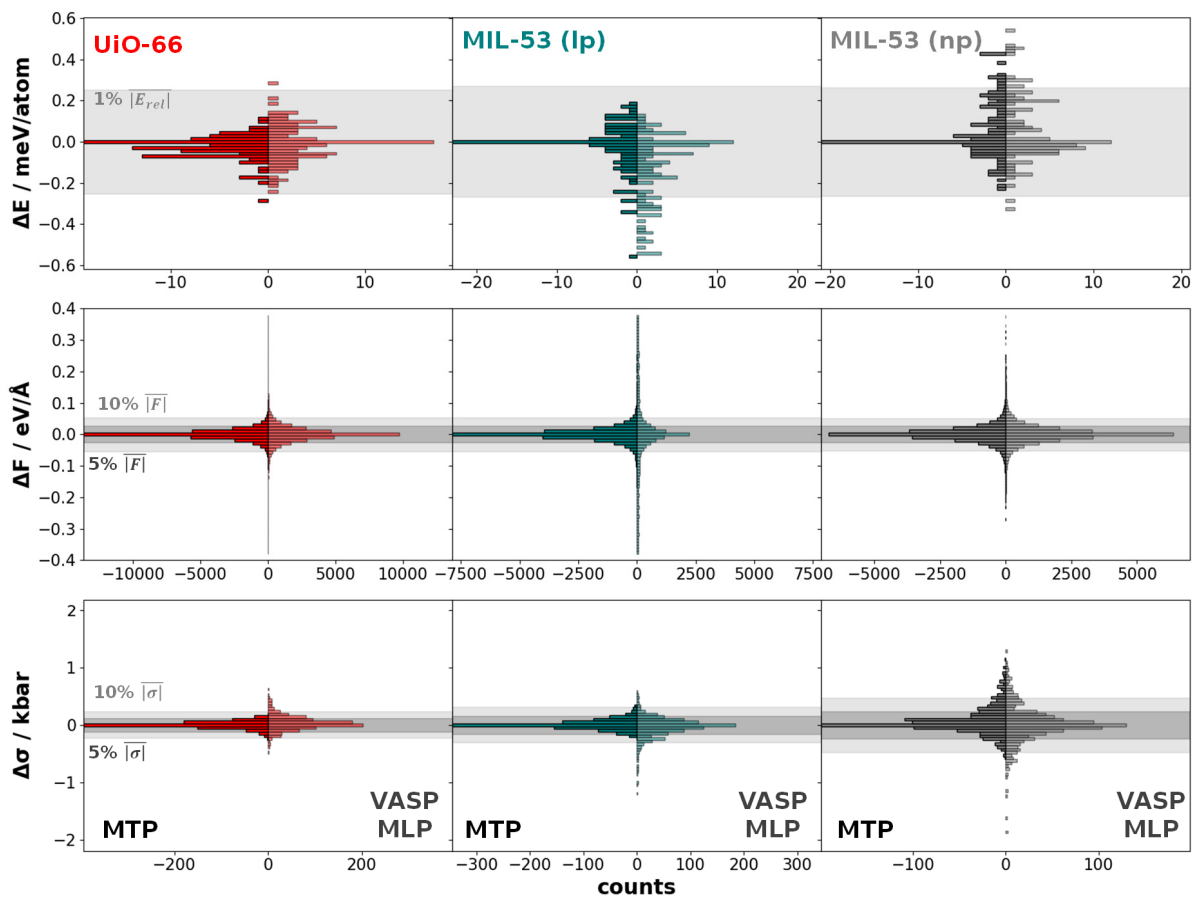

*Fig. S15: Histograms detailing the deviations (FF data subtracted from reference) of the energies (E), the components of the forces (F), and the components of the stresses (σ) obtained from DFT and from the VASP MLPs and the MTPs for UiO-66 (red), MIL-53 (lp) (teal) and MIL-53 (np) (grey). For the energies only the energy difference from the structure with the minimum energy is shown. The validation set structures were obtained from an active learning VASP MLP learning run at a temperature of 300 K. The areas shaded in gray indicate the ranges with errors within 5% and 10% of the absolute DFT values for forces and stresses and within 1 % for the energy differences.*

### S3.1.3 Phonon densities of states

Similarly, for the phonon densities of states, we show the machine learned potential versus DFT comparison for UiO-66 and both phases of MIL-53 in Fig. S16. Like for MOF-5 and MOF-74, the agreement of the MTPs with the DFT data is excellent with only minor deviations, which are mostly noticeable in the zoom into the low frequency region. The VASP MLPs show a slightly worse agreement in the lower frequency region, but overall, the agreement is still excellent.

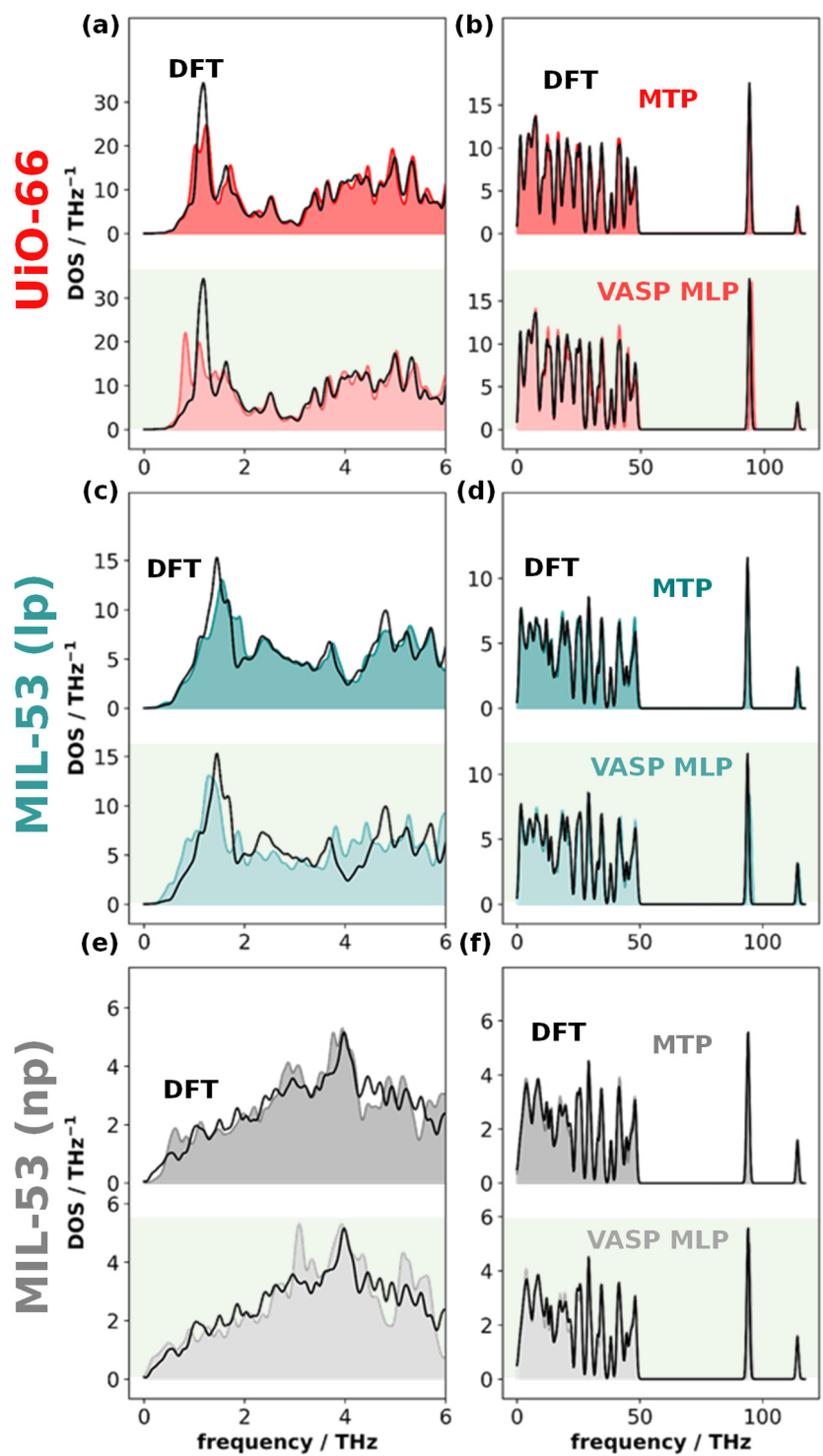

Fig. S16: Comparisons of the densities of states (DOS) obtained with DFT (indicated by black lines) and the machine learned force field potentials (shaded areas) for the low frequency phonon modes (a, c, e) and the entire frequency range (b, d, f). The comparisons are given for UiO-66 (a, b), MIL-53 (lp) (c, d) and MIL-53 (np) (e, f) for the MTP and the VASP MLP. The DOS was obtained for a 11x11x11 mesh sampled in the first Brillouin zone and a Gaussian smearing width of 0.05 THz was applied in (a, c, e) and of 0.5 THz in (b, d, f).

### S3.1.4 Elastic stiffness tensors

Fig. S17 shows a comparison of the elastic tensor elements obtained with DFT and the machine learned potentials for all systems. As can be seen, the agreement is generally good for both types of machine learned potentials for all system.

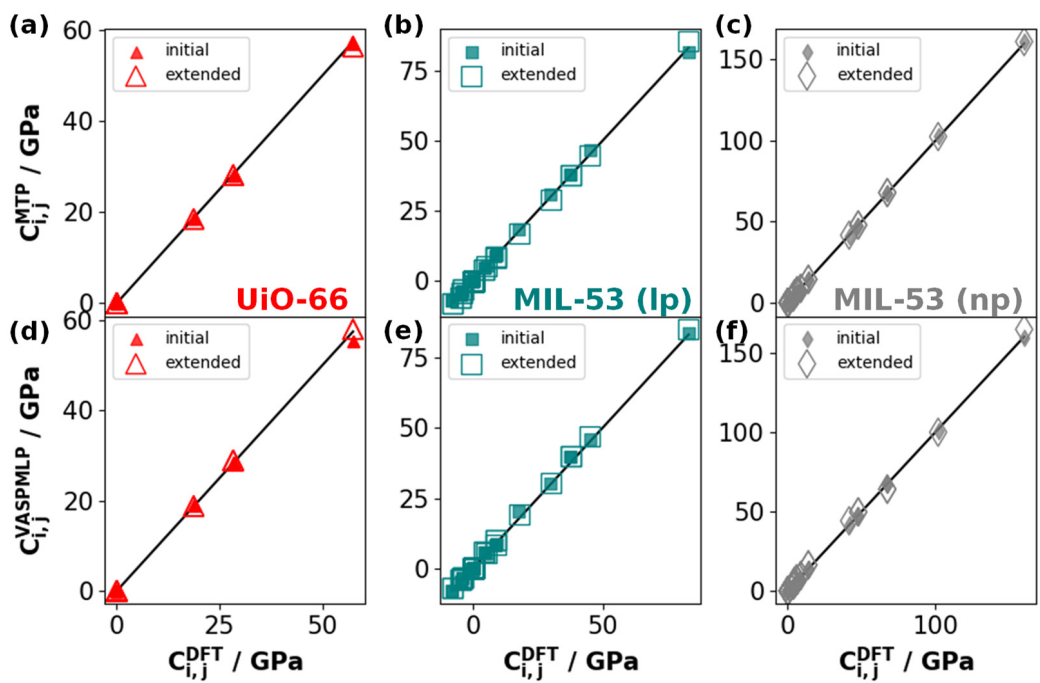

*Fig. S17: Comparison of the elements of the elastic stiffness tensor. Shown are the results calculated with DFT and with the MTPs in a, b, c, with DFT and with the VASP MLPs in d, e, f as well as with DFT). The plots shown here contain data for UiO-66 in a, d, MIL-53 (lp) in b, e and MIL-53 (np) in c, f. In the upper and central panels, small filled symbols indicate values obtained with potentials trained on the initial reference data set and large empty symbols refer to the values for potentials trained on the extended reference data set. The solid black line pass through the origin and have a slope of 1 such that they indicate an ideal agreement between force field and DFT simulations.*

The 21 elastic tensor elements for each system are given in Table S13, Table S14, and Table S15, to allow a fully quantitative comparisons of the elastic tensor elements calculated with DFT and with the force fields . The data are split up in multiple tables due to space limitations.

In particular, for MIL-53 (np) several values are slightly non-zero for the MTPs while they are zero for PBE and the VASP MLPs. The reason for this is that for the MTPs the elastic tensor was computed without applying any symmetry constraints, while for the VASP based calculations monoclinic symmetry was assumed (space groups Cc and C2/c).

*Table S13: 9 of the elements of the elastic tensor for each system evaluated by DFT, the VASP MLPs and the MTPs for the initial and the extended (ext.) reference data sets. The values of the tensor elements are given in GPa. Indices for C are given in Voigt notation which corresponds to the Cartesian directions as follows: 1=xx, 2=yy, 3=zz, 4=zy, 5=xz, 6=xy. The number of atom types in the used method is indicated with $n_{type}$.*

| System | Method | $n_{type}$ | $C_{11}$ | $C_{22}$ | $C_{33}$ | $C_{44}$ | $C_{55}$ | $C_{66}$ | $C_{12}$ | $C_{13}$ | $C_{23}$ |
|---|---|---|---|---|---|---|---|---|---|---|---|
| MOF-5 | VASP MLP | 4 | 25.39 | 25.39 | 25.39 | 0.94 | 0.94 | 0.94 | 10.95 | 10.95 | 10.95 |
| | VASP MLP | 7 | 25.35 | 25.34 | 25.35 | 0.87 | 0.87 | 0.87 | 10.73 | 10.73 | 10.73 |
| | VASP MLP ext. | 4 | 25.39 | 25.39 | 25.39 | 0.92 | 0.92 | 0.92 | 10.90 | 10.90 | 10.90 |
| | MTP | 7 | 25.99 | 25.99 | 25.99 | 0.95 | 0.95 | 0.95 | 11.25 | 11.25 | 11.25 |
| | MTP ext. | 7 | 25.70 | 25.70 | 25.70 | 0.89 | 0.89 | 0.89 | 11.31 | 11.31 | 11.31 |
| | PBE | 4 | 25.97 | 25.97 | 25.97 | 0.95 | 0.95 | 0.95 | 11.22 | 11.22 | 11.22 |
| UiO-66 | VASP MLP | 4 | 53.89 | 53.89 | 53.89 | 18.35 | 18.35 | 18.35 | 27.11 | 27.11 | 27.11 |
| | VASP MLP | 9 | 55.42 | 55.42 | 55.42 | 18.84 | 18.84 | 18.84 | 28.34 | 28.34 | 28.34 |
| | VASP MLP ext. | 4 | 58.00 | 58.00 | 58.00 | 18.82 | 18.82 | 18.82 | 28.95 | 28.95 | 28.95 |
| | MTP | 9 | 56.99 | 56.99 | 56.99 | 18.66 | 18.66 | 18.66 | 28.10 | 28.10 | 28.10 |
| | MTP ext. | 9 | 56.43 | 56.43 | 56.43 | 18.52 | 18.52 | 18.52 | 28.14 | 28.14 | 28.14 |
| | PBE | 4 | 57.50 | 57.50 | 57.50 | 18.72 | 18.72 | 18.72 | 28.35 | 28.35 | 28.35 |
| MOF-74 | VASP MLP | 4 | 27.48 | 27.48 | 13.93 | 16.53 | 16.53 | 2.19 | 23.10 | 3.65 | 3.65 |
| | VASP MLP | 8 | 23.92 | 23.92 | 10.58 | 16.26 | 16.26 | 2.25 | 19.42 | 2.68 | 2.68 |
| | VASP MLP ext. | 4 | 23.88 | 23.88 | 15.52 | 16.60 | 16.60 | 2.04 | 19.81 | 2.98 | 2.98 |
| | MTP | 8 | 24.67 | 24.68 | 14.18 | 16.35 | 16.34 | 2.35 | 19.92 | 3.72 | 3.72 |
| | MTP ext. | 8 | 23.82 | 23.85 | 13.05 | 16.19 | 16.19 | 2.29 | 19.22 | 2.97 | 2.98 |
| | PBE | 4 | 23.90 | 23.90 | 14.10 | 16.30 | 16.30 | 2.37 | 19.15 | 3.28 | 3.28 |
| MIL-53 (lp) | VASP MLP | 4 | 41.73 | 83.38 | 16.76 | 29.33 | 4.69 | 7.82 | 9.85 | 3.22 | 37.09 |
| | VASP MLP | 8 | 46.10 | 83.76 | 20.24 | 30.25 | 5.38 | 8.34 | 8.69 | 5.68 | 39.59 |
| | VASP MLP ext. | 4 | 47.06 | 85.07 | 19.23 | 30.24 | 5.57 | 8.41 | 10.02 | 5.84 | 39.77 |
| | MTP | 8 | 46.72 | 81.72 | 18.37 | 30.72 | 5.64 | 9.00 | 9.31 | 4.68 | 37.73 |
| | MTP ext. | 8 | 44.81 | 85.40 | 16.80 | 29.14 | 5.24 | 8.12 | 8.73 | 4.12 | 37.73 |
| | PBE | 4 | 45.02 | 83.20 | 17.62 | 29.95 | 5.27 | 8.71 | 8.92 | 4.34 | 37.57 |
| MIL53 (np) | VASP MLP | 4 | 167.04 | 7.73 | 100.76 | 3.44 | 42.92 | 5.58 | 9.93 | 51.63 | 7.70 |
| | VASP MLP | 8 | 159.50 | 6.71 | 100.21 | 2.88 | 41.63 | 3.41 | 7.15 | 47.51 | 5.47 |
| | VASP MLP ext. | 4 | 164.55 | 7.69 | 99.81 | 4.49 | 44.10 | 4.50 | 9.63 | 49.65 | 6.50 |
| | MTP | 8 | 161.01 | 5.41 | 102.47 | 3.96 | 40.83 | 4.49 | 7.36 | 47.63 | 5.71 |
| | MTP ext. | 8 | 161.02 | 6.04 | 102.37 | 3.48 | 41.49 | 4.89 | 8.68 | 47.79 | 7.14 |
| | PBE | 4 | 159.89 | 5.86 | 101.69 | 3.69 | 41.47 | 4.60 | 8.73 | 47.62 | 6.26 |

*Table S14: 6 of the elements of the elastic tensor for each system evaluated by DFT, the VASP MLPs and the MTPs for the initial and the extended (ext.) reference data sets. The values of the tensor elements are given in GPa. Indices for C are given in Voigt notation with 1=xx, 2=yy, 3=zz, 4=zy, 5=xz, 6=xy. Empty table entries indicate a value of zero. The number of atom types in the used method is indicated with $n_{type}$.*

| System | Method | $n_{type}$ | $C_{14}$ | $C_{15}$ | $C_{16}$ | $C_{24}$ | $C_{25}$ | $C_{26}$ |
|--------|--------|-----------|----------|----------|----------|----------|----------|----------|
| MOF-74 | VASP MLP | 4 | 3.28 | -0.58 | | -3.28 | 0.58 | |
| | VASP MLP | 8 | 3.26 | -0.66 | | -3.26 | 0.66 | |
| | VASP MLP ext. | 4 | 3.24 | -0.72 | | -3.24 | 0.72 | |
| | MTP | 8 | 3.27 | -0.71 | | -3.25 | 0.71 | |
| | MTP ext. | 8 | 3.36 | -0.69 | | -3.34 | 0.71 | |
| | PBE | 4 | 3.30 | -0.75 | | -3.30 | 0.75 | |
| MIL-53 | VASP MLP | 4 | | | -3.50 | | | -6.18 |
| (lp) | VASP MLP | 8 | | | -5.24 | | | -7.91 |
| | VASP MLP ext. | 4 | | | -4.14 | | | -6.84 |
| | MTP | 8 | | | -4.97 | | | -6.96 |
| | MTP ext. | 8 | | | -5.89 | | | -7.99 |
| | PBE | 4 | | | -4.59 | | | -8.00 |
| MIL-53 | VASP MLP | 4 | | 63.69 | | | 5.62 | |
| (np) | VASP MLP | 8 | | 66.68 | | | 4.05 | |
| | VASP MLP ext. | 4 | | 64.00 | | | 6.01 | |
| | MTP | 8 | -0.10 | 67.32 | -0.24 | -0.02 | 4.79 | -0.02 |
| | MTP ext. | 8 | -0.20 | 67.81 | -0.47 | -0.04 | 5.22 | -0.04 |
| | PBE | 4 | | 67.29 | | | 4.54 | |

Table S15: 6 of the elements of the elastic tensor for each system evaluated by DFT, the VASP MLPs and the MTPs for the initial and the extended (ext.) reference data sets. The values of the tensor elements are given in GPa. Indices for C are given in Voigt notation with 1=xx, 2=yy, 3=zz, 4=zy, 5=xz, 6=xy. Empty table entries indicate a value of zero. The number of atom types in the used method is indicated with $n_{type}$.

| System | Method | $n_{type}$ | $C_{14}$ | $C_{15}$ | $C_{16}$ | $C_{24}$ | $C_{25}$ | $C_{26}$ |
|---|---|---|---|---|---|---|---|---|
| MOF-74 | VASP MLP | 4 | | | | | 0.58 | 3.28 |
| | VASP MLP | 8 | | | | | 0.66 | 3.26 |
| | VASP MLP ext. | 4 | | | | | 0.72 | 3.24 |
| | MTP | 8 | | | | | 0.71 | 3.29 |
| | MTP ext. | 8 | | | | | 0.70 | 3.35 |
| | PBE | 4 | | | | | 0.75 | 3.30 |
| MIL-53 (lp) | VASP MLP | 4 | | | -3.36 | -3.80 | | |
| | VASP MLP | 8 | | | -4.62 | -3.36 | | |
| | VASP MLP ext. | 4 | | | -3.71 | -3.32 | | |
| | MTP | 8 | | | -4.24 | -3.70 | | |
| | MTP ext. | 8 | | | -4.23 | -3.22 | | |
| | PBE | 4 | | | -4.54 | -3.93 | | |
| MIL-53 (np) | VASP MLP | 4 | | 14.13 | | | 3.72 | |
| | VASP MLP | 8 | | 13.87 | | | 2.35 | |
| | VASP MLP ext. | 4 | | 16.37 | | | 3.65 | |
| | MTP | 8 | -0.03 | 13.96 | -0.07 | -0.06 | 2.86 | -0.10 |
| | MTP ext. | 8 | -0.05 | 14.22 | -0.13 | -0.13 | 3.46 | -0.20 |
| | PBE | 4 | | 13.88 | | | 3.66 | |

## S3.1.5 Further details regarding accuracy of the forces in the calculations relying on the UFF4MOF and Dreiding potentials

To put the performance of the machine learned potentials into perspective, we also evaluated the performance of UFF4MOF and Dreiding for the validation set. For the forces, we show the comparisons to DFT in Fig. S18 in the form of 2D Histograms. It can be seen that the deviations can be substantial and there is a trend to predict much higher forces by the UFF4MOF and Dreiding force fields than by DFT. For UFF4MOF in the case of MOF-5, there is still a somewhat reasonable agreement for a lot of the reference structure (but still much worse than any potential specifically fitted for MOF-5). For MOF-74 we see already a much worse agreement as the structure is substantially more complex. The worse performance primarily stems from the parameters used in UFF4MOF. One atom type has only one associated angle, which for the 4 coordinated Zn atoms (atom type Zn4+2 in UFF4MOF) amounts to 90°. This is close to the O-Zn-O angles in MOF-74 but there are still deviations of several degrees.

This leads to a distorted energy minimized structure for UFF4MOF with angles too close to 90°, but still resembling MOF-74 in its general shape. In that sense it is not surprising that UFF4MOF performs substantially worse for MOF-74. However, because of this one should be careful when applying these methods for high throughput screening. For the Dreiding potential the agreement is already extremely poor for MOF-5. Again, this is not at all surprising as that potential was never intended to be applied to MOFs. In fact, its creation even predates the discovery of the first MOFs.

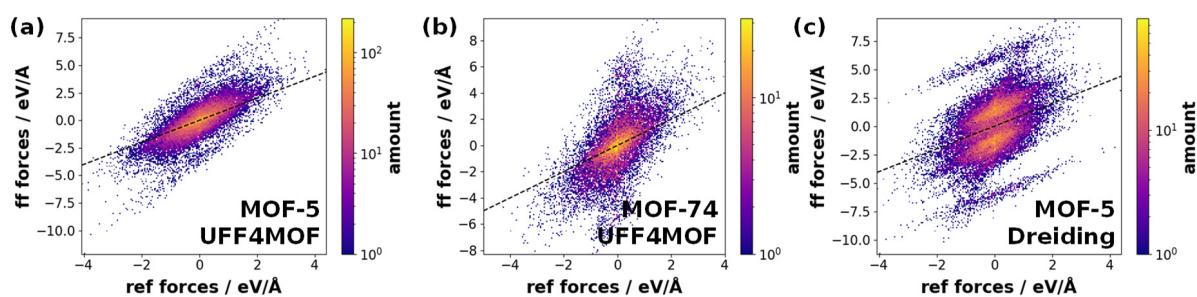

*Fig. S18: 2D Histograms showing the correlation of the forces between UFF4MOF (in a, b) and Dreiding (in c) force field calculations and DFT data (ref) for MOF-5 in a, c and for MOF-74 in b. The black dashed line with a slope of one serves as a guide to the eye indicating ideal agreement.*

## S3.2 Phonon eigenvector comparisons

Another aspect we checked for the vibrations and did not discuss in the main manuscript for it not to become too extensive is a quantitative evaluation of the actual equivalence of modes between the force field and DFT simulations. To test that, we calculated the dot products of the phonon mode eigenvectors, where values of 1 refer to a perfect agreement between the force-field and DFT calculated vibrations. This was done for all possible combinations of eigenvectors from both sets of modes. Then the agreement was optimized by solving an assignment problem to obtain the combination of eigenvectors with the highest overlap. In this way the actually matching modes – which might be ordered differently due to slight frequency shifts – are compared. A complication in that context is that many of the investigated MOFs belong to highly symmetric space groups. Consequently, they have large numbers of degenerate phonon modes, where any linear combination of the eigenvectors is a valid description of the motion, which prevents a meaningful evaluation of the dot product. MIL-53 is the only system, which does not show degenerate modes and where the comparison of all eigenvectors over the entire frequency range is sensible. Therefore, in the present comparison we focus on both phases of MIL-53, for which histograms of dot products are shown in

Fig. S19. For the MTPs and VASP MLPs, the agreement with DFT is almost perfect with the vast majority of the phonon modes being in the highest (0.95-1.00) bin.

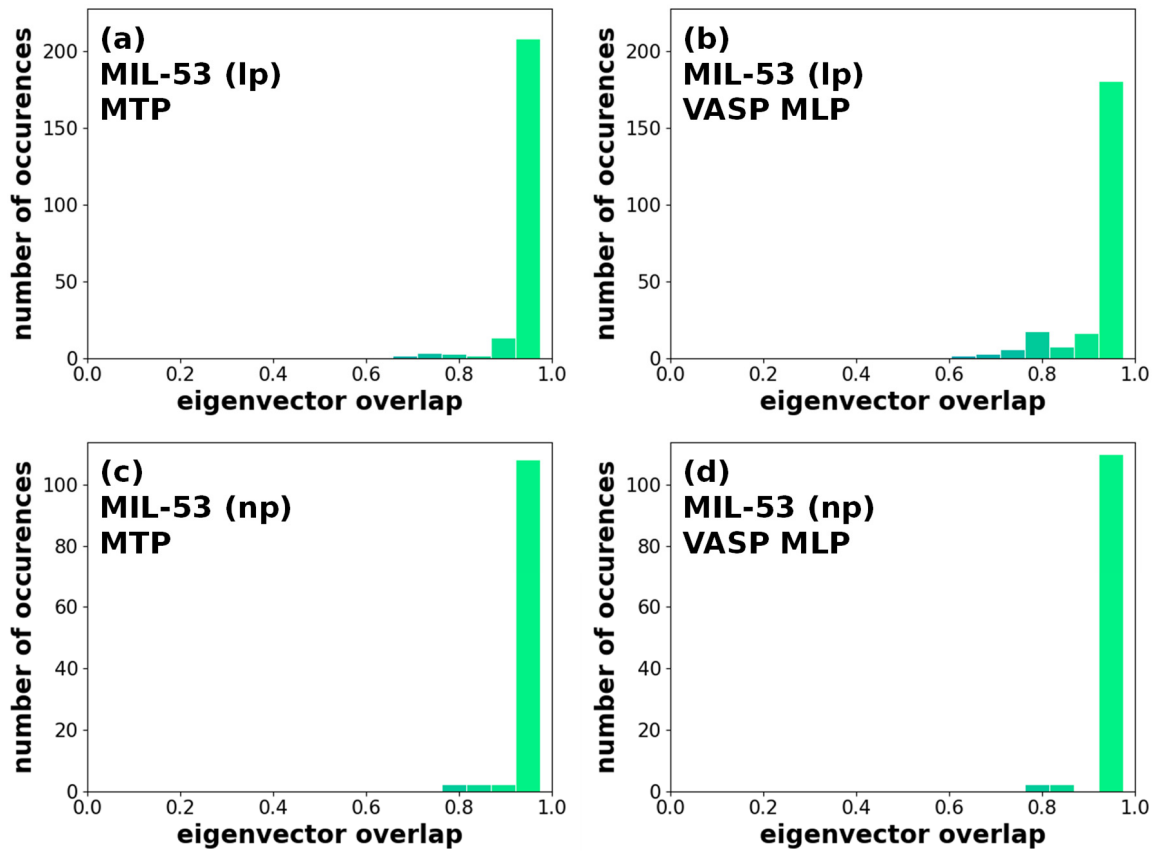

*Fig. S19: Histograms showing the optimized agreement of the overlap (among all possible combinations of eigenvectors from different modes) from the dot product of the phonon eigenvectors between DFT and MTP obtained phonon modes in a, c and between DFT and VASP MLP obtained phonon modes (b, d) for MIL-53 (lp) in a, b and MIL-53 (np) in c, d. The bins were distributed in intervals of 0.05. The range of the y axis was deliberately chosen to be equal to the total number of phonon modes in the system.*

The other materials have a higher degree of symmetry and feature many two or three-fold degenerate modes. As any linear combination of such degenerate eigenvectors is a proper solution to the eigenvalue problem, the displacements for these modes are ambiguous and a comparison between MLP and DFT becomes rather meaningless. However, it is possible to use phonopy[13] to identify such degenerate modes based on symmetry considerations by evaluating the irreducible representation. Using this (often in combination with a visual inspection of the eigenmodes required by the rather "messy" band structures), the degenerate modes can be eliminated and an evaluation of the agreement of the non-degenerate modes can be performed also for the other systems. The corresponding histograms for the comparison of the MTPs based on the initial reference data set with the DFT reference are shown in Fig. S20. As can be seen, the non-degenerate modes fit very well for the MTPs, especially for MOF-5 and UiO-66, where all modes are within the 0.95 to 1.0 bin. However, the total number of clearly non-degenerate modes is relatively low making the comparison far less conclusive.

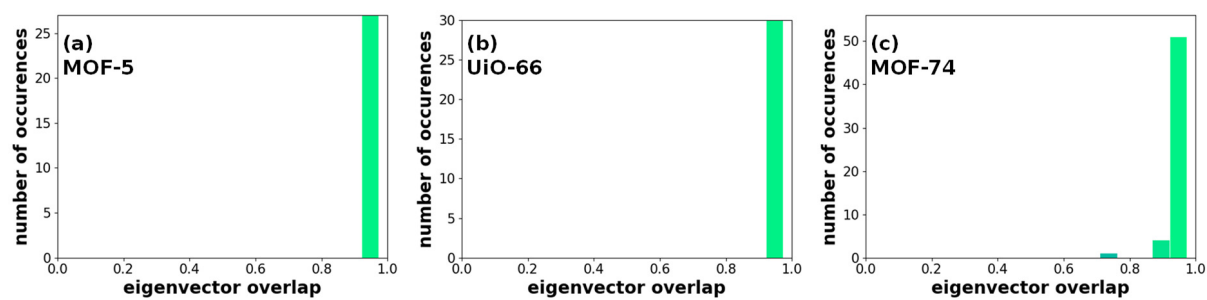

*Fig. S20: Histograms showing the optimized overlap (among all possible combinations of eigenvectors from different modes) of the eigenvectors obtained from the MTPs (initial) and DFT of only the non-degenerate phonon modes.*

## S3.3  Further analysis of the extended reference data set

To give a clearer overview of the impact of the extended reference data set, we provide a series of figures displaying relevant quantities in a style similar to the main manuscript for each system for both the MTP and the VASP MLPs. Generally, it should be noted that the VASP MLPs were only trained using 4 atom types for the extended reference data set. The exception for this represents MOF-74, for which results are shown in the main manuscript.

### S3.3.1  Deviation of forces, energies and stresses

Fig. S21 contains the comparisons of forces, stresses and energies with the validation set. Compared to the initial reference data set, it can be seen that the differences are rather small. Note, that for the VASP MLPs the agreement now looks slightly worse, since they were only trained for four atom types on the extended reference data set. Numerically we only observed small improvements as already indicated in Table S11.

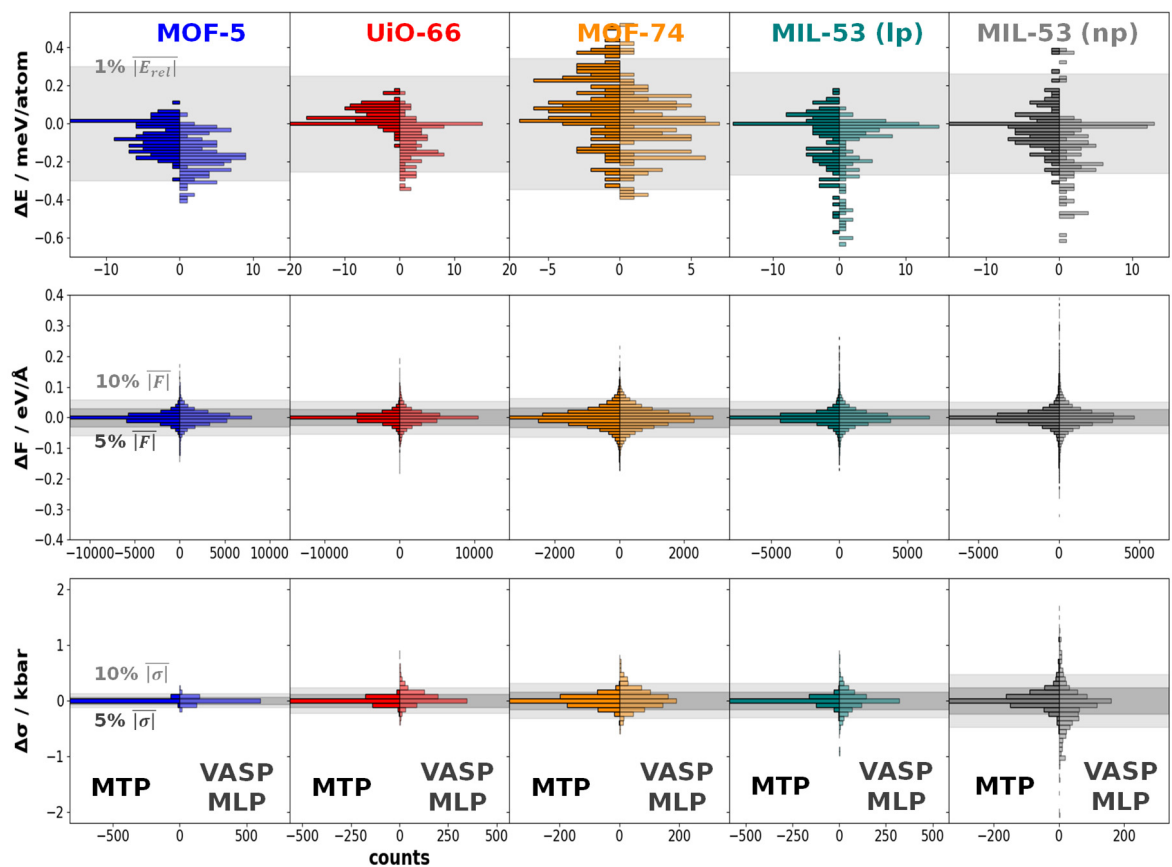

*Fig. S21: Histograms detailing the deviations (FF data subtracted from reference) of the energies (E), the components of the forces (F), and the components of the stresses (σ) obtained from DFT and from the VASP MLPs and the MTPs for MOF-5 (blue), UiO-66 (red), MOF-74 (orange), MIL-53 (lp) (teal) and MIL-53 (np) (grey). All force fields were obtained from the respective extended reference data set. For the VASP MLPs 4 atom types were used in this case. For the energies only the energy difference from the structure with the minimum energy is shown. The validation set structures were obtained from an active learning VASP MLP learning run at a temperature of 300 K. The areas shaded in gray indicate the ranges with errors within 5% and 10% of the absolute DFT values for forces and stresses and within 1 % for the energy differences.*

### S3.3.2  MTP calculated phonon band structures for the extended reference dataset

In *Fig. S22* one can see the phonon band structures as obtained from MTPs trained on the extended reference data set. Also here, it is apparent that the agreement is excellent and especially for MIL-53 (lp) some improvements can be discerned over the initial reference data set. However, the changes are mostly minor

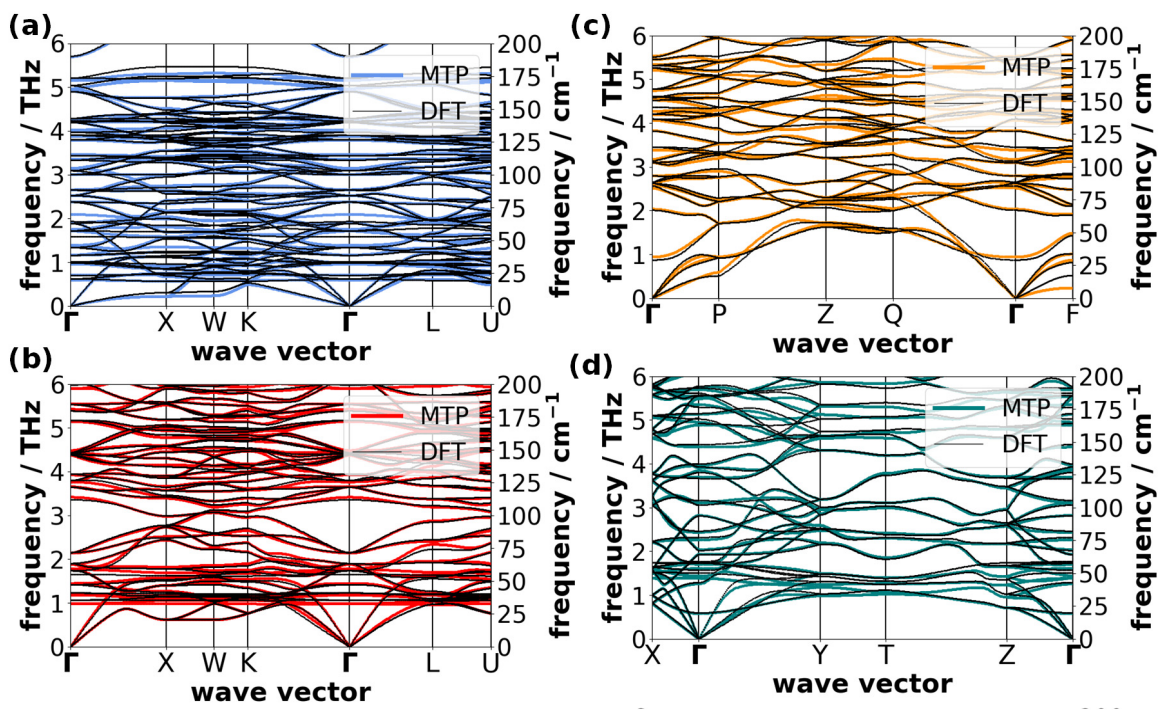

*Fig. S22: Low frequency phonon band structures computed using the MTPs parametrized on the extended reference data sets) and computed using DFT for the following systems: MOF-5 in a, UiO-66 in b, MOF-74 in c, and the large pore phase of MIL-53 in d.*

## S3.4 Convergence of the supercell size for the phonon band structures with the MTPs

When computing phonon band structures it is important to ensure a properly converged supercell. Fig. S23 shows the phonon band structures evaluated using MTPs for different supercells. Here, "primitive" always refers to the smallest possible unit cell and "conv" (conventional) means the

$$\begin{pmatrix} -1 & 1 & 1 \\ 1 & -1 & 1 \\ 1 & 1 & -1 \end{pmatrix}$$ supercell for MOF-5 and UiO-66 and the

$$\begin{pmatrix} 1 & 0 & 3 \\ -1 & 1 & 3 \\ 0 & -1 & 3 \end{pmatrix}$$ supercell for MOF-74. The bands are plotted on top of each other starting with the

first one included in the caption. If no bands with a specific color can be seen that means that they essentially coincide with the bands for larger supercells.

For MOF-5 even the primitive unit cell is sufficient to obtain fully converged phonon band structures for the MTPs. For UiO-66 there are still some minor differences between the primitive and cubic conventional unit cell. For MOF-74, the primitive unit cell is clearly insufficient, but the regular conventional cell is already converged. For MIL-53 (lp), the 1×2×1 supercell already shows no visible differences compared to the larger supercells. It should be noted that this is inconsistent with what was observed for DFT, where there were still substantial differences between 1×2×1 and 1×2×2 supercells for MIL-53 (lp). The reason for this is that the MTPs are cut off after already 5 Å, which should typically lead to a faster convergence behavior than DFT, where also longer-range interactions are included. For MIL-53 (np), we see convergence after using a 2×2×2 supercell with no imaginary frequencies (unlike DFT).

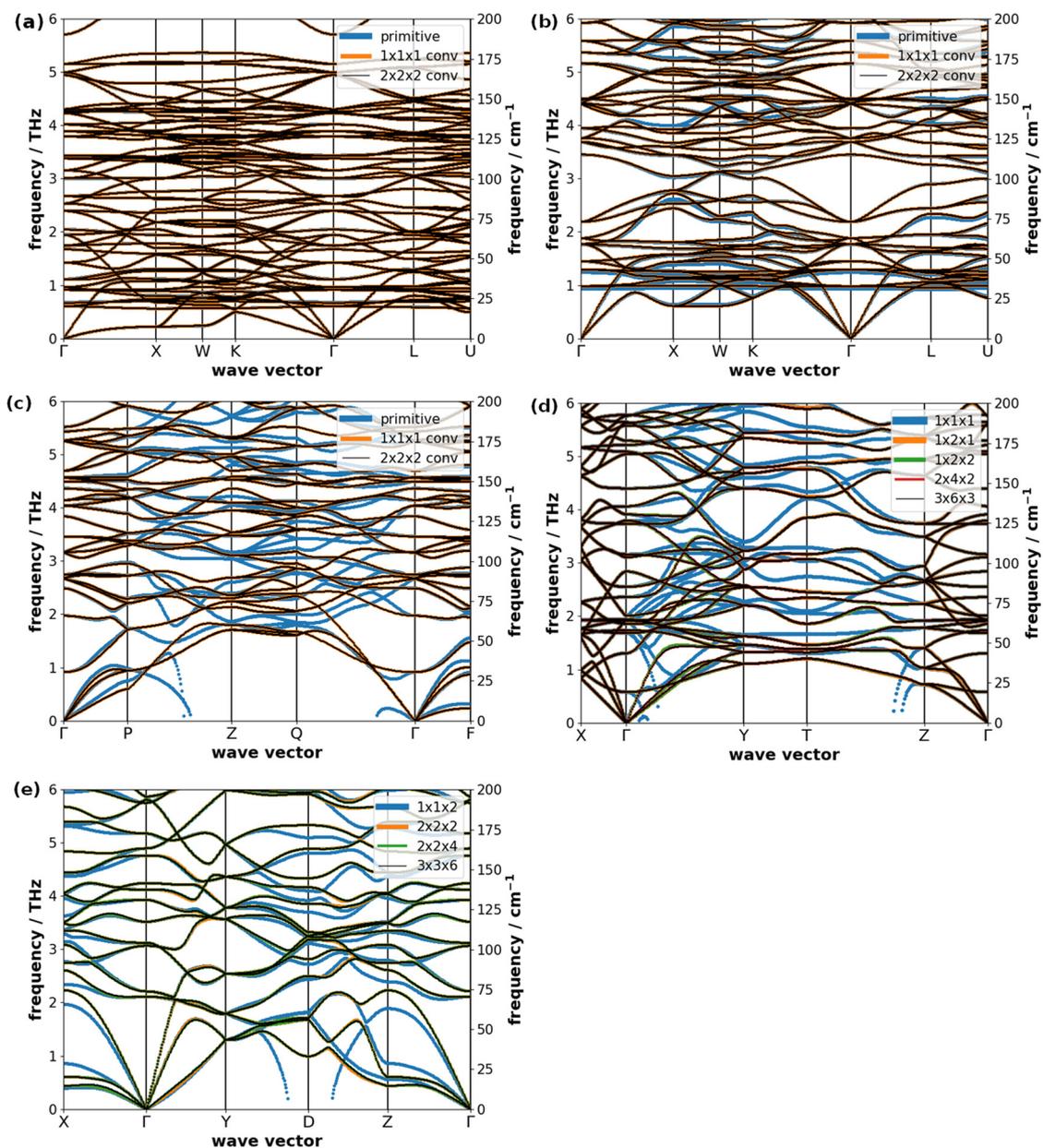

*Fig. S23: Phonon band structures evaluated for different supercells for the following systems: MOF-5 in a, UiO-66 in b, MOF-74 in c, MIL-53 (lp) in d, and MIL-53 (np) in e. The MTPs based on the initial reference data set were used. The last entry in each legend represents the respective top layer of the drawn band structures.*

## S3.5 Phonon band structures obtained with the VASP MLP and with UFF4MOF

In the main manuscript mostly the phonon band structures obtained with the MTPs are shown. For the VASP MLPs in particular only the phonon band structure for MOF-5 was shown. For the VASP MLPs trained on the initial reference data set, we provide the comparisons to the DFT-calculated band structures in Fig. S24 for 4 of the systems. Generally, the agreement is very good, maybe slightly worse than for the MTPs when compared with Fig. 4 in the main manuscript. There especially the optical mode around 1 THz in UiO-66 and some of the optical modes in the range of 2-5 THz for MOF-74 and MIL-53 show a somewhat better agreement for the MTPs. But as mentioned in the main manuscript: the VASP MLPs can still benefit substantially from an extension of the reference data set. This can be seen for MOF-74 in Fig. S25, where the phonon band structure is shown for a VASP MLP trained on the extended reference data set. There, the agreement is substantially improved compared to the initial reference data set and the somewhat poorly fitting modes around 2-4 THz present previously now show a good agreement.

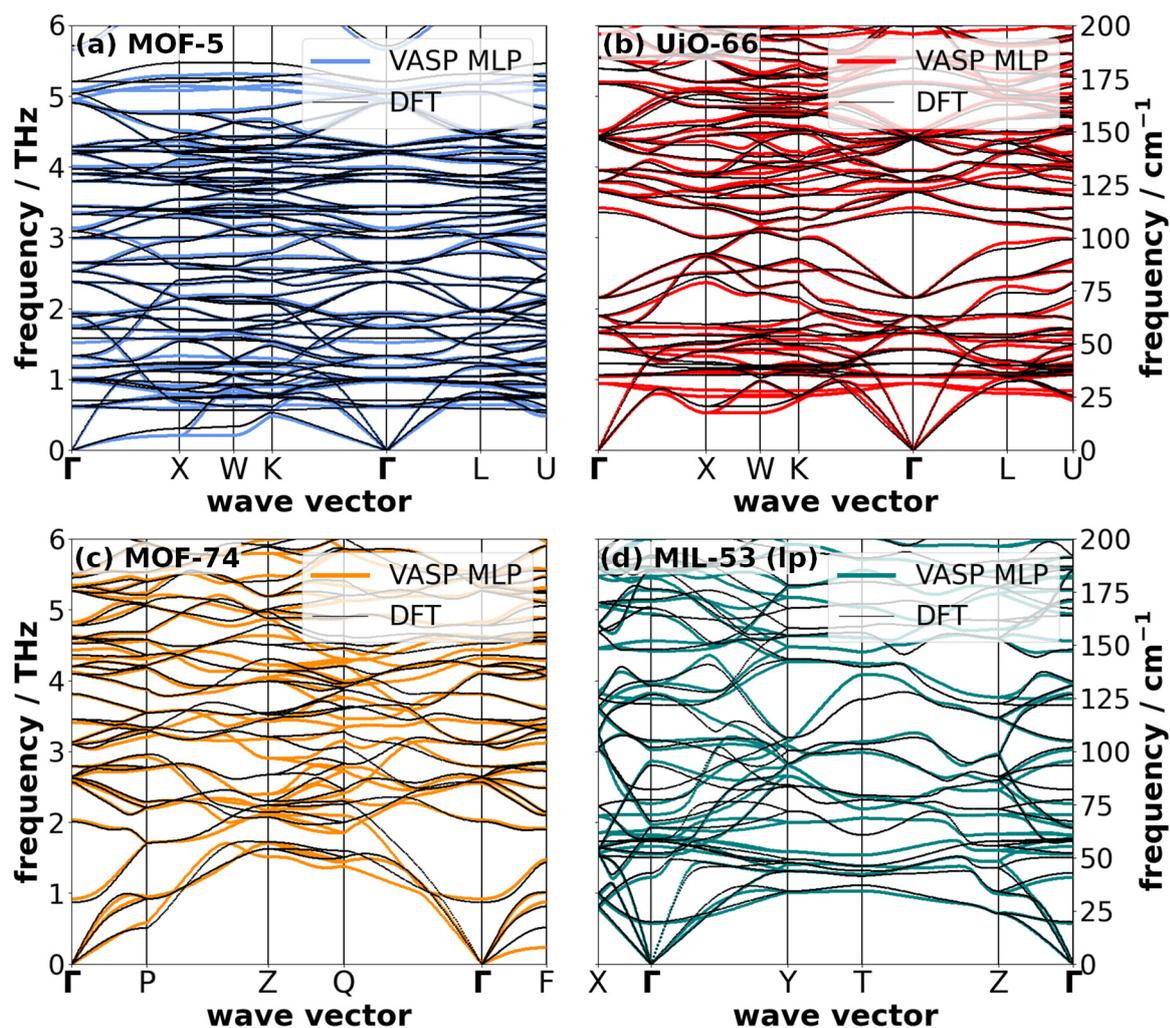

Fig. S24: *Low frequency phonon band structures computed using the VASP MLPs (from the initial reference data set) and from DFT for the following systems: MOF-5 in a, UiO-66 in b, MOF-74 in c and the large pore phase of MIL-53 in d.*

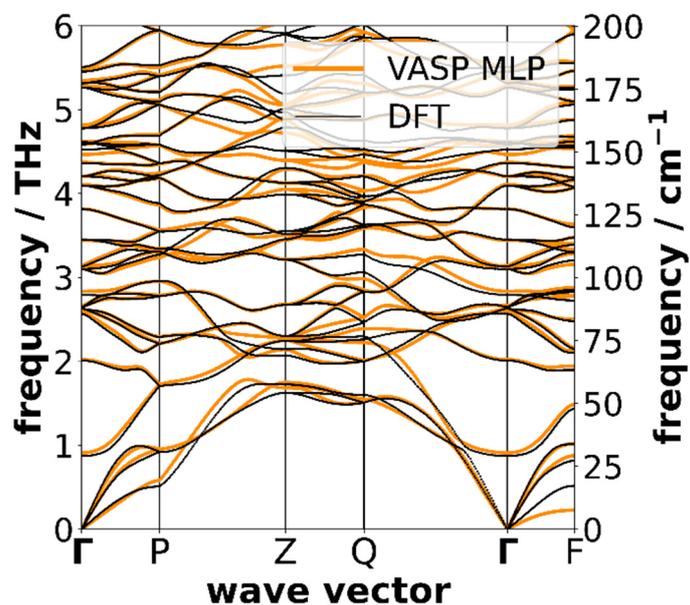

Fig. S25: *Low frequency phonon band structures computed using the VASP MLP trained on the extended reference data set (including separation of atom types) and DFT for MOF-74.*

For UFF4MOF, the low frequency phonon band structure of MOF-74 and its comparison with DFT is shown in Fig. S26. It can be seen that especially in this frequency region the agreement is extremely poor – even worse than for MOF-5.

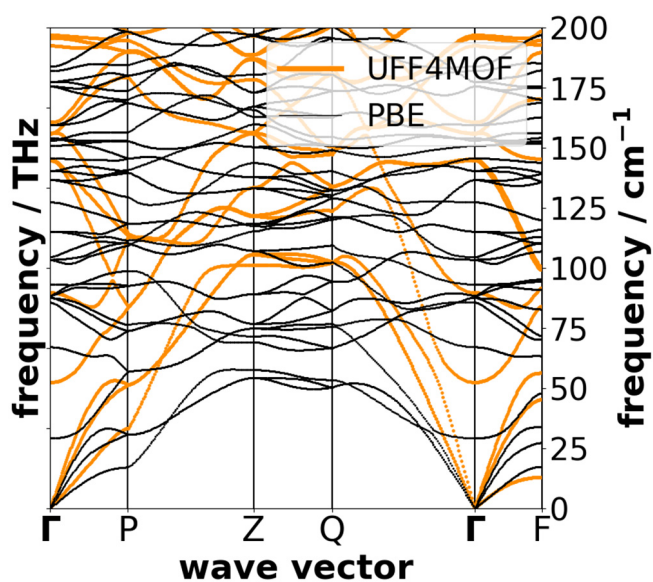

Fig. S26: *Low frequency phonon band structures computed using UFF4MOF and DFT for MOF-74.*

## S3.6 Details on the impact of the level of the MTPs

In this section, we provide additional information on how the level of the MTP impacts the calculation of various of the observables. The impact on the RMSD errors for forces, stresses and energies for validation set and on for the frequencies at the $\Gamma$-point are shown in Fig. S27. Generally, all of the errors show a steady decline with the MTP level. Especially for the phonon frequencies the RMSD value can be reduced by a factor of three for some of the systems. For the stresses, the differences between the systems merely results from different absolute values. Generally, there is a certain degree of noise throughout the data points. This is a result of the fact that only the primarily used level 22 potentials the "best" MTP of a set of fitted potentials was picked. For all other levels, only one MTP was fitted. This leads to a certain degree of variation in accuracy due to the stochastic initialization of the MTP fitting procedure. The energies show a larger degree of noise, which result from the errors already being low, even for the low level MTPs.

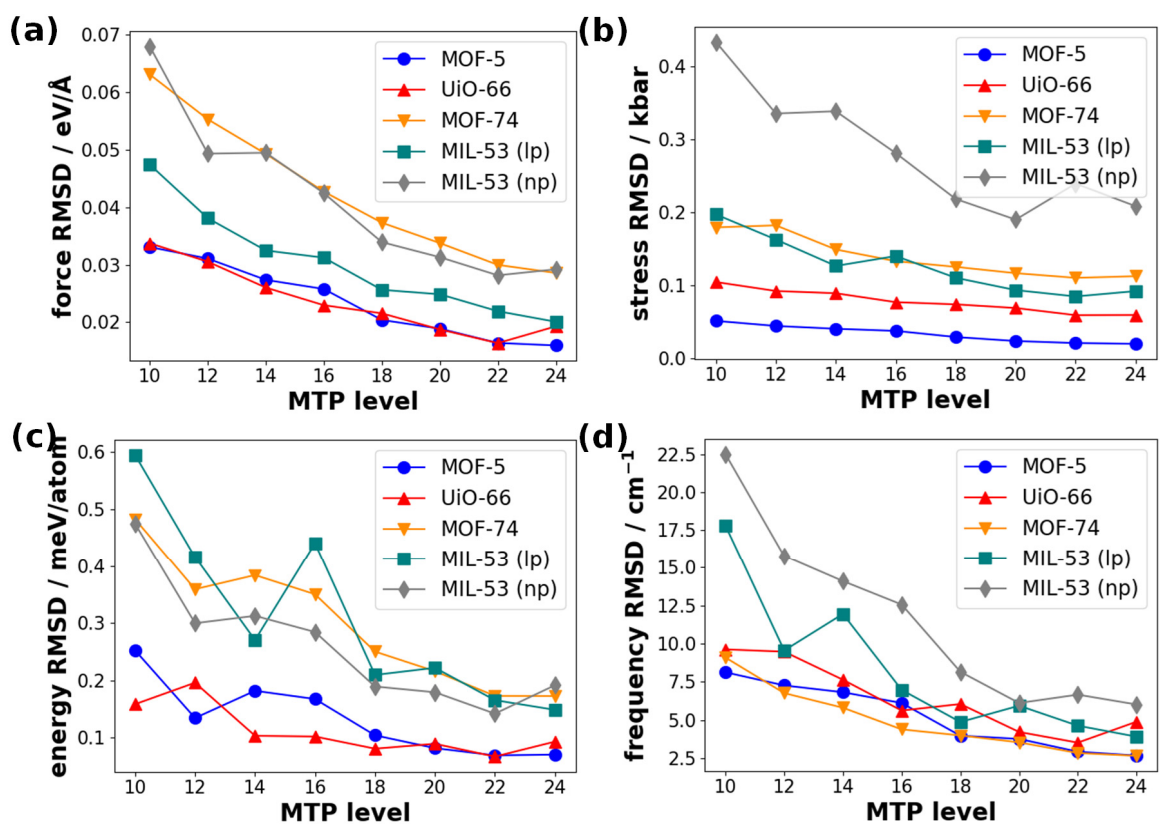

*Fig. S27: The effect of the MTP level on the RMSDs of the forces in a, stresses in b and the energies in c in the validation set and also on the phonon frequencies at the $\Gamma$-point (in d) for a comparison between calculations with MTPs at various levels and with DFT. The MTPs have been parametrized based on the initial reference data set. The studied systems comprise: MOF-5 (blue), UiO-66 (red), MOF-74 (orange) and MIL-53 (lp) (teal) and MIL-53 (np) (gray).*

Additionally, the comparisons for the elastic tensor elements are given in Fig. S28. In particular for MIL-53 (lp) and MOF-74 we see quite substantial deviations in the level 10-14 MTPs. These deviations do not occur for all the systems at the same levels, as the elastic tensor elements for MOF-5, UiO-66 and MIL-53 (np) are not that dissimilar from those for level 12 upwards. Overall, the results are quite system dependent and also depends on the particular tensor element. For example, for MOF-74, substantial deviations in $C_{33}$ can be seen even at high levels of 20-24. For MIL-53 this applies to most tensor elements.

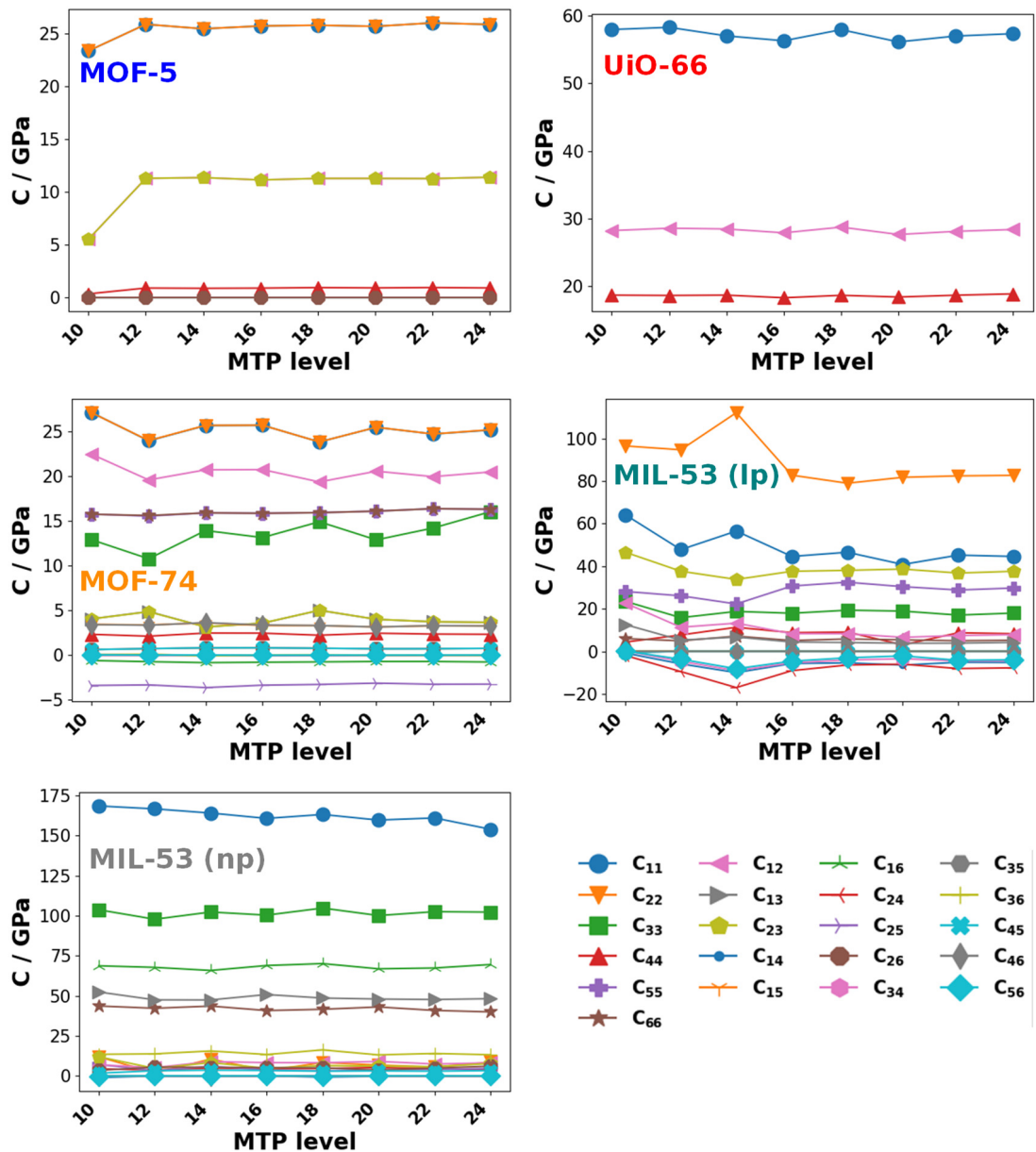

Fig. S28: Comparison of the elements of the elastic stiffness tensor for moment tensor potentials obtained for different levels (10-24) on the initial reference data set for each of the investigated systems.

## S3.7 Impact of VASP MLP training settings

In Fig. 7c in the main manuscript, the results for a number of different settings chosen for training MOF-74 VASP MLPs were shown. Here, we provide some additional information. Additionally, the role of some other settings, which have not resulted in large deviations, will be discussed. For this, Table S16 provides an overview of the RMSD values for the forces in the validation set for various settings in combination with the CPU time per time step and atom for MOF-74. Based on the high CPU time required it is clear that the potentials using ML_IALGO_LINREG 1 and 2 are not really suited for production runs. Therefore, one always should retrain the potential for a fast force field using either ML_IALGO_LINREG 3 or 4. The accuracy in this case even gets better due to the use of the singular value decomposition method. Additionally, changing the ML_SIGW0 parameter, which is a precision parameter when using the default ML_IALGO_LINREG=4, did not change the result substantially. Aside from the settings already discussed in the main manuscript, we also tested the effect of the ML_LMAX2 setting, which defines the maximum angular momentum quantum number of the spherical harmonics in the angular descriptor. However, the resulting potentials when setting the parameter to 2 performed essentially identical to the ones with the default value of 3.

Table S16: Evaluation of the force RMSDs (with the validation sets) and required CPU time for different VASP MLPs for MOF-74. The number of atom types is given as $n_{types}$ and the number of local reference configuration as $n_{conf}$. Unless specified otherwise, the baseline of the force fields is always given as the VASP 6.4.1 defaults with the specified setting modified.

| Data set | setting | $n_{types}$ | $n_{conf}$ | Force RMSD / eV/Å | CPU time / ms/(step·atom) |
|---|---|---|---|---|---|
| initial | VASP 6.3.0 defaults | 4 | 1877 | 0.080 | 310.62 |
| initial | 6.4.1 defaults | 4 | 4830 | 0.037 | 0.74 |
| initial | ML_IALGO_LINREG=1 | 4 | 4830 | 0.040 | 92.14 |
| initial | ML_IALGO_LINREG=2 | 4 | 4830 | 0.040 | 92.18 |
| initial | ML_IALGO_LINREG=3 | 4 | 4830 | 0.037 | 0.75 |
| initial | ML_SIGW0 = 1E-5 | 4 | 4830 | 0.038 | 0.74 |
| initial | ML_RCUT2=4 | 4 | 3947 | 0.038 | 0.64 |
| initial | ML_LMAX2=2 | 4 | 4795 | 0.037 | 0.74 |
| initial | ML_MB=3000 | 4 | 5859 | 0.034 | 0.85 |
| initial | 6.4.1 defaults | 8 | 8942 | 0.027 | 2.19 |
| initial | ML_IALGO_LINREG=1 | 8 | 8942 | 0.030 | 408.73 |
| initial | ML_IALGO_LINREG=2 | 8 | 8942 | 0.030 | 410.11 |
| initial | ML_IALGO_LINREG=3 | 8 | 8942 | 0.027 | 2.21 |
| initial | ML_SIGW0=1E-5 | 8 | 8942 | 0.027 | 2.30 |
| initial | ML_RCUT2=4 | 8 | 6333 | 0.029 | 1.81 |
| initial | ML_LMAX2=2 | 8 | 8942 | 0.027 | 2.24 |
| extended | 6.4.1 defaults | 4 | 5237 | 0.038 | 0.76 |
| extended | 6.4.1 defaults | 8 | 11350 | 0.022 | 2.53 |

# S4. Molecular dynamics

## S4.1 Constant temperature and pressure simulations to evaluate the average lattice parameters and thermal expansion coefficients

To obtain the dependent lattice parameters and thermal expansion coefficients, the same style of NPT simulations was used as already mentioned in section S2.2. The supercells employed for each system are listed in Table S17. The simulations were carried out in the temperature range between 100 and 700 K and all simulations for at least 100,000 time steps that concluded successfully (i.e., for which the system did not disintegrate) were used in the evaluation of the thermal expansion coefficient. However, some data points were still excluded occasionally (see discussion below for details and individual explanation of the reasoning) due to a non-linear evolution of the lattice parameters/unit-cell volumes with temperature, which would make the evaluation of linear thermal expansion coefficients difficult. These coefficients were obtained by a linear fit of the dependence of individual lattice parameters on temperature.

*Table S17: Used supercells to obtain the thermal expansion coefficients using NPT molecular dynamics simulations for both the VASP MLP and MTP approaches.*

| System | Supercell | Number of atoms |
|---|---|---|
| MOF-5 | 2x2x2 cubic conventional unit cells | 3392 |
| UiO-66 | 2x2x2 cubic conventional unit cells | 3648 |
| MOF-74 | 1x1x3 hexagonal conventional unit cells | 486 |
| MIL-53 (lp) | 2x2x4 unit cells | 1216 |
| MIL-53 (np) | 2x4x4 unit cells | 2432 |

To add to the cells used for the thermal expansion calculations, the speed tests were performed for substantially larger supercells to properly test the parallelization on a high performance computation system. For MOF-5 4×4×4 conventional supercells (27,136 atoms) and for MOF-74 4×4×12 hexagonal conventional unit cells (31,104 atoms) were used. As an additional note, it should be said that in general for the tests of the computational efficiency time values were taken from the actual execution of the MD run steps. In LAMMPs this means that the time elapsed was taken from the performance breakdown after the "run" command. In VASP the value specified as "pes_ff" in the table of the Accumulative profile is taken, which is only output when the code was compiled using the "-DPROFILING" compiler option. This option also excludes the time lost due to outputting various quantities during the simulation, which can be a time bottleneck otherwise. Both strategies serve to exclude the initialization overhead from the actual evaluation of the required computation time per time step. This is the reason why we only performed the efficiency tests for 2000 time steps to save computation time for simulations on such large supercells.

### S4.1.1  NPT simulations with the MTPs

Fig. S29 and Fig. S30 show the linear thermal expansion fits for the lattice parameters and the tilt angles for all the systems obtained from NPT simulations carried out by the MTPs based on the initial reference data set. For the cubic systems MOF-5, MIL-53 (np) and UiO-66 the trends are clearly linear and the thermal expansion can be obtained straightforwardly. For MOF-74, we see some noisy datapoints in a and b direction which is mostly due to the thermal expansion being very small. For MIL-53 (lp), we see an odd step in the b parameter and a non-orthorhombic angle at low temperatures. This is related to the minimum structure at 0 K showing a slight deviation from the orthorhombic symmetry. This deviation amounts to two minima at +/- 1 ° with a weak energy barrier in between. At high temperatures, kinetic energies are high enough to overcome that barrier and one obtains an on average orthorhombic structure, which is consistent with experiment. Due to this, the thermal expansion for MIL-53 (lp) was only evaluated for temperatures of 150 K upwards. For MIL-53 (np) we also see a significant increase of the tilt angle $\gamma$ with temperature amounting to 30.6 K$^{-1}$.

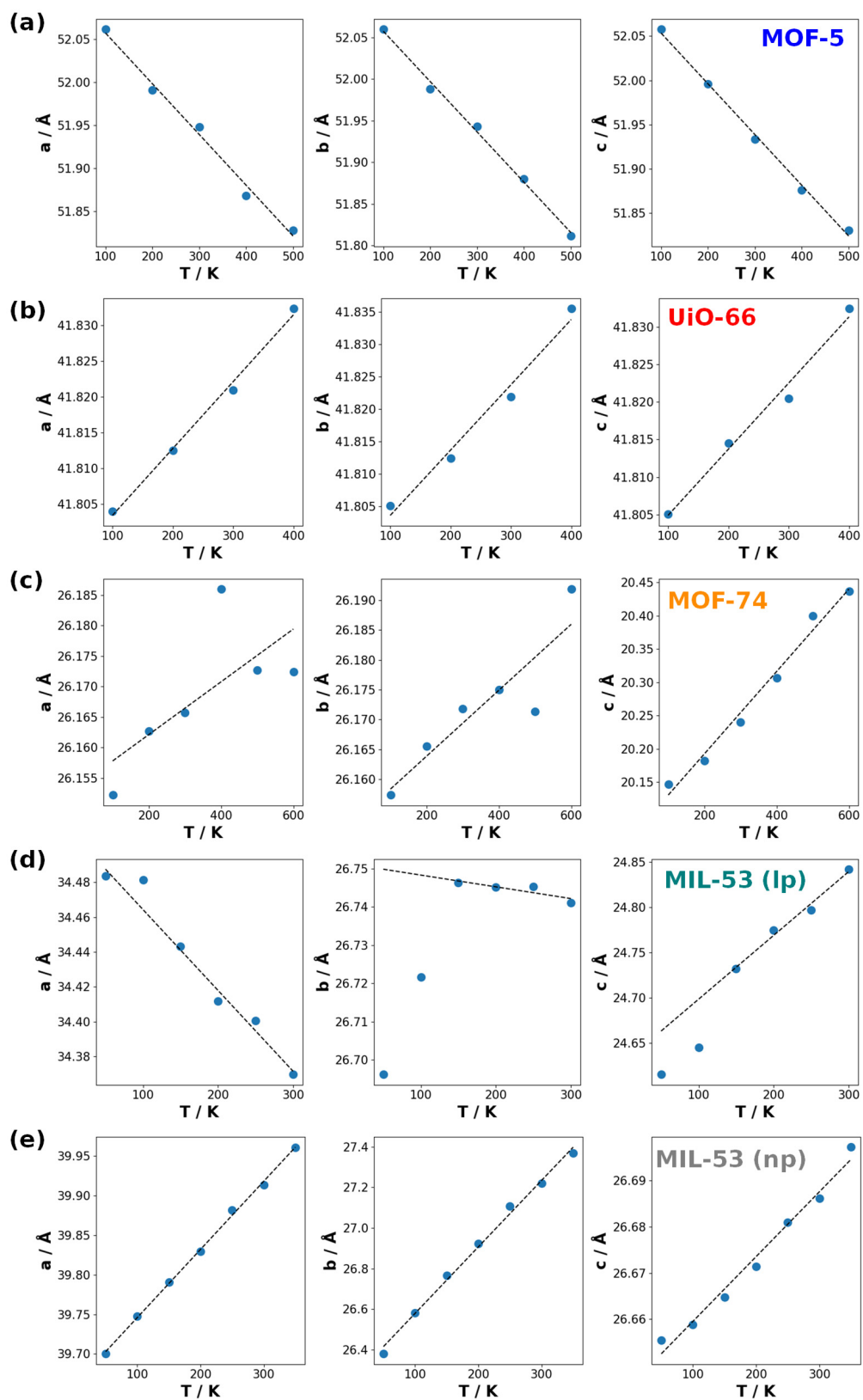

Fig. S29: Averaged lattice parameters of the investigated systems – MOF-5 (in a), UiO-66 (in b), MOF-74 (in c), MIL-53 (lp) (in d) and MIL-53 (np) (in e) – as a function of temperature (blue dots) obtained from NPT simulations using the MTPs based on the initial reference data sets. The black dashed line represents the linear fit used to obtain the thermal expansion coefficients.

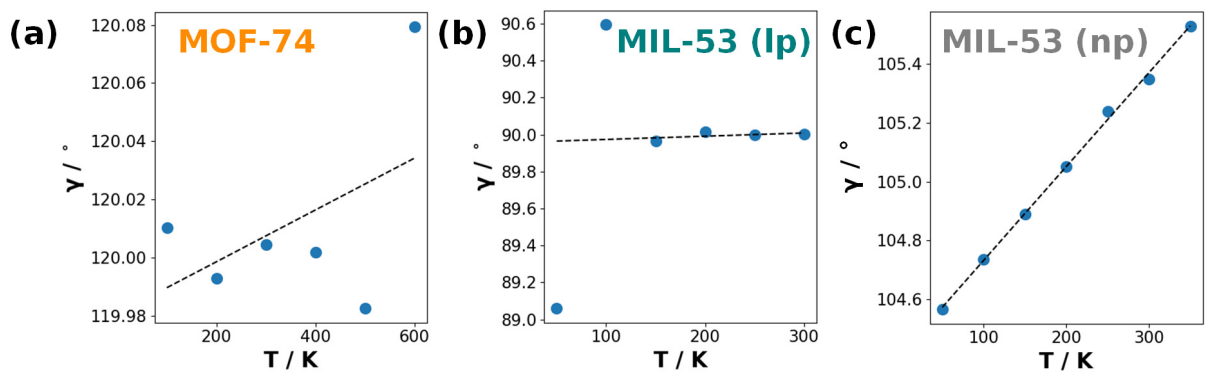

*Fig. S30: Averaged tilt angles of the investigated systems – MOF-74 (in a), MIL-53 (lp) (in b) and MIL-53 (np) (in c) – as a function of temperature (blue dots) obtained from NPT simulations using the MTP based on the initial reference data sets. The black dashed line represents the linear fit used to obtain the change of the tilt angles with temperature.*

It should be noted, that for MOF-74, it has been shown for DFT calculations and the Grüneisen theory of thermal expansion that a thermal expansion coefficient similar to the experiment is provided only by the PBEsol and not by the PBE functional [8]. Therefore, we also used this functional to train separate MTPs with an analogous approach as done throughout this work employing the PBE functional. A reference data set of 1034 DFT calculations was produced using a VASP active learning approach with temperatures reaching up to 700 K. However, the obtained MTPs in NPT simulations yielded thermal expansion coefficients as the PBE-trained MTP (e.g., thermal expansion coefficients for a level 22 MTP with a radial basis set size of 10: 5.5 $10^{-6}$ $K^{-1}$ in linker direction and 28.8 $10^{-6}$ $K^{-1}$ in stacking direction). Therefore, we conclude that the functional is not the primary reason for the discrepancy.

### S4.1.2 NPT simulations with the VASP MLPs

Also for the VASP MLPs, we performed NPT simulations up to a temperature of 700 K using the potential variant trained on the initial reference data set including separation of atom types. The lattice parameters as a function of temperature including the linear fits to obtain the thermal expansion coefficients, can be seen in *Fig. S31*. For the VASP MLPs all systems were technically stable up to a temperature of 700 K, but as can be clearly seen in the figure, at high temperatures large distortions of the unit cells occur for a lot of the systems. This is the reason why the linear fits were only performed up to a temperature, where the trend was still essentially linear. These temperatures were 600 K for MOF-5, 400 K for UiO-66, 400 K for MOF-74, 300 K for MIL-53 (lp) and 300 K for MIL-53 (np). For MOF-74 we only imposed the limit due to the nonlinearity of the lattice parameters as a function of temperature and not due to discontinuities. For the other systems the behavior after the given temperature is most likely not physical.

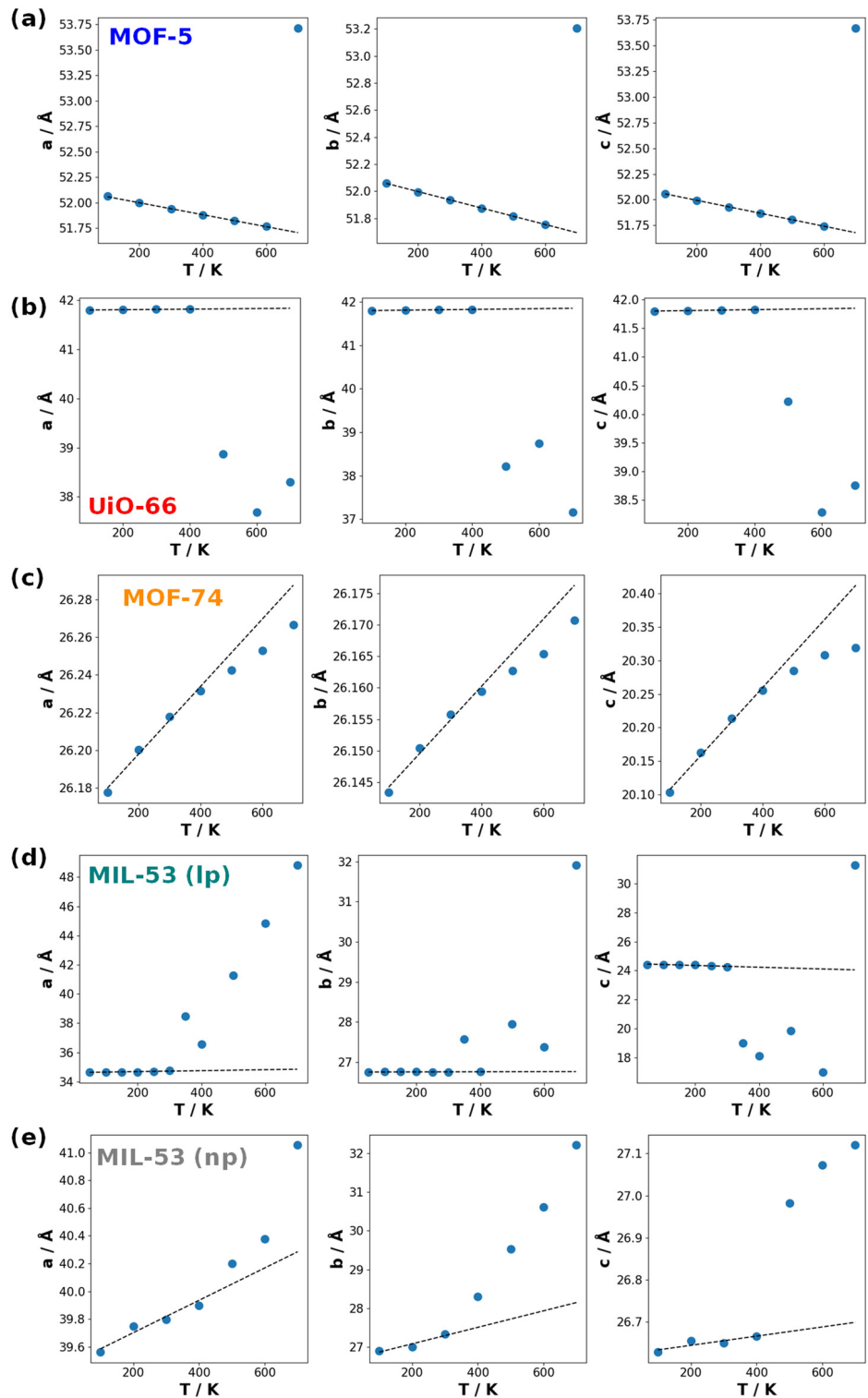

*Fig. S31: Averaged lattice parameters of the investigated systems – MOF-5 (in a), UiO-66 (in b), MOF-74 (in c), MIL-53 (lp) (in d) and MIL-53 (np) (in e) – as a function of temperature (blue dots) obtained*

*from NPT simulations using the VASP MLPs based on the initial reference data sets and using separation of atom types for the potentials. The black dashed line represents the linear fit used to obtain the thermal expansion coefficients.*

### S4.1.3 Lattice parameters at room temperature and thermal expansion coefficients

Table S18 provides a comparison of the calculated MTP and VASP MLP lattice parameters at room temperature (300 K for the simulations) with experiments. Generally, the values for the machine learned approaches are rather similar. For MOF-5, UiO-66 and MOF-74 the differences compared to experiment are relatively small. For MIL-53 (lp) we see a difference of up to 0.5 Å compared to the relatively consistent experimental data. This must be related to inaccuracies from the underlying DFT approach, as this large deviation from experiment was already present for the relaxed DFT unit cells. For MIL-53 (np) we also see large deviations between the MTP calculations and the experiment. However, for that system the available experimental data are also not particularly consistent.

Table S19 summarizes the linear thermal expansion coefficients obtained with the MTPs and VASP MLPs for all systems as obtained by the procedure described in the preceding section.

*Table S18: Room temperature lattice parameters obtained with MTPs and VASP MLPs (employing NPT simulations) and obtained from experiments.*

| System | Method | a / Å | b / Å | c / Å | gamma / ° | ref. |
|---|---|---|---|---|---|---|
| MOF-5 | MTP | 25.97 | 25.97 | 25.97 | 90.00 | |
| | VASP MLP | 25.97 | 25.97 | 25.96 | 90.00 | |
| | exp. | 25.82 | 25.82 | 25.82 | 90.00 | [18] |
| | | 25.78 | 25.78 | 25.78 | 90.00 | [19] |
| | | 25.82 | 25.82 | 25.82 | 90.00 | [20] |
| UiO-66 | MTP | 20.91 | 20.91 | 20.91 | 90.00 | |
| | VASP MLP | 20.91 | 20.91 | 20.90 | 90.00 | |
| | exp. | 20.77 | 20.77 | 20.77 | 90.00 | [20] |
| | | 20.70 | 20.70 | 20.70 | 90.00 | [21] |
| | | 20.75 | 20.75 | 20.75 | 90.00 | [22] |
| | | 20.67 | 20.67 | 20.67 | 90.00 | [23] |
| | | 20.70 | 20.70 | 20.70 | 90.00 | [23] |
| | | 20.96 | 20.96 | 20.96 | 90.00 | [23] |
| | | 20.74 | 20.74 | 20.74 | 90.00 | [23] |
| | | 20.76 | 20.76 | 20.76 | 90.00 | [23] |
| | | 20.76 | 20.76 | 20.76 | 90.00 | [23] |
| | | 20.82 | 20.82 | 20.82 | 90.00 | [23] |
| | | 20.84 | 20.84 | 20.84 | 90.00 | [23] |
| MOF-74 | MTP | 26.17 | 26.17 | 6.75 | 120.00 | |
| | VASP MLP | 26.22 | 26.16 | 6.74 | 120.00 | |
| | exp. | 25.93 | 25.93 | 6.80 | 120.00 | [8] |
| | | 25.93 | 25.93 | 6.84 | 120.00 | [24] |
| | | 25.93 | 25.93 | 6.83 | 120.00 | [25] |
| | | 26.16 | 26.16 | 6.83 | 120.00 | [26] |
| MIL-53 (lp) | MTP | 17.18 | 6.69 | 12.42 | 90.00 | |
| | VASP MLP | 17.37 | 6.69 | 12.12 | 90.00 | |
| | exp. | 16.76 | 6.64 | 12.84 | 90.00 | [11] |
| | | 16.76 | 6.63 | 12.79 | 90.00 | [27] |
| | | 16.73 | 6.63 | 12.84 | 90.00 | [28] |
| MIL-53 (np) | MTP | 19.96 | 6.67 | 6.81 | 105.35 | |
| | VASP MLP | 19.90 | 6.66 | 6.83 | 105.24 | |
| | exp. | 20.76 | 6.61 | 7.06 | 113.58 | [11] |
| | | 19.51 | 6.58 | 7.61 | 104.24 | [12] |
| | | 19.63 | 6.56 | 7.16 | 104.7 | [28] |

*Table S19: Thermal expansion coefficients for the investigated systems obtained from NPT simulations performed with the MTPs and VASP MLPs compared to experimental values at room temperature.*

| System | Method | $\alpha_a$ / $10^{-6}$ K$^{-1}$ | $\alpha_b$ / $10^{-6}$ K$^{-1}$ | $\alpha_c$ / $10^{-6}$ K$^{-1}$ | $\alpha_V$ / $10^{-6}$ K$^{-1}$ |
|---|---|---|---|---|---|
| MOF-5 | MTP | -11.0 | -11.3 | -11.6 | -33.9 |
| | VASP MLP | -11.7 | -11.3 | -11.6 | -35.0 |
| UiO-66 | MTP | 2.2 | 2.4 | 2.1 | 6.7 |
| | VASP MLP | 1.4 | 2.1 | 1.9 | 5.4 |
| MOF-74 | MTP | 1.6 | 1.2 | 31.0 | 33.6 |
| | VASP MLP | 6.9 | 2.0 | 25.3 | 31.2 |
| MIL-53 (lp) | MTP | -15.5 | 4.4 | 40.2 | 28.6 |
| | VASP MLP | 9.3 | 0.5 | -24.9 | -15.7 |
| MIL-53 (np) | MTP | 21.7 | 5.3 | 124.7 | 137.2 |
| | VASP MLP | 29.6 | 4.1 | 80.0 | 101.3 |

### S4.1.4  Multi-process scaling of the VASP MLPs and of the MTPs

In the main manuscript, we discussed in the context of Fig. 7 that a major reason why the VASP MLPs are slower than the MTPs is the unfavorable parallelization over multiple cores of a processing node. The values in the main manuscript are given for 64 cores contained in a dual socket AMD EPYC 7713 (Milan) node (i.e., using half of the node). Fig. S32 shows the required CPU time of a level 22 MTP, a VASP MLP trained with 4 atom types and a VASP MLP trained with 7 atom types for a 4×4×4 conventional supercell of MOF-5 (27136 atoms) starting from just using a single core up to the full 128 cores available on each node. It can be seen that in this comparison the difference in speed between the VASP MLPs and the MTPs is rather small (VASP MLPs with 4 species only about 50% slower) up to using 16 cores. However, the performance deteriorates for more cores. The improved parallelization of LAMMPS when performing molecular dynamics simulations is not particularly surprising, as the code was developed solely for performing such calculations. Conversely, VASP is primarily a DFT code. In passing we note that especially for the VASP MLP the scaling gets even much worse when using smaller unit cells.

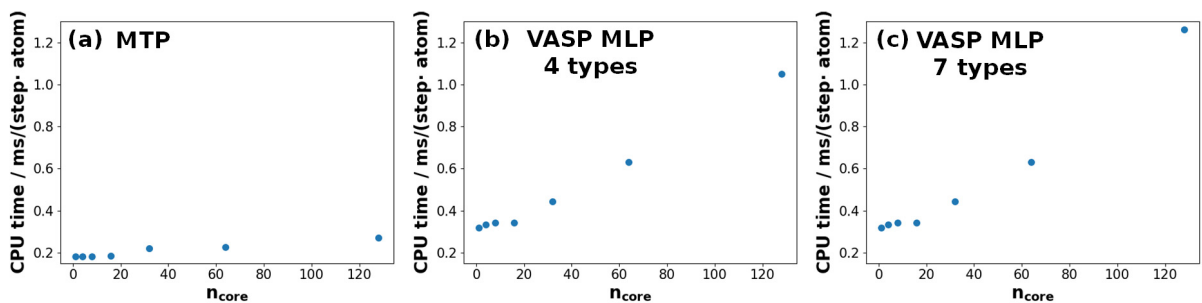

*Fig. S32: CPU time per step and atom as a function of the number of cores used tested. The tests have been performed for a level 22 MTP in a, a VASP MLP using 4 different atom types in b and a VASP MLP using 7 different atom types in c. The test was performed on a dual socket AMD EPYC 7713 (Milan) node for a 4×4×4 supercell of the conventional MOF-5 unit cell comprising 27136 atoms in an NPT ensemble.*

## S4.2 Thermal conductivity

### S4.2.1 Non-equilibrium molecular dynamics

To compute the thermal conductivity, non-equilibrium molecular dynamics (NEMD) simulations were performed for MOF-5. Before the NEMD simulations were conducted, the simulation box, which is based on the unit cell parameters at 300 K obtained from the previously described NPT simulations, was equilibrated at the target temperature of 300 K in an NVT ensemble for 25.5 ps (using a Langevin thermostat with a damping parameter of 100 fs). Afterwards, additional time steps were performed until the concurrent time step showed a total energy corresponding to the average energy during the preceding 25 ps. This is done to prevent any adverse effects due to temperature oscillations caused by the thermostat and to precisely reach a temperature of 300 K for the rest of the simulation. Then, the thermostat was turned off and the simulation was continued in an NVE ensemble. To perform the actual NEMD simulation, a temperature gradient was imposed on a super cell of the material substantially elongated in one direction which leads to a heat flux from the hot to the cold region. The thermal conductivity can then be obtained after reaching steady state by employing Fourier's law. Due to the limited size of simulation boxes, there are still significant finite size effects present in this type of simulation, massively affecting the result [29]. These can be corrected to some extent by performing NEMD simulations for several different cell lengths to obtain a length-dependent thermal conductivity. The inverse of that length-dependent thermal conductivity can then be related to the inverse of the length. The thermal conductivity in the infinite size limit is then obtained by performing a linear fit through the available data points and obtaining the intercept value at an inverse length of 0.

As described already in the main manuscript, due to the sizable computational cost of the slowly converging NEMD simulations, lower level MTP (level 18) was used to carry out the entire finite size correction procedure. To show that the results for the slightly less accurate potential are consistent, three cell sizes were also computed using the same level 22 MTPs, which were used in the other parts of in this work. The required temperature gradients were created employing the Müller-Plathe heat exchange algorithm [30]. Here, the simulation box is divided into N slabs that correspond to twice the number of conventional unit cells of MOF-5 so that each slaps encompasses exactly 1 node and 1 linker. The first slab defines the cold region and the (N/2+1)st slab defines the hot region. Periodic boundary conditions were applied in all directions. During the simulation every 4800 time steps the kinetic energies of the 16 slowest (coldest) atoms in the hot slab were swapped with the kinetic energies of the 16 fastest (hottest) atoms in the cold slab. The reason for the infrequent energy swapping is to minimize the slight energy drift emerging throughout the simulation. A center-of-mass motion emerging in the system due to the arbitrary directions of the modified velocities of the atoms due to the Müller-Plathe algorithm is prevented by uniformly rescaling the velocities of all the atoms every 300 time steps while maintaining the relative velocities of the atoms relative to each other. The simulations were then performed for 5 ns in which the temperature profile over the simulation box and also the exchanged energy was recorded. The temperature profile and average heat flux for such a run using the level 18 MTP for an 8×2×2 supercell of the conventional unit cell (207×52×52 Å) of MOF-5 is depicted in Fig. S33. The heat flux, obtained from the exchanged energy from the Müller-Plathe algorithm, was averaged starting at the end of the total simulation time (so that, for example, the value at 10 % of the time is averaged over the remaining 90 %) to visualize and allow the easy exclusion of the initial time period, where the simulation has not reached steady state yet. This can clearly be seen by the different slope of the heat flux within the first ns of simulation time. Then, the actual value of the heat flux was determined by averaging over the last 3.5 ns of the simulation time. Fig. S33 (b) also reveals a step in temperatures at the boundaries between the thermalized regions and the bulk of the material. This arises due to phonon scattering at these boundaries. To obtain the thermal conductivity there are two approaches to determine the temperature gradient: to use the temperature gradient in the linear region in the bulk of the material or to use the absolute temperature difference between the thermalized regions. The former method has been traditionally employed in the majority of the literature [29] while the latter has recently been shown to be more consistent with results from more sophisticated methodologies that do not suffer from scattering at the thermostat boundaries, like the HNEMD (homogeneous non-equilibrium molecular dynamics) method [31]. This approach was tested for a slightly different style of NEMD simulation using two thermostats (Langevin or Nosé-Hoover) and was shown to be particularly accurate with the Langevin thermostat. However, there is currently a technical issue of the implementation of the energy tallying using the MLIP-lammps interface in conjunction with the Langevin thermostats leading to nonsensical energy tallying in LAMMPS. Therefore, as described above, the Müller-Plathe algorithm was instead chosen to generate the temperature difference. Since the scattering effects at the thermostat boundaries were similar as what we already observed for the Langevin approach using traditional force field potentials [32], we still started by using the method of Li et al. [31] to obtain the temperature gradient, even though that method was specifically developed for Langevin or Nosé-Hoover thermostats.

Using this approach, we obtained the thermal conductivity values for the level 18 and level 22 MTPs, whose inverse are shown in Fig. S34 in tandem with the extrapolation to the infinite size limit. This led to a thermal conductivity of 0.260 ± 0.011 W/mK for the level 18 MTP and 0.261 ± 0.002 W/mK for the level 22 MTP. Here, the provided error values reflect only the statistical standard error of the linear fit and should by no means be treated as the total error of the simulation. These values are somewhat below the experimental value of 0.32 W/mK [33] and also below the value of 0.29 W/mK we obtained with our traditional MOF-FF potentials [32].

A possible reason for the underestimation of the thermal conductivity in the present calculations could be that the approach by Li et al. [31] is not well justified for the Müller-Plathe algorithm. Therefore, as a second approach, we also determined the temperature gradient in the region of constant slope in the plots of the position dependent averaged temperatures away from the thermalized regions, which has been the more traditional approach for NEMD simulations. For the level 18 MTP this resulted in a finite size corrected thermal conductivity of 0.32 ± 0.02 W/mK and for the level 22 MTP of 0.33 ± 0.02 W/mK. Again, both MTPs agree well with each other, and the values are now extremely close to the experimentally obtained value at room temperature.

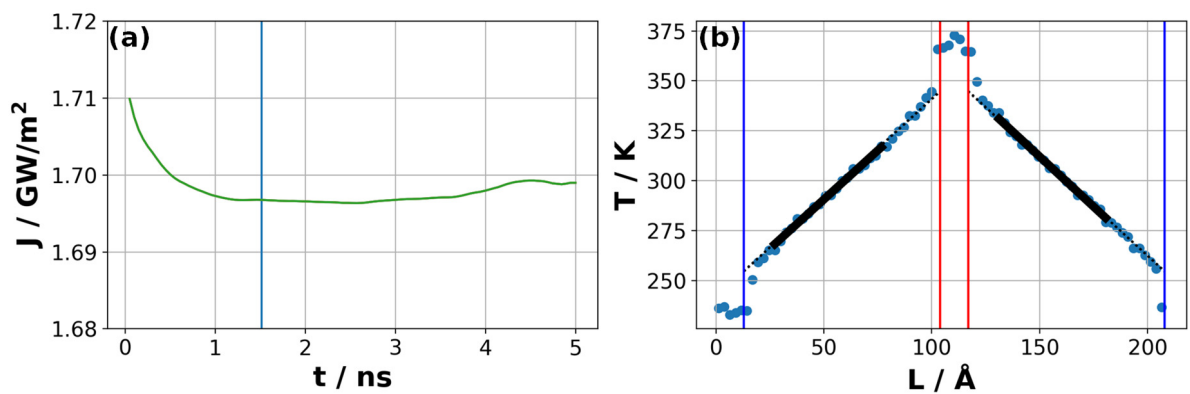

*Fig. S33: Evaluation of a NEMD simulation for a 8×2×2 supercell of MOF-5 using the level 18 MTP trained on the initial reference data set. The evolution of the averaged heat flux as a function of time*

*is depicted as the green line in a, where here the average was formed over all values starting at the end of the simulation (so that, for example, the value at 10 % of the time is averaged over the remaining 90 %). This serves to easily determine the amount of time it takes to reach the steady state. The blue vertical line indicates the time at which the averaged heat flux value was taken that was actually used for the evaluation of the thermal conductivity in Fourier's law (i.e., the heat flux was averaged between 1.5 ns and 5ns). In b the time averaged temperature profile is shown. The red vertical lines enclose the hot region and the blue vertical lines the cold region where the Müller-Plathe algorithm was invoked.*

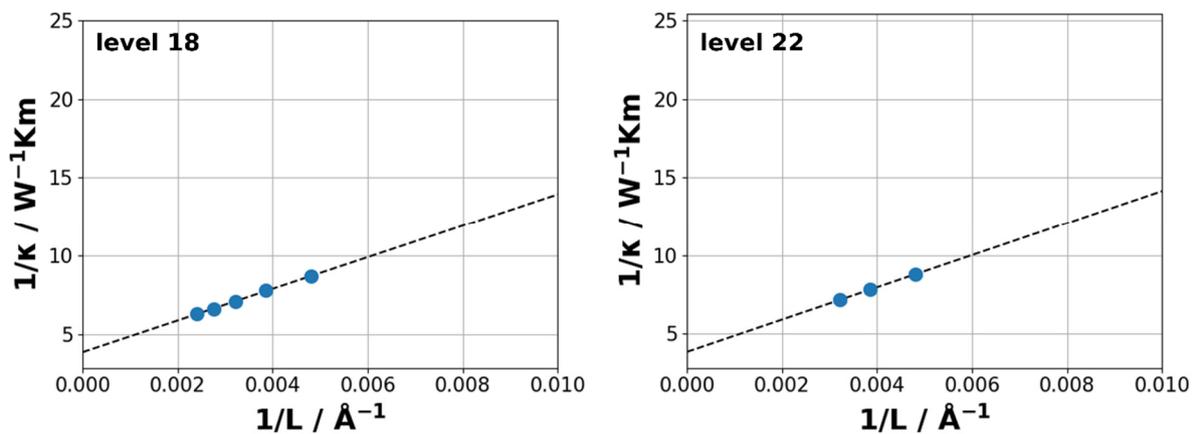

*Fig. S34: Extrapolation of the inverse thermal conductivity to the infinite size limit for MOF-5 using level 18 and level 22 MTPs to carry out the individual NEMD simulations. The supercells used for the level 18 MTP are 8, 10, 12, 14, 16×2×2 and for the level 22 MTP 8, 10, 12×2×2 times the conventional cubic unit cells of MOF-5.*

## S4.2.2 Approach to equilibrium molecular dynamics

Additionally, approach to equilibrium molecular dynamics (AEMD) simulations were performed for a level 18 MTP of MOF-5. Conceptually, the approach is rather similar: one uses long supercells (we used conventional supercells of MOF-5 ranging from 4×2×2 to 20×2×2) of which one half is heated, while the other half is cooled down compared to the target temperature. Convergence of the AEMD simulations was reached (meaning no difference in the evaluated thermal diffusivity as a function of time) at a simulation time of 1 ns for all system lengths. For cells smaller than 35 nm, we considered the temperature evolution over 0.5 ns, as the equilibrium was reached substantially faster. The instantaneous temperature difference $\Delta T$ between the halves of the system is tracked as a function of time $t$. Generally, the temperatures were computed classically from the average kinetic energies of the atoms contained in the respective halves. This evolution is plotted in Fig. S35. Subsequently the following function is fitted to obtain the thermal diffusivity $\bar{\kappa}$[34]:

$$\Delta T(t) = \sum_{n=1}^{\infty} C_n e^{-\alpha_n^2 \bar{\kappa} t}$$

with

$$C_n = 8(T_1 - T_2) \frac{\left[\cos\left(\frac{\alpha_n L_z}{2}\right) - 1\right]^2}{\alpha_n^2 L_z^2}$$

and

$$\alpha_n = \frac{2\pi n}{L_z}.$$

Here, $L_z$ is the length of the unit cell, $T_1$ and $T_2$ are the starting temperatures of each half of the unit cell and $n$ is the order of the exponential. Since it is impossible to fit this curve up to an infinite order, it is important to check the convergence of the series expansion as is done in Fig. S36. There, one can see that the dependence on exponential order is extremely small. Nonetheless, we use an order of 10 (i.e., including n between 1 and 10) for all our calculations to be on the safe side.

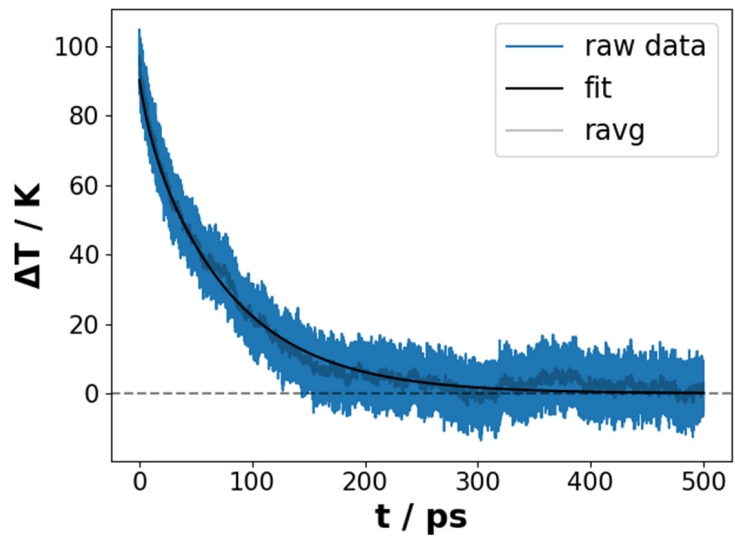

Fig. S35: Evolution of the temperature difference of the hot half and the cold half of a system during an AEMD simulation for an 8×2×2 conventional supercell of MOF-5. The blue line represents the raw data tracked at every time step, the grey line is the running average over the raw data using the closest 500 data points and the black line is the fit to obtain the thermal diffusivity.

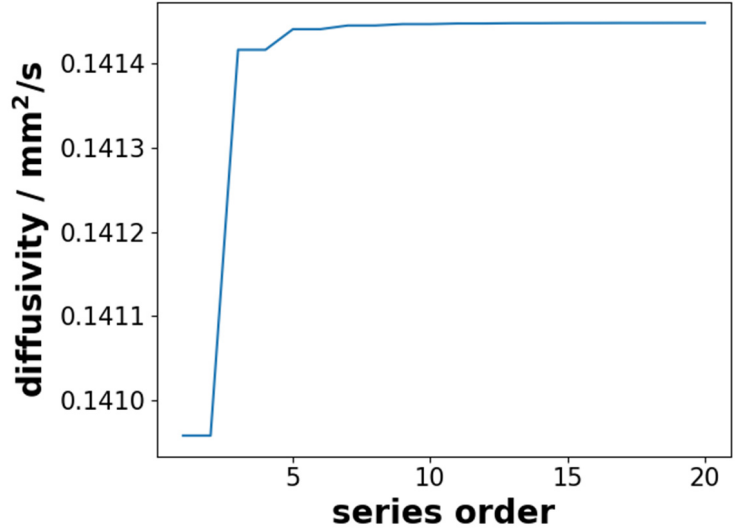

*Fig. S36: Dependence of the diffusivity for an 8×2×2 conventional supercell of MOF-5 on the exponential series order used to perform the fit over the temperature difference in an AEMD simulation.*

From the fit, we obtain the thermal diffusivity $\bar{\kappa}$ for each supercell. From the thermal diffusivity, the thermal conductivity $\kappa$ can be obtained via

$$\kappa = \frac{\bar{\kappa} C_p}{V},$$

with $C_p$ being the heat capacity and $V$ the volume of the supercell. Since we are doing classical MD simulations, we will always be in the Dulong-Petit limit, so the heat capacity at constant pressure when neglecting thermal expansion is

$$C_p \approx 3 N_A k_B$$

With $k_B$ the Boltzmann constant and $N_A$ the number of atoms in the system. Technically quantum corrections can now be applied due to the limited phonon occupation at the target temperature. The Dulong-Petit limit is definitely far from being reached in any MOF at 300 K. However, only the low frequency modes significantly contribute to the thermal conductivity due to the generally flat bands in the high frequency region (meaning a low group velocity) and the massive decay of phonon lifetimes at higher frequencies in many materials. Such a decay has previously observed in MOFs [35]. As it is mostly the high-frequency phonons that would be affected by quantum effects at higher temperatures, employing purely classical considerations appears reasonable.

As a final step, also in the AEMD simulations finite size corrections have to be employed. The linear finite size correction we used in NEMD [29],

$$\frac{1}{\kappa(L)} = \frac{1}{\kappa_{\infty, linear}} \left( 1 + \frac{\lambda}{L} \right)$$

where $L$ is the length of the simulation box in heat transport direction, $\lambda$ is a fitting parameter and $\kappa_\infty$ is the thermal conductivity in the infinite size limit, was originally justified due to scattering at the thermostat boundaries. However, there are no thermostat boundaries in an AEMD simulation. Therefore, Zaoui et al. developed a new finite size correction scheme[36], which is derived from the Boltzmann transport equation:

$$\kappa(L) = \kappa_{\infty,Zaoui}\left(1 - \sqrt{\frac{\Lambda}{L}}\right).$$

Here $\Lambda$ is also a fitting parameter. We employed this approach to obtain the thermal conductivity in the infinite size limit and the fitting curve with all the data points for different supercell sizes can be seen in Fig. S37. Here, we also added the linear fitting approach in the visualization. The results are actually extremely similar with $\kappa_{\infty,linear} = 0.302$ W/mK and $\kappa_{\infty,linear} = 0.317$ W/mK.

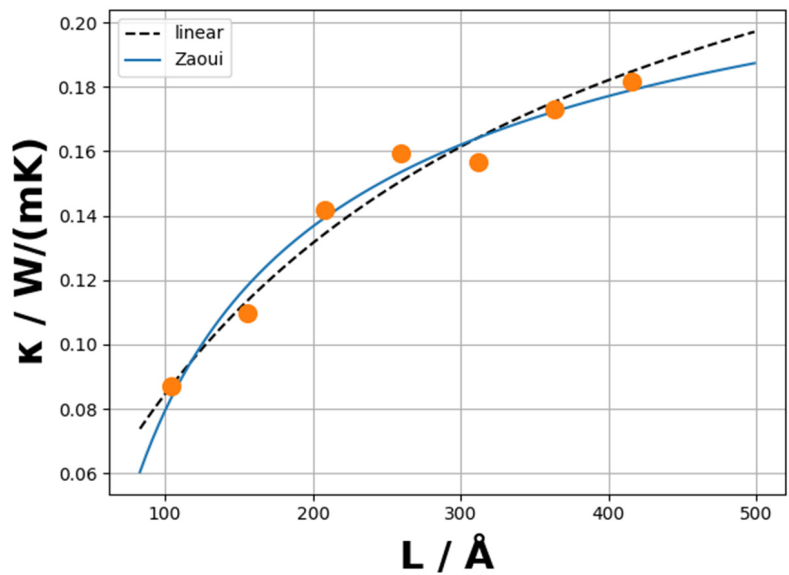

Fig. S37: Extrapolation of the thermal conductivity to the infinite size limit for the AEMD simulations of MOF-5 using the (inversely) linear extrapolation approach (black dashed line) and the approach by Zaoui et al. which was specifically developed for AEMD (blue line). The orange points represent the results from the individual AEMD runs.

**S4.2.3 Widths of the supercells in the heat-transport simulations**

In the above discussion, we were primarily concerned with the effect of the cell length. However, the cell thickness perpendicular to the heat transport direction is also important to converge NEMD and AEMD simulations [32]. We already discussed the convergence for NEMD and MOF-5 in [32]. There we found that using ×2×2 conventional supercells (×52×52 Å) is sufficient. However, we are now using a different force field and also the alternative AEMD approach. It would be reasonable to assume that AEMD shows the same convergence behavior as NEMD due to the similarity of the approaches. Additionally, since the MTPs used in this work actually have a shorter cutoff for the atom-atom interactions than the MOF-FF potentials used in ref. [32], it is unlikely that the difference in convergence behavior is very different. Especially given that the MOF-FF potentials already modeled the phonons in MOF-5 very well.

Nonetheless, to be on the safe side, we perform some convergence tests for using thicker cells perpendicular to the transport directions in AEMD simulations. Table S20 shows the thermal conductivities from AEMD simulations carried out on 16×2×2, 16×4×4, 32×2×2 and 32×4×4 supercells. For both of the cell lengths the difference in thermal conductivities between the thinner and thicker cells is extremely small. This is why we conclude that a 52 Å thick 2×2 cell is also sufficient for the AEMD simulations for MOF-5.

However, it should be mentioned that recently Green-Kubo simulations obtained a much higher value for the thermal conductivity of MOF-5 with 0.61 W/(mK) [37] also using highly accurate machine learned potentials. There, they needed to use 5×5×5 supercells to achieve convergence, but this was the case for Green-Kubo (equilibrium molecular dynamics) [29] and homogeneous non-equilibrium molecular dynamics [38] approaches, which are fundamentally different from NEMD and AEMD simulations. As can be seen in Table S20, the difference in the results between their and our work cannot be explained by different cell thicknesses. The exact origin of this difference needs to be investigated further and it could be connected to difference between non-equilibrium and equilibrium based methods.

*Table S20: Thermal conductivities $\kappa$ obtained from AEMD simulations for MOF-5 for different supercell lengths $n_{fluxdir}$ (in Å: $L_{fluxdir}$) in heat transport direction and for different supercell thicknesses $n_{perpend.}$ (in Å: $L_{perpend}$) in both perpendicular directions.*

| $n_{fluxdir}$ | $L_{fluxdir}$ / Å | $n_{perpend.}$ | $L_{perpend.}$ / Å | $\kappa$ / W/(mK) |
|---|---|---|---|---|
| 16 | 415.5 | 2 | 51.9 | 0.182 |
| 16 | 415.5 | 4 | 103.9 | 0.185 |
| 32 | 831.0 | 2 | 51.9 | 0.240 |
| 32 | 831.0 | 4 | 103.9 | 0.239 |